\documentclass{elsart}

\usepackage{graphicx} 
 \usepackage{epsfig}
\pdfoutput=1
\usepackage{amsfonts,amssymb,amsmath}
\usepackage{graphicx}

\usepackage{multicol}
\usepackage{mathrsfs}
\usepackage{latexsym}
\usepackage{alltt}
\usepackage[small,bf]{caption}
\usepackage{url}
\usepackage{array}
\usepackage{subfigure}
\usepackage{dcolumn}
\usepackage{color}
\usepackage{url}

\newcommand{\nc}{\newcommand}
\newtheorem{Def}{Definition}
\nc{\bseq}{\begin{subequations}}
\nc{\eseq}{\end{subequations}}
\nc{\be}{\begin{equation}}
\nc{\ee}{\end{equation}}
\nc{\red}[1]{\textcolor{red}{#1}}
\nc{\Veps}{\mathcal E}
\nc{\F}{\mathcal F}

\nc{\cale}{\mathcal{E}}
\nc{\calv}{\mathcal{V}}
\nc{\bq}{\begin{equation}}
\nc{\eq}{\end{equation}}
\nc{\p}{\partial}

\begin{document}

\begin{frontmatter}



\title{A discontinuous Galerkin method for the Vlasov-Poisson system}

\author[label1]{R. E. Heath,}
\author[label2]{I. M. Gamba,}
\author[label3]{P. J. Morrison,}
\author[label4]{and C. Michler}
\address[label1]{Applied Research Laboratories, University of Texas at Austin, TX 78758, USA. {\em Present Address:}
GE Global Research,
1 Research Circle,
Niskayuna, NY 12309; 
Email: heathr@ge.com
}
\address[label2]{Department of Mathematics and ICES, University of Texas at Austin, TX 78712, USA;  Email: gamba@math.utexas.edu}
\address[label3]{Department of Physics and Institute for Fusion Studies, University of Texas at Austin, TX 78712, USA; Email: morrison@physics.utexas.edu}
\address[label4]{Institute for Computational Engineering and Sciences (ICES)
The University of Texas at Austin
Austin, TX 78712
U.S.A., {\em Present Address:}
Computing Laboratory,
The University of Oxford,
Wolfson Building, Parks Road
Oxford OX1 3QD,
U.K.;
Email:  christian.michler@comlab.ox.ac.uk}


\begin{abstract} A discontinuous Galerkin method for approximating the Vlasov-Poisson system of
equations describing the time evolution of a collisionless plasma is proposed.  The method is mass
conservative and, in the case that piecewise constant functions are used as a basis, the method
preserves the positivity of the electron distribution function and weakly enforces continuity of the electric 
field through mesh interfaces and boundary conditions.  The performance of the method is
investigated by computing several examples and error estimates associated  system's approximation are stated.  
In particular, computed results are benchmarked
against established theoretical results for linear advection and the phenomenon of linear Landau damping
for both the Maxwell and Lorentz distributions.  Moreover, two nonlinear problems are considered:
nonlinear Landau damping and a version of  the two-stream instability are  computed.  For the latter,   
fine scale details of the  resulting long-time BGK-like state are presented.  Conservation laws are examined and various 
comparisons to theory are made.  The results  obtained
 demonstrate that the discontinuous Galerkin method is a viable option for integrating the Vlasov-Poisson
system.
\end{abstract}


\end{frontmatter}

\section{Introduction}
 \label{Sect:Intro}

The single species Vlasov-Poisson system is a nonlinear kinetic system that models time
evolution of a collisionless plasma consisting of electrons and a uniform background of fixed ions
under the effects of a self-consistent electrostatic field and possibly an externally supplied field.  The
Vlasov equation models the transport of the electrons that are coupled to 
the electrostatic potential through Poisson's equation.  

The unknown electron distribution function,
a phase space density,  is denoted by $f=f(x,v,t),$ where the independent variables $x, v, t$ are
position, velocity, and time, respectively.  For a given $t$, the quantity $f(x,v,t)\,dxdv$ denotes the number
of electrons contained  in the infinitesimal phase space volume element $dxdv$ centered
about $(x,v)$ at time $t$.  Upon a proper renormalization, $f$ can be interpreted as a probability
distribution function for the electrons over the phase space.

The Vlasov-Poisson system  
has solutions that can exhibit a  variety of dynamical phenomena \cite{vk,balescu,Lifs-Pita}.  
One of the most well-known effects is filamentation or `phase mixing'  as it is sometimes called, 
which occurs when different characteristics surfaces associated to the nonlinear transport (Vlasov) 
equation wrap in the phase space.  This effect results in stiff gradients, since $f$ generally takes on
disparate values along different characteristics.   

Related to filamentation is the  interesting phenomenon of Landau damping \cite{Landau}:  
electrostatic disturbances can be interpreted as  the interaction between plasma waves and  electrons with a  resulting net energy transfer from the wave to  electrons leading to an exponential collisionless damping of the  electric field modes in time.  Because of such phenomena,  the Vlasov-Poisson equation can be challenging to simulate numerically.

Existing numerical techniques for solving the Vlasov-Poisson system can be divided into
two groups: (i) those that approximate the system in the phase space directly
and (ii) those that transform the system into a different coordinate space.
The numerical approaches that treat the phase space directly do not, however, usually involve
discretizing the Vlasov equation.  Rather most of these methods
take advantage of the characteristic structure of the Vlasov equation, which
implies that the plasma particles evolve along trajectories that
satisfy a given set of ordinary differential equations.  The most widely used
particle scheme is the Particle-in-Cell (PIC) method \cite{BirLan,HocEas,Daw},  which
represents ensembles micro-particles as a finite number of macro-particles.  Each
macro-particle is then assumed to evolve along a characteristic trajectory, where the
electric field defining the trajectory is computed via any standard scheme.
The PIC method seems to give reasonable results in cases where the tail of the
distribution is negligible and a large number of particles is not necessary.  Otherwise, the
method suffers from numerical noise that is proportional to $1/\sqrt{n},$ where
$n$ is the number of particles.

Other methods based on the discretization of the phase space have also been proposed.  In \cite{CheKno},
an operator splitting method was introduced and shown to be both efficient and accurate
for solving a wide range of problems.  A continuous finite element method was developed in
\cite{ZakGarBoy1,ZakGarBoy2} and was shown to achieve results  similar to  those obtained in
\cite{CheKno}.  A positive and mass conservative scheme was employed
in \cite{FilSonBer}  to solve both linear and nonlinear
damping problems  in one- and two-dimensional physical space.  This method is defined at a given time step by first building a piecewise constant approximation over a mesh of the phase space using the approximation obtained from
the previous time step and two correction terms whose values are found by solving two
fixed point problems.  The piecewise approximation is then used in conjunction with a slope limiter
to reconstruct a local polynomial approximation of $f$ for each cell in the mesh.

Transform methods based on Fourier or Hermite series have also been used, e.g.\ in \cite{GraFei,armstrong,DenKru} and more recently in
\cite{Eli}.  In \cite{Kli,KliFar}, the phase space was transformed using a Fourier transform and a splitting method was employed to
advance the approximation in time.  This method also included a filamentation filtering
step for the purpose of smoothening the filaments.  The numerical results obtained using this
method seem reasonable only for problems in which filamentation  is not a dominant effect,
where perhaps the nonlinearity either slows down or prevents the onset of filamentation.  However, this method may be  inadequate for
problems where the physics of interest depends upon the filamentary nature of the distribution, such as is the case for Landau damping.

The objective of this paper is to propose a coupled Upwind Penalty Galerkin (UPG) method for the approximation
of the Vlasov-Poisson system and to evaluate its numerical efficacy.   Our UPG method gives a unified approach for approximating both the hyperbolic and
elliptic parts of the Vlasov-Poisson system.  Specifically, the Vlasov equation is discretized using the standard Upwind 
Galerkin (UG) scheme for conservation laws \cite{CocShu1,CocLinShu,CocHouShu}  and the Poisson equation
is discretized using one of the three Discontinuous Galerkin (DG) interior penalty available schemes \cite{Wheeler,RivWheGir,Sun}. 
Stability and convergence estimates for the UPG method were presented in \cite{Heath} for the six-dimensional phase space
In the same reference, the method was shown to be both locally and globally mass conservative.

More specifically, the semi-discrete UPG approximation  $f_h$  to the 
electron distribution function  is defined to be the solution 
of a first-order, nonlinear, ordinary differential equation (ODE) system.  
Moreover, it has been shown that the method  preserves positivity of $f_h$  
when piecewise constants basis functions are used to approximate the solution
to the Vlasov-Poisson system.

In this manuscript we show the numerical efficacy of the DG method by performing accuracy, convergence, 
and conservation tests on computed UPG approximate solutions to a variety of linear and nonlinear  problems where sufficiently data or information of the true nonlinear 
solutions has been established.  The computed results for these
problems are benchmarked against known theoretical results and are compared to results obtained using
established methods.

For  computing plasma problems the UPG   method offers several advantages. In particular, the local nature of the method facilitates adaptive mesh refinements  with  easy adaptation to parallelization techniques. 
By taking advantage of these benefits,   regions of the phase space where the electron distribution experiences strong filamentation or boundary layer effects can be resolved by local mesh refinements.  The discontinuous nature of the method also helps to resolve the stiff gradients associated with filamentation, since requiring the approximation to be continuous in these cases  can be too restrictive and   typically lead to excessive numerical diffusion and oscillatory behavior. Due to the fact that the method imposes boundary conditions weakly, a variety of boundary conditions can easily be accommodated. 
 
Recently,  an alternative DG formulation for the Boltzmann-Poisson system was introduced in  \cite{CGMS-1, CGMS-2, CGMS-cmame2009,CGMS-3, CGMS-4} for simulations of hot electron transport for one and two space  dimensional nanometer  scale devices,  where kinetic corrections are known to be very significant.  In \cite{Cheng-gamba-proft-10}  a DG scheme was constructed for the Vlasov-Boltzmann equation  by means of a maximum principle satisfying a limiter for conservation laws \cite{zhang-shu10-1,zhang-shu10-2}, and it was shown to be high order accurate  and  positivity-preserving, not only for piecewise constant basis functions  but also for higher order polynomial approximations.  Consequently a new approximation of the Vlasov   system based on a UPG scheme for higher order polynomial basis functions is  currently being developed and tested  \cite{cheng-gamba-morrison11,cheng-gamba-11}.
 Thus, the DG method is well-suited to approximate a range of plasmas spanning from the collisionless to the highly collisional regimes.

For completeness of this introduction we include some historical remarks regarding DG  methods. The initial development of these numerical approximating schemes for hyperbolic and elliptic equations occurred independently, but nearly
simultaneously.  In 1973, the first DG scheme for linear hyperbolic equations was introduced by Reed and Hill for
approximating  a neutron transport equation \cite{ReeHil}.  This work was followed by Lesaint and Raviart \cite{LesRav} in 1974, where \emph{a priori} error estimates were proved for the DG method applied to two-dimensional, linear
hyperbolic problems.  The DG schemes for hyperbolic problems were further studied in the series of
papers \cite{CocShu1,CocLinShu,CocHouShu}, which culminated in the introduction of the local
discontinuous Galerkin (LDG) method \cite{CocShu2}.  The generality of the LDG
method was further extended to the multidimensional setting under more relaxed assumptions on the data
in \cite{CocDaw}.
One of the first DG schemes for approximating the solutions to second-order elliptic equations was introduced
in 1971 by Nitsche \cite{Nitsche} where Dirichlet boundary conditions were enforced weakly
rather than strongly through the use of a penalty term.  Shortly thereafter,
applications of the penalty method to Laplace's equation were proposed by Babu\v{s}ka et al.\  in \cite{Babuska1,Babuska2,BabZla}.  

The use of penalty terms across interior
faces as a means of enforcing continuity among adjacent elements was introduced in
\cite{Wheeler} and \cite{WhePer} using a symmetric interior penalty Galerkin
(SIPG) finite element method.  A non-symmetric interior penalty method (NIPG) similar to
the SIPG method was proposed and analyzed in \cite{RivWheGir}.  The incomplete interior penalty
method (IIPG) was introduced in \cite{Sun,DawSunWhe,SunWhe} and is very similar to the SIPG and
NIPG methods.

The outline of this paper is as follows.  In Section \ref{sect:VP}, we describe the Vlasov-Poisson system and give a brief discussion of Landau damping.  In Section \ref{sect:MF}, the UPG method for the approximation of the Vlasov-Poisson system is derived and the error estimates associated to the approximation of the system are stated. In Section \ref{sect:NR}, several numerical simulation results are presented and analyzed, including a study of the free streaming operator (simple advection), Landau damping
for Maxwellian and Lorenztian equilibria, strong nonlinear Landau damping and a careful study of a  symmetric two-stream instability, all for the case of repulsive
potential forces.  In Section \ref{sect:conclusions},  we comment on the efficiency  of the UPG method, draw some conclusions, and  remark on  future work.  Finally,  an Appendix dedicated to
the analysis of dispersion relations for the Lorentzian and two-stream equilibria is included.

\section{The Vlasov-Poisson System}
 \label{sect:VP}

The Vlasov-Poisson system considered in this work has been scaled by the usual
characteristic time and length scales, i.e., time is scaled by the inverse plasma frequency~$\omega^{-1}_p$,
length by the Debye length~$\lambda_D$, and velocity accordingly
by a thermal velocity $v_{th}=\omega_p \lambda_D$.

Using this nondimensionalization, the Vlasov-Poisson problem is as follows:
For the divergence free vector in $(x,v)$ phase space
\begin{equation}\label{alpha-vector}
 \alpha(x,v,t) = (v,-E(v,t) ) \, \qquad  \mathrm{for}\ (x,v,t)\in   \Omega\times[0,T] \, ,
\end{equation}
find the electron distribution function $f(x,v,t)$ and the electric field $E(x,t)$ with corresponding  electrostatic 
potential $\Phi(x,t)$ such that, for fixed $T>0$,
\begin{subequations}
\begin{align}
0 &\ =\ \partial_t f + \mathrm{div}(\alpha\, f) \ =\ 
\partial_t f +v\cdot\nabla_x f - E\cdot\nabla_v f\, , &\ \text{in}\   \Omega\times(0,T]\,,\ \,
\label{eq:vlasov.full} \\
E  &\ =  -\nabla_x\Phi \, ; \ \ \   -\Delta_x \Phi \ =\  1 - \int_{\mathbb{R}^n}\,f\,dv\, , \ &\ \text{in}
\ \Omega_x\times(0,T] \,, \label{eq:poisson.full}
\end{align}
\end{subequations}
subject to an initial condition on $f$ and boundary conditions on $f$ and $\Phi.$  
The domain of definition of the initial boundary value problem  $\Omega:=\Omega_x\times\mathbb{R}^n$, where the physical domain
$\Omega_x\subset\mathbb{R}^n$ can either be bounded or all of $\mathbb{R}^n$ with $n=1,2,3$.
The boundary condition given for $f$ depends on $\Omega_x.$  If $\Omega_x=\mathbb{R}^n,$ then the
condition $f\rightarrow0$ must be enforced both as $|x|\rightarrow\infty$ and as $|v|\rightarrow\infty.$  If
$\Omega_x$ is bounded, then a condition must be imposed on $f$ along the inflow boundary $\Gamma_{_I},$ defined by
\begin{equation}
\Gamma_{_I}\ =\ \{ (x,v)\in\partial\Omega_x\times\mathbb{R}^n \,|\, v\cdot\nu_{_x}<0 \} \,,
\label{eq:inflow boundary}
\end{equation}
with $\nu_{_x}$ being the unit outward normal vector to $\partial\Omega_x.$  Often   $f_{_I}$ is given in some parts of the boundary and 
may be  periodic
in other parts of the boundary region  $\Gamma_{_I}$.  
In this manuscript we assume periodic boundary conditions in space and the
decaying boundary condition in velocity.  

The Poisson equation must also be endowed with spatial boundary conditions,  either Dirichlet, Neumann,  Robin,  or periodic, on  different disjoint regions of the boundary $\partial\Omega_x$.   We denote the Dirichlet portion of the boundary by $\partial\Omega_{x,D}$. 
If the measure of the Dirichlet boundary is zero, i.e.  $|\partial \Omega_{x,D}| =0$, and one has homogeneous or periodic boundary conditions such that   $\int_{\partial \Omega_x} \nabla\Phi\cdot \nu_x =0$, then in order to maintain a  well-posed problem that keeps the existence and uniqueness of the corresponding Poisson boundary value problem, one needs to add to the solution space the compatibility (neutrality) 
condition $\int_{\Omega_x}(1 - \int_{\mathbb{R}^n}\,f\,dv\,) dx =0$,  or equivalently  $ \int_{\Omega_x} \int_{\mathbb{R}^n}\,f\,dx\,dv
=|\Omega_x| $ on each connected component of the spatial domain $\Omega$.

Macroscopic fluid quantities of interest are easily computed from $f$.
The electron density $\rho=\rho(x,t)$, current density $j=j(x,t)$, kinetic energy density $\mathcal{E}_k(x,t)$ and electrostatic
energy density $\mathcal{E}_e(x,t)$ are defined by
\begin{align}
\rho(x,t)\ &=\  \int\limits_{\mathbb{R}^n} f(x,v,t)\,dv \,,
\label{eq:density}\\
j(x,t)\ &=\  \int\limits_{\mathbb{R}^n} vf(x,v,t)\,dv \,,
\label{eq:current}\\
\mathcal{E}_k(x,t)\ &=\  \frac{1}{2}\,\int\limits_{\mathbb{R}^n} |v|^2 f(x,v,t)\,dv \,,
\label{eq:kin.energy}\\
\mathcal{E}_e(x,t)\ &=\  \frac{1}{2}\,|E(x,t)|^2 \,.  \label{eq:pot.energy}
\end{align}
These quantities satisfy a number of  respective conservation laws (see e.g.\ \cite{Rein}).  In particular,
it is well-known that the Vlasov-Poisson system conserves total particle number, momentum, energy, and the Casimir invariants,  which are given, respectively,   by
\begin{align}
N\ &=\ \int_{\Omega_x\times{\mathbb{R}^n}} f(x,v,t) \, dx\,dv=\int_{\Omega_x}  \, \rho(x,t)\,dx\,,
\label{eq:number}\\
P\ &=\ \int_{\Omega_x\times{\mathbb{R}^n}}   vf(x,v,t)  \, dx\,dv=\int_{\Omega_x}  j(x,t)\,dx\,,
\label{eq:momentum}\\
H\ &=\  \frac{1}{2}\int_{\Omega_x\times{\mathbb{R}^n}}   |v|^2 f(x,v,t) \, dx\,dv + \frac{1}{2}\,\int_{\Omega_x}   |E(x,t)|^2 \, dx
\label{eq:energy}\\
\ & {\ }\  \hspace{.25 in} = \int_{\Omega_x}  \big( \mathcal{E}_k(x,t) + \mathcal{E}_e(x,t) \big) \, dx \,,
\nonumber\\
C\ &=\ \int_{\Omega_x\times{\mathbb{R}^n}}\,  \mathcal{C}(f) \, dx\,dv  \,,
\label{eq:casimir}
\end{align}
The notation  $\mathcal{C}(f) $ in (\ref{eq:casimir}) refers to an  arbitrary function of $f$ and  includes 
the  `enstrophy'  when $\mathcal{C}(f)=f^2$,  entropy when $\mathcal{C}(f)=-f\ln f$, or particle number, as in  
(\ref{eq:number}), when
 $\mathcal{C}(f)=f$.     
 We will check the invariance of these quantities  in our nonlinear computations, 
particularly in Section \ref{sec:Nonlin.two.stream}.

Interesting properties of the Vlasov-Poisson system result by considering a linear perturbation~$\delta f(x,v,t)$ to an
equilibrium distribution $f_{eq}(v)$ over the 2D-phase space $[0,L]\times(-\infty,\infty),\ L>0.$  Specifically, suppose
that $f = f_{eq} + \delta f$, where $f_{eq}$ is a given equilibrium probability distribution, $\delta f$ and $\Phi$
are $L$-periodic in $x$, and the initial average value of $\delta f$ over $\Omega$ is zero.
Equations (\ref{eq:vlasov.full})-(\ref{eq:poisson.full}) imply that $\delta f$ satisfies
\begin{subequations}
\begin{align}
\partial_t(\delta f)+v(\delta f)_x - E(\delta f)_v &\ =\ E f_{eq}' &[0,L]\times(-\infty,\infty)\times(0,T]\,,
\label{eq:vlasov.pb}\\
E &\ =\ -\Phi_x &[0,L]\times(0,T]\,,
\label{eq:efield.pb}\\
\Phi_{xx} &\ =\ \int_{\mathbb{R}^n}\,\delta f\,dv &[0,L]\times(0,T]\,. \label{eq:poisson.pb}
\end{align}
\end{subequations}
Supposing $|E(\delta f)_v|<<1$ and dropping this term from (\ref{eq:vlasov.pb}) leads to
\begin{subequations}
\begin{align}
\partial_t(\delta f)+v(\delta f)_x &\ =\ E f_{eq}' &[0,L]\times(-\infty,\infty)\times(0,T]\,,
\label{eq:lin.vlasov.pb}\\
E &\ =\ -\Phi_x &[0,L]\times(0,T]\,,
\label{eq:lin.efield.pb}\\
\Phi_{xx} &\ =\ \int_{\mathbb{R}^n}\,\delta f\,dv &[0,L]\times(0,T]\,.
\label{eq:lin.poisson.pb}
\end{align}
\label{eq:lin vp system}
\end{subequations}
$\!\!\!$The linear system of (\ref{eq:lin.vlasov.pb})-(\ref{eq:lin.poisson.pb}) was analyzed by using the Laplace transform in the
 famous paper of  Landau \cite{Landau},  by expansion in terms of continuum eigenfunctions in \cite{vk}, and by a tailored integral transform
introduced in \cite{mp,m}.   Landau showed  that an electric field mode $E_k(\omega,t)$ decays exponentially in the long-time limit,  which we investigate in  Section~\ref{sec:Lin.Landau} for  two well-known equilibria:  the  Maxwellian
\be
f_{M}=(2\pi T)^{-n/2}\,e^{-v^2/2T} \,,
\label{Meq}
\ee
and the Lorentzian,
\be \label{Leq}
f_{L}(v) =\ \frac{1}{\pi}\,\frac{\gamma}{v^2  + \gamma^2} \,.
\ee


\section{Method Formulation}
\label{sect:MF}

In this section, we derive the UPG method for  the Vlasov-Poisson system.  The derivation 
proceeds by first discretizing the Vlasov equation using the standard UG  discretization for
transport equations \cite{CocDaw}.  Here it is assumed that the electric field is given and hence the divergence-free flow field   $\alpha(x,v,t)= (v,-E(x,t))$, defined in \eqref{alpha-vector} for the Vlasov  equation  \eqref{eq:vlasov.full},  is known.  Afterwards, the DG discretization for the Poisson equation is considered. 

It is assumed that any mesh for the phase space, where by mesh we mean a partitioning of the phase space into convex sets called
elements, is the Cartesian cross product of a mesh for the physical domain and a mesh
for the velocity domain.  Under this assumption, the physical and velocity domains can be independently
refined.  Given a mesh for the phase space, the UG method then defines an approximate solution to the true
Vlasov solution in such a way that at any given time the approximate solution restricted to each element of the mesh
is a polynomial function.  However, the approximate solution is not required to be continuous across the
intersections of any two adjacent elements, so it is a piecewise defined polynomial
function with respect to the mesh at any given time.

In order to compute the approximation to the electrostatic potential from  Poisson's equation \eqref{eq:poisson.full},  for a given distribution function,  one may use one of  three interior penalty methods that  weakly enforce both approximate continuity across the interior mesh faces and  Dirichlet boundary regions.   These three alternative methods, 
 symmetric interior penalty Galerkin (SIPG) \cite{Wheeler,WhePer},  non-symmetric interior penalty Galerkin (NIPG) 
 \cite{RivWheGir}, and 
incomplete interior penalty Galerkin (IIPG) \cite{Sun,SunWhe}  are discussed in detail and  \emph{a priori} error estimates for each of them are given in the respective references.  The only difference among the
three methods is in the value of one specific parameter that arises in the weak formulation that is common to each
of them.  
Thus, for a given space charge function,  each penalty Galerkin method  defines a piecewise polynomial
approximation of the true solution to the Poisson equation using a mesh in the spatial domain.

The spatial domain mesh used in the discretization of  Poisson's equation is required to be the same as that used in the UG  discretization of the Vlasov equation.  
This requirement is a practical one, both in terms of analysis and implementation.  However, the polynomial degree of the
potential approximation on a given element of the spatial mesh is not required to equal the degree, with respect
to $x$, of the polynomial approximation of the distribution $f$.

Thus, the UPG method of approximation to the Vlasov-Poisson system is defined by coupling together the UG method
of approximation to the Vlasov equation with interior penalty methods of approximation to the Poisson
equation.  This nonlinear semi-discrete approximation  results in a 
first-order  nonlinear  ODE system, the solution of which determines the approximation $f_h$ of $f$.  
The resulting ODE system is readily solved using
an explicit conservative time-integrator such as the Runge-Kutta method.  Moreover, in the process of computing $f_h$, both
the approximation $E_h$ of the electric field $E$ and the approximation $\Phi_h$ of the potential $\Phi$ are
computed by one of the penalty methods for the elliptic equation.

\subsection{Preliminaries}\label{prelim}

We assume the computational domain in velocity space is a bounded set $\Omega_v$ and that the approximate solution
 to $f(x,v,t)$ is assumed to vanish in $\partial\Omega_v$.
It is then, implicitly for our simulation,  assumed that the velocity support of the approximation to the true solution $f$ 
of the Vlasov-Poisson system is contained in
$\Omega_v$ for all times.  This is a reasonable assumption for problems with spatial periodic boundary conditions, as it is expected 
that most of the density   associated  with the  approximation to $f$ will be contained in a sufficiently large fixed set $\Omega_v$.
The error due to this assumption only depends  on the  density of the true solution 
$f$ computed on the complement of $\Omega_v$ in $\mathbb{R}^n$.

Further, the conservative nature of the transport equation \eqref{eq:vlasov.full} with the space dependent divergence-free flow field $\alpha$, motivates us to choose the computational scheme as follows:

Let $\{ \mathcal{T}_{h_x} \}_{h_x>0}$ be a sequence of successively refined meshes of the bounded domain
$\Omega_x\subset\mathbb{R}^n$ and let $\{ \mathcal{T}_{h_v} \}_{h_v>0}$ be a sequence of successively
refined meshes of the bounded domain $\Omega_v\subset\mathbb{R}^n$, where $n=1,2,3.$  
Given the meshes
$\mathcal{T}_{h_x}=\{K_{j_x}\}_{j_x=1}^{N_{h_x}}$ and $\mathcal{T}_{h_v}=\{K_{j_v}\}_{j_v=1}^{N_{h_v}}$,
the elements $K_{j_x}$ and $K_{j_v}$ comprising each of the respective meshes are sets of the following
types: intervals, if $n=1$; triangles or quadrilaterals, if $n=2$; and tetrahedra, prisms or hexahedra, if $n=3$.
The corresponding  spatial refinement level $h_x$ and
 velocity refinement level $h_v$ are defined by $h_x=\max_{j_x}\,\{\mathrm{diam}(K_{j_x})\}$ and
$h_v=\max_{j_v}\,\{\mathrm{diam}(K_{j_v})\}$,  respectively.

A sequence of successively refined meshes $\{ \mathcal{T}_h \}_{h>0}$ of the, now, computational domain
$\Omega=\Omega_x\times\Omega_v$ is generated by defining each mesh $\mathcal{T}_h=\{K_j\}_{j=1}^{N_h}$
to be $\mathcal{T}_{h}=\mathcal{T}_{h_x}\times\mathcal{T}_{h_v},$ where the refinement level  is   $h=(h_x,h_v)$.  Thus, for any given element $K_j\in\mathcal{T}_h$ there exists a unique pair of
elements $K_{j_x}\in\mathcal{T}_{h_x}$ and $K_{j_x}\in\mathcal{T}_{h_x}$ such that
$K_j=K_{j_v}\times K_{j_v}$, which is equivalent to the existence of an invertible mapping
from $j\in\{1,\ldots,N_h\}$ to $(j_x,j_v)\in\{1,\ldots,N_{h_x}\}\times\{1,\ldots,N_{h_v}\}$, where
$N_h=N_{h_x}N_{h_v}$.

The derivation of the following UPG method requires the use of the broken Sobolev space
$H^s\{ \mathcal{T}_h \}$, $s>1/2$, which is defined as follows:
\begin{equation} \label{def:brokensobolev}
H^s\{ \mathcal{T}_h \}\ =\ \{\,w\in L^2(\Omega)\ |\ w_{_{|K_j}}\in H^s\{K_j\},\ \forall\,K_j\in\mathcal{T}_{h}\, \} \,,
\end{equation}
i.e., $H^s\{ \mathcal{T}_h \}$ is the space of those functions that have elementwise weak derivatives up to, and
including, the order $s$.  Then for  nonnegative integers $r_x$ and $r_v$, the discontinuous approximation space
$D^{r_x,r_v}( \mathcal{T}_h )\subset H^s\{ \mathcal{T}_h \}$ is given by 
\begin{equation} \label{def:brokenapproximation}
D^{r_x,r_v}( \mathcal{T}_h )\ =\ \{\,w\in H^s\{ \mathcal{T}_h \}\ |\
w_{_{|K_j}}\in \mathbb{Q}^{r_x}\{K_{j_x}\}\times\mathbb{Q}^{r_v}\{K_{j_v}\},\ \forall\,K_j\in\mathcal{T}_{h}\, \} \,,
\end{equation}
where $\mathbb{Q}^r(K)$ denotes the space of polynomials on a set $K$ with degree less than or equal to $r$
in each variable.  Thus, $\mathbb{P}^r(K)\subset\mathbb{Q}^r(K)$, where $\mathbb{P}^r(K)$ denotes the
space of polynomials satisfying that the sum of the degrees of all the variables is less than or equal to $r$.

The choice of $\mathbb{Q}^r(K)$   for basis functions is suitable for Cartesian meshes in both 
 $x$-space and $v$-space,  respectively, where trace and inverse inequalities  that are derived by mapping 
to the reference element in the approximating framework are possible, 
as first introduced in \cite{Heath}.

However, one may use triangles in two-dimensions, and prisms or hexahedra in three-dimensions, 
for both $x$-space and $v$-space, 
for which the natural choice of 
polynomial space would be   $\mathbb{P}^{r_x}\{K_{j_x}\}\times\mathbb{P}^{r_v}\{K_{j_v}\}$. 
We point out that these selections of approximating spaces is consistent with the divergence free, 
linear,  and conservation  form of the Vlasov equation,  and the fact that  the choice of the mesh associated with the computational domain  $\Omega=\Omega_x\times\Omega_v$ is a   set of  product mesh elements $K_j=K_{j_v}\times K_{j_v}$,  for $j\in\{1,\ldots,N_h\}$,  $(j_x,j_v)\in\{1,\ldots,N_{h_x}\}\times\{1,\ldots,N_{h_v}\}$ and  $N_h=N_{h_x}N_{h_v}$.  Such  is clearly preserved by the Vlasov flow and the corresponding 
  approximation results for $\mathbb{P}^r(K)$ will be valid.   This choice may be preferable in 
higher dimensions since the number of degrees of freedom  of the basis functions in  $\mathbb{P}^r(K)$  is $rn +1$,  while  for $\mathbb{Q}^r(K)$ is $(r+1)^n$, 
for $n$-dimensional calculations.

The discontinuous nature of the space $H^s\{ \mathcal{T}_h \}$ needs the introduction of mesh faces.
If $K_j$ is a boundary element, then $F_k=\partial K_j\cap\partial\Omega$ is called a boundary mesh face.
If $K_1$ and $K_2$ are two intersecting elements whose
common intersection lies in the interior of $\Omega$, then $F_k=\partial K_1\cap \partial K_2$ is said to be an interior
mesh face.  The set of all mesh faces is denoted by $\mathcal{F}_{h}=\{F_1,\ldots,F_{P_h}, F_{P_h+1},\ldots,F_{M_h}\}$,
where $F_k$ is an interior face if $1\le k\le P_{h}$ and a boundary face if $P_{h}+1\le k\le M_{h}$.  Each face
$F_k\in\mathcal{F}_{h}$ is associated with a unit normal vector $\nu_{k}$.  For $k>P_{h}$, $\nu_{k}$ is chosen to
be the outward unit normal to $\partial\Omega$.  For $1\le k\le P_{h}$, we fix $\nu_{k}$ to be one of the two unit
normal vectors to $F_k.$  For every interior face $F_k,$ the elements $K_1$ and $K_2$ will always
be used to denote the two unique elements such that $F_k=K_1\cap K_2.$  Moreover, it is always assumed that
$K_1$ satisfies $\nu_{_{K_1}}=\nu_{k}$ on $F_k$, where $\nu_{_{K_1}}$ denotes the outward unit
normal to $\partial K_1.$  Then $\nu_{_{K_2}}=-\nu_{k}$ on $F_k$.

The fact that each mesh $\mathcal{T}_h$ is the  product of a spatial mesh $ \mathcal{T}_{h_x}$ and a
velocity mesh $\mathcal{T}_{h_v}$ gives a specific structure to the boundaries of the elements and to the
set of mesh faces $\mathcal{F}_{h}$.  It follows that for $K_j=K_{j_x}\times K_{j_v}$ we have  
$\partial K_j=(\partial K_{j_x}\times K_{j_v})\cup(K_{j_x}\times \partial K_{j_v})$.  We denote the set of mesh
faces for $\mathcal{T}_{h_x}$ and $\mathcal{T}_{h_v}$ by
$\mathcal{F}_{h_x}=\{F_1^x,\ldots,F_{P_{h_x}}^x, F_{P_{h_x}+1}^x,\ldots,F_{M_{h_x}}^x\}$ and
$\mathcal{F}_{h_v}=\{F_1^v,\ldots,F_{P_{h_v}}^v, F_{P_{h_v}+1}^v,\ldots,F_{M_{h_v}}^v\}$, respectively,
where $F_{k_x}^x$ is an interior face if $1\le k_x\le P_{h_x}$ and a boundary face if
$P_{h_x}+1\le k_x\le M_{h_x}$, and $F_{k_v}^v$ is an interior face if $1\le k_v\le P_{h_v}$ and a boundary
face if $P_{h_v}+1\le k_v\le M_{h_v}$.  Then,  given any arbitrary $F_k\in\mathcal{F}_{h}$, either there exist
an $F_{k_x}^x\in\mathcal{F}_{h_x}$ and a $K_{j_v}\in\mathcal{T}_{h_v}$ such that $F_k=F_{k_x}^x\cup K_{j_v}$
or there exist a $K_{j_x}\in\mathcal{T}_{h_x}$ and an  $F_{k_v}^v\in\mathcal{F}_{h_v}$ such that
$F_k=K_{j_x}\cup F_{k_v}^v$.

When considering functions in $H^s\{ \mathcal{T}_h \}$, we use the usual average and jump operators, 
respectively, 
 defined for $w\in H^s\{ \mathcal{T}_h \}$ along an interior face $F_k$ by
\begin{equation} \label{interface-jump}
\overline{w}\ =\ \frac{1}{2}\,\left(\, (w_{|K_1})_{|F_k}\ +\ (w_{|K_2})_{|F_k} \,\right) \,,\qquad
[\,w\,]\ =\ (w_{|K_1})_{|F_k}\ -\ (w_{|K_2})_{|F_k} \,.
\end{equation}
The above definitions are also valid for vector-valued functions $w\in [H^s\{ \mathcal{T}_h \}]^{2n}$, in which case it follows that
 $[w]\cdot\nu_{k}= (w_{|K_1})_{|F_k}\cdot\nu_{K_1}+ (w_{|K_2})_{|F_k}\cdot\nu_{K_2}$.

\subsection{Upwind Galerkin Approximation of the Vlasov Equation}
\label{sec:DFUG Vlasov}

Here we describe the  UG  scheme for the Vlasov equation in full generality for a  $2n$-dimensional ($n=1,2$ or $3$) phase space with inflow boundary conditions and piecewise polynomials of arbitrary degree  approximating $f$.  The simpler derivation of the method for a two-dimensional phase space with  periodic boundary conditions in $x$  and  a piecewise constant    approximation to  $f$     is given explicitly in Section \ref{ssec:2d}, since this method  was used for  the numerics  of Section \ref{sect:NR}.  Note,  for this simpler version  the derivation does not require use of the set of mesh faces $\mathcal{F}_h$.

For a given time $T>0$ and  data trio~$(\alpha,f_{_0},f_{_I})$, the Vlasov equation along with the corresponding initial and boundary conditions are 
\bseq
\begin{eqnarray}
\label{eq:vlasov}
\partial_t f + \alpha\cdot\nabla f \ &=\  0  \qquad & \Omega \times (0,T] \,,
\\ \label{eq:initialc}
f(t=0) \ &=\ f_{_0}  \qquad & \Omega \,,
\\ \label{eq:inflowbc}
f \ &=\ f_{_I}  \qquad & \Gamma_{_I} \times (0,T] \,,
\end{eqnarray}
\eseq
where $\alpha=(v,-E)\in\mathbb{R}^{2n}$, $E\in\mathbb{R}^{n}$ is assumed given, $\nabla=(\nabla_x,\nabla_v)$,
and $\Omega=\Omega_x\times\Omega_v$.
It is assumed that $\Omega_x=\Pi_{i=1}^n[0,X_i]$ and $\Omega_v=[-V_c,V_c]^n,$ where $X_1,\ldots,X_n,V_c>0$ are fixed.
Following \eqref{eq:inflow boundary}, we also define (keeping the same notation without loss of generality)
 the  computational  inflow boundary $\Gamma_{_I}$, associated with  the computational domain $\Omega$, by
\begin{equation}
\Gamma_{_I}\ =\ \{ (x,v)\in\partial\Omega\ |\ \alpha\cdot\nu<0 \} \,,
\end{equation}
where $\nu$ is the outward unit normal to $\partial\Omega$.  Then,  $\Gamma_{_O}=\partial\Omega\setminus\Gamma_{_I}$.



For domains $K\subset\mathbb{R}^{2n}$, let $(\cdot,\cdot)_K$ denote the $L^2(K)$-inner product.  To distinguish integration
over domains $K\subset\mathbb{R}^{2n-1}$, we use the notation $\langle\cdot,\cdot\rangle_{_K}$.  A weak formulation
for (\ref{eq:vlasov})-(\ref{eq:inflowbc}) is derived by multiplying equation (\ref{eq:vlasov}) by an arbitrary test function $w \in H^1( \mathcal{T}_h )$ and
integrating  by parts over an arbitrary $K_j \in \mathcal{T}_{h}$.  This yields
\begin{equation}
\label{eq:weak_vlasov1}
(\partial_t f,w)_{K_j} - ( f ,\alpha\cdot \nabla w )_{K_j} +
\langle  f w,\alpha\cdot\nu_{K_j} \rangle_{\partial K_j} \ =\ 0 \,, \ \forall\  t \in ( 0, T ] \,,
\end{equation}
with $\nu_{K_j}$ being the outward unit normal to $\partial K_j$ and  $w^-$ denoting the interior trace of  $K_j$, i.e.,  $w^-(x,v) = \lim_{s\downarrow 0^-} w \left( (x,v) + s \nu_{K_j}\right)$.

Upon summing equation (\ref{eq:weak_vlasov1}) over all $K_j$
and weakly enforcing the inflow boundary condition, we get
\begin{multline} \label{eq:weak_vlasov2}
(\partial_t f,w)_{\Omega} - \sum\limits_{{j=1}}^{N_{h}} ( f, \alpha\cdot\nabla w )_{K_j} +
\sum\limits_{k=1}^{P_{h}} \langle f [w],\alpha\cdot\nu_{k} \rangle_{F_k}
+ \sum\limits_{F_k\in\Gamma_{_O}} \langle fw,\alpha\cdot\nu_{k} \rangle_{F_k}
\\
=\ -\sum\limits_{F_k\in\Gamma_{_I}} \langle f_{_I}\,w,\alpha\cdot\nu_{k} \rangle_{F_k} \,.
\end{multline}
Approximating $f$ by a function $f_h$, which may be discontinuous across the interior faces, requires the introduction of a numerical 
flux $f^u$.  A standard technique is to
replace $f$ by its ``upwind value'' $f^u$ on the interior faces \cite{CocDaw}, where $f^u$ along each interior face
$F_k$ is defined  by the given advecting flow field $\alpha(x,v,t)$ according to 
\begin{equation} \label{eq:upwind}
\begin{split}
f^u(v,v,t;\alpha) &\ = \lim_{s\downarrow 0} f \left( (x,v,t) + s \alpha(x,v,t))\right)\ =  \\
 &\ =\ \left\{ \begin{array}
{r@{\quad,\quad}l}
f_{|K_1}(x,v,t) \ \qquad &\mathrm{if\ }\alpha(x,v,t)\cdot\nu_{k}\ \ge\ 0\,,
\\
f_{|K_2}(x,v,t) \ \  &\mathrm{if\ }\alpha(x,v,t)\cdot\nu_{k}\ <\ 0\,.
\end{array} \right.  \notag
\end{split}\end{equation}
Consistency  follows from condition  (\ref{eq:upwind}), since $f^u(\alpha)=f$ on $F_k$ whenever $f$ is continuous across each interior face $F_k$.  We also note that $f^u$ depends nonlinearly on $\alpha$ as, in general, if $\alpha_1$ and $\alpha_2$ are two
flow fields having different values on $F_k$ and if $g\in H^s\{ \mathcal{T}_h \}$ is discontinuous across $F_k$, then
$g^u(\alpha_1+\alpha_{_2}) \neq g^u(\alpha_1)+g^u(\alpha_{_2})$.  Replacing $f$ on the interior faces in (\ref{eq:weak_vlasov2})
by its upwind value $f^u$ leads to
\begin{multline*} 
(\partial_t f,w)_{\Omega} - \sum\limits_{j=1}^{N_{h}} ( f, \alpha\cdot\nabla w )_{K_j} +
\sum\limits_{k=1}^{P_{h}} \langle f^u [w],\alpha\cdot\nu_{k} \rangle_{F_k}
+ \sum\limits_{F_k\in\Gamma_{_O}} \langle fw,\alpha\cdot\nu_{k} \rangle_{F_k}
\\
= -\sum\limits_{F_k\in\Gamma_{_I}} \langle f_{_I}\,w,\alpha\cdot\nu_{k} \rangle_{F_k} \, ,
\end{multline*}
which is the UG scheme for the Vlasov equation.
Finally, defining the bilinear operator $\mathcal{A}$ by 
\begin{multline} \label{eq:def.A}
\mathcal{A}(f,w;\alpha) \ =\ -\sum\limits_{j=1}^{N_{h}} ( f, \alpha\cdot\nabla w )_{K_j} +
\sum\limits_{k=1}^{P_{h}} \langle f^u [w],\alpha\cdot\nu_{k} \rangle_{F_k}
+ \sum\limits_{F_k\in\Gamma_{_O}} \langle fw,\alpha\cdot\nu_{k} \rangle_{F_k} 
\end{multline}
and the corresponding linear operator $\mathcal{L}$, depending on the inflow data $f_I$,  by 
\begin{equation} \label{eq:def.L}
\mathcal{L}(w;\alpha, f_{_I}) \ =\
-\sum\limits_{F_k\in\Gamma_{_I}} \langle f_{_I}\,w,\alpha\cdot\nu_{k} \rangle_{F_k} \,,
\end{equation}
yields a variational formulation  for the  semi-discrete problem
of finding the $f_h\in C^1([0,T],D^{r_x, r_v}( \mathcal{T}_h ) )$ approximation
to $f$, satisfying, 
\begin{eqnarray} 
&\, (\partial_{_t}\,f_{h},w_{h})_{\Omega}\ +\ \mathcal{A}(f_{h},w_{h};\alpha) = \mathcal{L}(w_{h};\alpha, 
f_{_I}) \qquad\ &\forall \ t\in(0,T]\, , \label{def:ug2}\\
&\ &\ \notag \\
&\,  ( f_{h}(x,v,0), w_{h} )_{K_j} \ =\ ( f_{_0},w_{h} )_{K_j}  \quad&\forall \ K_j\in\mathcal{T}_{h}, \label{def:ug1}
\end{eqnarray} 
for all $w\in D^{r_x, r_v}( \mathcal{T}_h )$, 
with $f_0$ and $f_I$  approximations 
of the initial data $f(x,v,0)$ and the inflow boundary data on $\Gamma_I$,  respectively. 

We note that (\ref{def:ug2}) produces  a first-order ODE system to be described below.
Indeed, for each $K_i=K_{i_x}\times K_{i_v}\in\mathcal{T}_h,$ let $\psi^i_1,\ldots,\psi^i_{n_b}$ be a basis for
$Q^{r_x}(K_{i_x})\times Q^{r_v}(K_{i_v})$, where $n_b=(r_x+1)^n\times(r_v+1)^n$ and then extend the domain of these functions
to $\Omega$ by defining each to be identically zero in $\Omega\setminus K_i$.  If 
$\beta(t)=(\beta^1_1(t),\ldots,\beta^1_{n_b}(t),\ldots,\beta^{N_h}_1(t),\ldots,\beta^{N_h}_{n_b}(t))$ denotes the unique vector such that the UG approximation $f_h$ satisfies
\begin{equation} \label{eq:fh.representation}
f_{h}(x,v,t)\ =\ \sum\limits_{j=1}^{N_{h}}\sum\limits_{m=1}^{n_b}\,\beta^j_m(t)\,\psi^j_m(x,v) \,,
\end{equation}
then  $f_{h_{{|K_j}}}=\sum\limits_{m=1}^{n_b}\,\beta^j_m\,\psi^j_m$.  Inserting (\ref{eq:fh.representation}) into (\ref{def:ug2}) yields
\begin{multline} \label{def:ug2.1}
\sum\limits_{j=1}^{N_{h}}\sum\limits_{m=1}^{n_b}\,\dot{\beta}^j_m(t)\,(\psi^j_m,w_{h})_{\Omega}
\ +\ \sum\limits_{j=1}^{N_{h}}\sum\limits_{m=1}^{n_b}\,\beta^j_m(t)\,\mathcal{A}(\psi^j_m ,w_{h};\alpha)
\\
\ =\ \mathcal{L}(w_{h};\alpha, f_{_I}) \,,
\qquad\forall\,w_{h}\in D^{r_x, r_v}( \mathcal{T}_h ) \,.
\end{multline}
Finally, since $\{\psi^i_p\}_{p=1,i=1}^{n_b,N_h}$ is a basis for $D^{r_x, r_v}( \mathcal{T}_h )$, then
(\ref{def:ug2.1}) is equivalent to
\begin{multline} \label{def:ug2.2}
\sum\limits_{m=1}^{n_b}\,\dot{\beta}^i_m(t)\,(\psi^i_m,\psi^i_p)_{K_i}
\ +\ \sum\limits_{j\in N(i)}\sum\limits_{m=1}^{n_b}\,\beta^j_m(t)\,\mathcal{A}(\psi^j_m ,\psi^i_p;\alpha)
\\
\ =\ \mathcal{L}(\psi^i_p;\alpha, f_{_I}) \,,
\qquad \forall i\in\{1,\dots,N_{h}\},\ \forall\,p\in\{1,\dots,n_b\} \,,
\end{multline}
where $N(i)$  contains the indices of all neighboring elements of $K_i$.

Equation~(\ref{def:ug2.2}) is seen to generate an equivalent matrix system, where $n_b$ rows of the matrix are generated
at a time by sequentially taking $i$ to equal $1,\ldots,N_h$ and for each $i$ sequentially taking $p$ to equal
$1,\ldots,n_b$ in Eq.~(\ref{def:ug2.2}).  This procedure results in the matrix ODE system
\begin{equation} \label{def:ug2.3}
A_1\dot{\beta}(t)\ +\ A_{_2}(\alpha)\beta(t)\ =\ L(\alpha, f_{_I}) \,,
\end{equation}
where $A_1$ is a constant matrix and $A_{_2}(\alpha)$ is the corresponding
sparse matrix, both of which are of dimension $n_b N_{h}\times n_b N_{h}$,  and $L(\alpha,f_{_I})$ is a vector of length $n_b N_{h}$.

Since the support of the functions $\psi^i_1,\ldots,\psi^i_{n_b}$ is $K_i$, $\forall i\in\{1,\ldots,N_h\}$, it follows that $A_1$
is a block-diagonal matrix, where each block is an  $n_b\times n_b$ matrix.  This means  that the inverse of $A_1$ is easily computed.
Thus, the UG approximation $f_{h}$ is equivalently defined to be the unique solution to
\begin{equation} \label{eq:final.ugode}
\dot{\beta}(t)\ =\ -A_1^{-1}A_{_2}(\alpha)\beta(t)\ +\ A_1^{-1}L(\alpha, f_{_I}) \,,
\end{equation}
where the initial condition $\beta(0)$ is uniquely determined by (\ref{def:ug1}).  To solve this system in time,
a conservative explicit time stepping method such as the Runge-Kutta method can be used.

\subsection{Interior Penalty Approximations of the Poisson Equation}

In order to make this manuscript self contained, we also discuss the interior penalty formulations 
for the Poisson equation that weakly enforces approximate continuity across interior mesh faces 
and Dirichlet boundary regions.
As already noted, the mesh $\mathcal{T}_{h_x}$ used to discretize the 
Poisson equation must be the same as that  for 
the spatial domain used for  the Vlasov equation, 
but  the polynomial degree $r_x$ for  the 
Poisson equation need not be equal to the degree in $x$ for $f_h$.  

Hence, for a given i) source $G\in L^2(\Omega_x)$,  ii) boundary data $\Phi_{_D}\in L^2(\partial\Omega_{x,D})$ in the portion of the boundary referred as the Dirichlet boundary $\partial\Omega_{x,D}$,    and
iii)  $\nabla \Phi\cdot\nu=0$ (homogeneous Neumann)  or periodic boundary conditions on $\partial\Omega_{x}\setminus\partial\Omega_{x,D}$, 
the boundary complement of the Dirichlet region, then the more general form of   
Poisson equation with a positive definite permittivity term $a$  discretized using 
the symmetric interior penalty (SIPG) method \cite{Wheeler,WhePer},  
incomplete interior penalty (IIPG) method \cite{Sun,SunWhe,DawSunWhe}, or  non-symmetric interior penalty (NIPG) \cite{RivWheGir} method is
\bseq
\begin{align} \label{eq:poisson}
-\nabla_{x}\cdot \left(a\nabla_x \Phi\right) &\ =\ G \qquad\ \Omega_x \,,
\\ \label{eq:dirichletbc}
\mathcal{B}_x\Phi &\ =\ \Phi_{_D} \qquad\partial\Omega_{x,D} \,.
\end{align}
\eseq
where $a(x)$ is any positive-definite continuous function in $(C^1(\Omega_x))^{n\times n}$.
 
For $K\subset\mathbb{R}^{n}$, $n$=1,2, or 3, recall $(\cdot,\cdot)_K$ is the $L^2(K)$-inner product
with integration over $K\subset\mathbb{R}^{n-1}$, while the notation $\langle\cdot,\cdot\rangle_{_K}$ is for boundary  integrals. In addition,  we use the identity
\begin{align}\label{eq:weakpoisson1}
\sum_{j_x=1}^{N_{h_x}} \langle a \nabla_x\Phi\cdot\nu_{K_{j_x}},\theta^- \rangle_{\partial K_{j_x}}
&=\sum_{k_x=1}^{P_{h_x}}\langle\, \overline{a\nabla_x\Phi}\,[\theta]
+[\nabla_x\Phi]\,\overline{a\theta},\nu_{_{k_x}} \rangle_{F_{k_x}}
\notag\\
&\ +\sum_{F_{k_x}\in\partial\Omega_x} \langle \nabla_x\Phi\cdot\nu_{_{k_x}},\theta \rangle_{F_{k_x}} \, ,
\end{align}
where $\nu_{K_{j_x}}$ is the outward unit normal to $\partial K_{j_x}$.

Thus,  the corresponding schemes SIPG, IIPG, and NIPG   are all derived by multiplying (\ref{eq:poisson}) 
by an arbitrary test function $\theta\in H^{s}\{ \mathcal{T}_{h_x} \}$, $s>1/2$, 
integrating by parts on each $K_{j_x}\in\mathcal{T}_{h_x},$ and  summing each
of the resulting local equations.  Whence, one obtains the non-symmetric variational formulation given by  
\begin{multline} \label{eq:weakpoisson2}
\sum_{j_x=1}^{N_{h_x}}(a\nabla_x\Phi,\nabla_x\theta)_{K_{j_x}}
-\sum_{k_x=1}^{P_{h_x}}\langle\, \overline{a\nabla_x\Phi}\,[\theta]
+[a\nabla_x\Phi]\,\overline{\theta},\nu_{_{k_x}} \rangle_{F_{k_x}}
\\
-\sum_{F_{k_x}\in\partial\Omega_x}
\langle a\nabla_x\Phi\cdot\nu_{_{k_x}},\theta \rangle_{F_{k_x}}
= (G,\theta)_{\Omega_x} \,, 
\end{multline}
where the identity  (\ref{eq:weakpoisson1}) was used to represent the inner boundary integrals.

In particular, since the true solution for a bounded right-hand-side has at least the regularity 
$\Phi\in H^1(\Omega_x)\cap H^2\{ \mathcal{T}_{h_x} \}$,  it follows that  
$[\Phi]\equiv0$ and $[\nabla_x\Phi]\equiv0$ along every interior face $F_{k_x}$ \cite{Evans}.  Thus,
In order to get a good approximation for a regular solution satisfying these two jump conditions,  one adds 
`interior' penalty terms that vanishes on the true solution $\Phi$. These terms are of the form
\begin{equation}
\label{def:symm.term}
c_s\sum_{k_x=1}^{P_{h_x}}\langle\, \overline{a\nabla_x\theta}\,[\Phi],\nu_{_{k_x}} \,\rangle_{F_{k_x}} \,,
\end{equation}
where the values selected for $c_s$ for the SIPG, IIPG and
NIPG methods are -1, 0 and 1, respectively. Similarly, such penalization is also required for the 
Dirichlet boundary region $\partial\Omega_{x,D}$ as follows. 

Indeed,  we add to the bilinear form the interior penalty terms 
\begin{equation}\label{penal_param}
\sum_{k_x=1}^{P_{h_x}} \frac{\sigma}{(h_{j_x})^{n/2}}\,\langle [\Phi],[\theta] \rangle_{F_{k_x}}\,,
\end{equation}
which are now  completed by weakly enforcing both approximate continuity across the interior
mesh faces and  Dirichlet boundary regions $\partial\Omega_{x.D}$,  by including the penalization
\begin{equation}\label{penal_param_bdry}
\sum_{F_{k_x}\in\partial\Omega_{x,D}}\frac{\sigma}{(h_{j_x})^{n/2}}\,\langle \Phi-\Phi_{_D},\theta \rangle_{F_{k_x}} \,,
\end{equation}
where $\sigma>0$ is an arbitrary penalty parameter that is usually set equal to unity.  
Note that homogeneous or periodic boundary conditions vanish on boundary terms of the corresponding 
bilinear structure. Consequently,  they do not  require the boundary penalization term.

Usually the penalization terms are denoted by the following  non-symmetric bilinear form
 \begin{equation}
 \label{penal_param_bilinear}
J_\sigma(\Phi, \theta)= \sum_{k_x=1}^{P_{h_x}} \frac{\sigma}{(h_{j_x})^{n/2}}\,\langle [\Phi],[\theta] \rangle_{F_{k_x}}
+\sum_{F_{k_x}\in\partial\Omega_{x,D}}\frac{\sigma}{(h_{j_x})^{n/2}}\,\langle \Phi, \theta \rangle_{F_{k_x}} \, .
\end{equation}

Adding both penalty terms and
(\ref{def:symm.term})  to the left-hand-side of (\ref{eq:weakpoisson2}) results in
\begin{multline} 
\label{eq:weakpoisson5}
\sum_{j_x=1}^{N_{h_x}}(a\nabla_x\Phi,\nabla_x\theta)_{K_{j_x}}
-\sum_{k_x=1}^{P_{h_x}}\langle\, \overline{a\nabla_x\Phi}\,[\theta]+c_s\overline{a\nabla_x\theta}\,[\Phi],\nu_{_{k_x}} \,\rangle_{F_{k_x}}
\\
\ -\sum_{F_{k_x}\in\partial\Omega_x}
\langle\,a \nabla_x\Phi\cdot\nu_{_{k_x}},\theta \,\rangle_{F_{k_x}} 
  +\sum_{k_x=1}^{P_{h_x}}\frac{\sigma}{(h_{j_x})^{n/2}}\,\langle [\Phi],[\theta] \rangle_{F_{k_x}} \\
+
\sum_{F_{k_x}\in\partial\Omega_{x,D}}\frac{\sigma}{(h_{j_x})^{n/2}}\,\langle \Phi,\theta \rangle_{F_{k_x}}
= (G,\theta)_{\Omega_x}+\sum_{F_{k_x}\in\partial\Omega_{x,D}}\frac{\sigma}{(h_{j_x})^{n/2}}\,\langle \Phi_{_D},\theta \rangle_{F_{k_x}} \,.
\end{multline}
Equation~(\ref{eq:weakpoisson5}) completely defines each of the three interior penalty schemes.

Setting $A_{c_s}(\Phi,\theta)$   equal to the first four terms on the left-hand-side of (\ref{eq:weakpoisson5}), 
where dependence on the parameter $c_s$ from \eqref{def:symm.term} is noted as a subscript, we let the bilinear operator
$\mathcal{B}_{c_s}(\Phi,\theta) := A_{c_s}(\Phi,\theta) + J_\sigma(\Phi,\theta)$   and  the linear
operator $\mathcal{H}(\theta;G,\Phi_{_D})$   be equal to the  two terms on the right-hand-side of (\ref{eq:weakpoisson5}). 
Then the function $\Phi_h\in D^{r_x}(\mathcal{T}_{h_x} )$ is the corresponding
interior penalty Galerkin approximation to the Poisson solution $\Phi$,  if
\begin{equation} \label{def:nipg}
\mathcal{B}_{{c_s}}(\Phi_{h},\theta_{h}) \ = \mathcal{H}(\theta_{h};G,\Phi_{_D})\, , \ \ \ \
\forall\,\theta_{h}\in D^{r_x}\{ \mathcal{T}_{h_x}\} \,.
\end{equation}

Note,   $\mathcal{B}_{{c_s}}$ is positive definite (or coercive) even though each of  $A_{c_s}(\Phi,\theta)$ and $J_\sigma(\Phi,\theta)$
separately are only positive semi-definite \cite{DawSunWhe,Heath, RivWheGir, Sun,SunWhe}. In particular, $\mathcal{B}_{{c_s}}$ generates an equivalent norm for the Hilbert space $H^1(\mathcal{T}_h)$,
\begin{equation}\label{NIPG-norm}
 \| \theta \|^2_{NIPG} = A_{c_s}(\theta,\theta) + J_\sigma(\theta,\theta)\,,   \qquad \ \  \theta\in H^1(\mathcal{T}_h) \, ,
\end{equation}
if the measure of the Dirichlet boundary $|\partial\Omega_{x,D}|>0.$ In the case of periodic or homogeneous Neumann boundary conditions 
we need the compatibility condition $ \int_{\Omega_x} \int_{\mathbb{R}^n}\,f\,dx\,dv
=|\Omega_x|$,   for  each connected component $\Omega_x$ of  the spatial domain.  In fact, the approximation to the Vlasov equation is done with a conservative UPG scheme and  high order Runge-Kutta schemes, and in particular, for the case of periodic boundary conditions the compatibility condition  at the numerical level is satisfied as well.

Finally it is possible to see that the bilinearity and positive definiteness   of $\mathcal{B}_{c_s}$  
for either a portion of Dirichlet or full Neumann or periodic boundary conditions, implies that 
(\ref{def:nipg}) is equivalent to a uniquely solvable 
matrix system described next.
Let $n_b=(r_x+1)^n$ and
$\mu=\{\mu^1_1,\ldots,\mu^1_{n_b},\ldots,\mu^{N_{h_x}}_1,\ldots,\mu^{N_{h_x}}_{n_b} \}$ be the
unique vector such that
\begin{equation} \label{eq:phih.representation}
\Phi_h = \sum\limits_{j_x=1}^{N_{h_x}} \sum\limits_{m=1}^{n_b}\, \mu^{j_x}_{m}\, \theta^{j_x}_{m}(x) \,.
\end{equation}
Upon substituting this representation into (\ref{def:nipg}), we conclude that (\ref{def:nipg}) is equivalent to
\begin{equation} \label{def:nipg1}
B\mu = H \,,
\end{equation}
where $B$ is an $(n_b+1)N_{h_x}\times(n_b+1)N_{h_x}$ invertible sparse matrix and $H(G,\Phi_D)$ is a vector of
length $(n_b+1)N_{h_x}$.  However, B is not block diagonal,  as was the case for $A_1$ in the Vlasov ODE system.
Thus, if the spatial domain is two- or three-dimensional,  then using an iterative solver is in general  the most efficient
means of computing the solution $\mu$ in (\ref{def:nipg1}).  However, for the one-dimensional spatial domain, $\mu$
can be computed by using an LU-matrix decomposition algorithm to factor B.  For convenience, we write
$\mu=B^{-1}H$, even though in practice $\mu$ might be computed using an iterative method.


\subsection{Discontinuous Galerkin Approximation of the Vlasov-Poisson System}

The UPG  method for the Vlasov-Poisson system 
results from combining the UG approximation of the Vlasov equation together with the interior penalty
approximation of the Poisson equation. Thus,  the approximation $f_h(t)$ to the solution $f(x,v,t)$ of the Vlasov-Poisson 
system (\ref{eq:vlasov.full})-(\ref{eq:poisson.full}), at time $t$,  results from an iteration as follows. 

Let  $\tilde f_h(t)$ be given, where $\tilde f_h (0)$ is the approximation of the initial distribution function.
Then,  an approximation   $\alpha_{h}(t)=(v,-E_h(t))$ to $\alpha(x,v,t)=(v,-E(x,t))$ is
determined by computing the corresponding approximation to  $-E_h(t)$, 
using one of the interior penalty approximation schemes to compute an approximation $\Phi_h(t)$ to the
potential $\Phi_D (x,t)$ via the formula $\mu=B^{-1}H(G,\Phi_D)$, where $G$ is defined to be $1-\int \tilde f_h(t) dv$.

Next,  the approximate local field  is computed by taking the local spatial potential gradients  $(E_h)_{|K_{j_x}}
=-\nabla_x(\Phi_h)_{|K_{j_x}}$ on each $K_{j_x}$,
which implies that $E_h(t)=E_{h}(\tilde f_{h} (t))$ is discontinuous across the interior faces of $\mathcal{T}_{h_x}$.
Consequently, it follows that the approximate  flow field $\alpha_{h}(t)=\alpha_{h}(\tilde f_{h}(t))$, or equivalently $\alpha_{h}(t)=\alpha_{h}(\beta(t))$, 
where $\beta(t)$ is a well defined given function.

Summarizing,   for any  given $\tilde f_h(t)$,  first compute the approximate   $\alpha_h(t)$, and then  the approximate solution $f_h(t)$ to the Vlasov-Poisson system  by solving the ODE  system (\ref{eq:final.ugode}) with $\alpha(t)$ replaced by $\alpha_{h}(\beta(t))$.  This leads to
the following definition:
\begin{Def} \label{def:dgnipgvlasovpoisson}
The semi-discrete function $f_h(\beta(t))\in C^1([0,T],D^{r_x, r_v}( \mathcal{T}_h ) )$ is the
UPG approximation to Vlasov-Poisson solution $f$,  if $f_h(\beta(t)):= f_h$ as defined in
 \eqref{eq:fh.representation} where $\beta(t)$ satisfies the nonlinear system of ODEs   
\begin{eqnarray} 
&(i) \  \ & ( f_{h}(\beta(0)), w_{h} )_{K_j} \ =\ ( f_{_0},w_{h} )_{K_j}\, ,\ \quad\forall \ K_j\in\mathcal{T}_{h},\ 
\forall\,w\in D^{r_x, r_v}( \mathcal{T}_h )\, ,  \notag \\
&(ii) \  & \ \alpha_{h}(\beta(t)) =(v,\nabla_x \Phi_{h} (\mu(t))) \    \qquad\mathrm{with}\ \  \mu(t) = B^{-1} H(\beta(t))\, ,  	 
\notag \\
&(iii)&  \ 
\dot{\beta}(t)\ =  -A_1^{-1}A_{_2}(\, \alpha_{h}(\beta(t)) \,)\,\beta(t)\ +\ A_1^{-1}L(\alpha_{h}(\beta(t)), f_{_I}) \, , 
\label{def:ugnipg2} 
\end{eqnarray}
for all $t\in(0,T]$. 
\end{Def}
System  (\ref{def:ugnipg2}) can be solved for $\beta(t)$ using any explicit time-integrator, following a classical Gummel map type iteration, where at any given approximation  $\beta^{n-1}$, the   $\beta(t^{n-1})$ step (ii) is first computed since it does not involve time variation.  Thus,   one obtains $\alpha_h(\beta(t^{t-1}))$,  which allows for the calculation of  $\beta^n$,  the approximation to  $\beta(t^n)$ by step (iii).    In our simulations we use a conservative high  order Runge-Kutta time integrator. 

 This very same iteration scheme was previously proposed in \cite{CGMS-cmame2009} for the calculation of Boltzmann-Poisson solvers for semiconductors, and related work by the same authors cited below. There the calculation of the Poisson equation is done by a LDG scheme. 

We also point out that an error estimate for this nonlinear scheme can be found in \cite{Heath},  which we state here in a concise form.  Let  $(f,\Phi)$ be a solution pair of the Vlasov Poisson system  \eqref{eq:poisson.full},   with boundary and initial conditions as described above,     potential $\Phi(t, \cdot)\in H^{\bar s}(\Omega_x)$ for $\Omega_x \subset \mathbb{R}^n$,  and    distribution function $f \in C^1([0,T], H^{2s}(\mathcal{T}_h))$   for  $\Omega \subset \mathbb{R}^{2n}$,  for both $\bar s\, ,s>n$.  Also, let  
 $F_k$ denote the interior faces associated with element $K_k$ in $\Omega$, with $F^x_{k_x}$ denoting the corresponding  one associated  with the elements in the $x$-space  $\Omega_x$, as it was defined at the end of Section~\ref{prelim}.

Further, recall $\mathbb{Q}^{r} (K)$ is the   space of polynomials on a set $K$ of degree less
than or equal to $r$ (cf.\  (\ref{def:brokensobolev}) and (\ref{def:brokenapproximation})), and  $r_x$  and $r_v$  the degrees
 in $x$-space and  $v$-space, respectively.    Let the parameter $\lambda$ be the first eigenvalue to the Poisson equation in $\Omega_x$, 
  $\tilde \Phi_h $ be the distributional solution to the perturbed Poisson equation for a source term $\rho_h$, the charge density ($v$-average) associated with $f_h$, and let $\mu_x$ and $\mu_v$ be defined by 
\begin{equation}
\mu_x = \min\{r_x+1,\bar s\}  \quad  \ \mathrm{and }\quad \  \mu_v = \min\{r_v+1, s\}\,.  
\end{equation}
Then,   we obtained the following error estimate for the 
$2n$-dimensional   semi-discrete formulation of the Vlasov Poisson system  
in terms the difference  of suitable norms of potentials $\Phi-\Phi_h$, fields  $E-E_h= -\nabla\Phi+\nabla\Phi_h$, 
and particle distribution functions $f-f_h$: 
\begin{eqnarray}\label{error_estimate}
&\| \Phi -\Phi_h \|^2_{NIPG}  \ &\le\ {\lambda^{-1}} \| \rho-\rho_h  \|^2_{L^{2}(\Omega_{x})} +  c\frac{h^{2\mu_x-2}}{r_x^{2\bar s-2}} \|\tilde \Phi_h \|^2_{L^{2}(\Omega_x)}\,, \notag\\
 &\ \ &\   \notag \\
&\| \nabla\Phi -\nabla\Phi_h \|^2_{L^{2}(\Omega_{x})} &+  \sum_{k_{x}=1}^{P_{h_{x}}} \frac{r_v\sigma}{|h_{j_{x}}|^{n/2}} \| \Phi -\Phi_h \|^2_{L^2(F_{k_{x}})} +
\sum_{F_{k_{x}} \in \Omega_{x,D}}  \frac{ r_x\sigma}{|h_{j_{x}}|^{n/2}} \| \Phi -\Phi_h \|^2_{L^2(F_{k_{x}})}  \notag\\
&\ \  &\ \ \notag\\
&\qquad \le\ &{\lambda^{-1}} \| \rho-\rho_h  \|^2_{L^{2}(\Omega_{x})} +  c\frac{h^{2 \mu_x-2}}{r_x^{2  \bar s -2}} \|\tilde \Phi_h 
\|^2_{L^{2}(\Omega_x)}\,, \\
&\ \  &\ \ \notag\\
 &\|f(T)- f_h(T)  \|^2_{L^{2}(\Omega)} &+ \int_0^T \sum_{k=1}^{P_h} \| | \overline{\alpha_h}\cdot\nu_k |^{1/2} [f-f_h]  \|^2_{L^2(F_{k})}
 \notag\\
&\ &\  \notag\\
&+ \int_0^T  \| | \alpha_h\cdot\nu_k|^{1/2} &[f-f_h]  \|^2_{0,{\Gamma_{0}}} +\int_0^T  \| | \alpha_h\cdot\nu_k|^{1/2} [f-f_h]  \|^2_{0,\Gamma_{I}} \notag\\
&\ &\  \notag\\
& \qquad&\qquad\qquad\qquad\qquad\le \ C h^{2\mu_v -1} + \mathit{o}_{\{h,\mu_x, \mu_v\} } (h^{2\mu_v -1}) \,,\notag
\end{eqnarray}
where  $\sigma$ is  the penalization parameter of (\ref{penal_param}) and (\ref{penal_param_bdry}) and the  $\| \cdot \|^2_{NIPG}$
was defined in the previous subsection at  \eqref{NIPG-norm}.
In addition,  for a  sufficiently smooth potential $\Phi(t,\cdot)\in H^{\bar s}(\Omega_x)$ for $\Omega_x\subset \mathbb{R}^n$ and distribution function $f(t,\cdot)\in H^{2s}(\Omega)$ for $\Omega\subset\mathbb{R}^{2n}$, where the order of smoothness is given by the parameters $s$ and $\bar s$,  this estimate is optimal.

We close this subsection by noting that very recently our iteration scheme was  reproduced in \cite{Ayuso1,Ayuso2}  with a different Poisson solver.   These authors perform error estimates for quadratic basis functions that preserve  energy,   but do not preserve the positivity of $f$.  Numerical simulations have yet to be performed for their scheme and the amount of degradation cause by the lack of positivity remains to be ascertained.


\subsection{Two-Dimensional Phase Space with Piecewise Constant Approximation}
\label{ssec:2d}

We end this section with a description of the  simplified scheme for  a two-dimensional phase space using   piecewise constant approximations to $f$.  As noted above, this is  a positivity preserving (monotone) scheme that was used for the plasma simulations presented in Section~\ref{sect:NR}. Such a  piecewise constant basis function scheme  can be easily extended to higher dimensions.   We point out that, in  work currently under preparation \cite{cheng-gamba-morrison11,CGLM11},   we extend the positivity condition to higher order basis functions by  new  limiter techniques inspired by \cite{Cheng-gamba-proft-10, zhang-shu10-1, zhang-shu10-2}, which  are maximum principle preserving and can be applied to both the Vlasov-Poisson and Vlasov-Maxwell systems.  It remains a challenge to find a proper scheme that would preserved positivity and  higher order moments,  like momentum and energy.  In the future we hope to compare our approach with extensive existing work on Vlasov-Ma
 xwell system \cite{MaCaTr,VaVeMa,VaTrCaHeMa}.

Here, the simplified  spatial and velocity domains are  $\Omega_x=[0,L]$ and $\Omega_v=[-V_c,V_c]$, with mesh points $0=x_0<x_1<\ldots<x_{N_{h_x}-1}<x_{N_{h_x}}=L$ and
$-V_c=v_0<v_1<\ldots<v_{N_{h_v}-1}<v_{N_{h_v}}=V_c$, where $N_{h_x},N_{h_v}\in\mathbb{N}$.  Then for
$j_x=1,\ldots,N_{h_x}$ and $j_v=1,\ldots,N_{h_v}$, take $\mathcal{T}_{h_x}=\{K_{j_x}\}_{j_x=1}^{N_{h_x}}$ and
$\mathcal{T}_{h_v}=\{K_{j_v}\}_{j_v=1}^{N_{h_v}}$ by defining each spatial element $K_{j_x}=[x_{j_x-1},x_{j_x}]$ of size
  $h_{j_x}=x_{j_x}-x_{j_x-1}$, 
and each velocity element $K_{j_v}=[v_{j_v-1},v_{j_v}]$ of size $h_{j_x}=v_{j_v}-v_{j_v-1}$, respectively.
A mesh $\mathcal{T}_h=\{K_j\}_{j=1}^{N_h}$ of the phase space domain $\Omega$ is now generated according to
$K_j=K_{j_x}\times K_{j_v}$, where the index $j$ is defined by the element ordering $j=(j_v-1)N_{h_x}+j_x$, for
$j_x=1,\ldots,N_{h_x}$ and $j_v=1,\ldots,N_{h_v}$, so  that $\mathcal{T}_h$ contains a total of
$N_h=N_{h_x}N_{h_v}$ elements. The corresponding piecewise basis function is then given by
setting  $\theta^{i_x}(x)=1$, for $x\in K_{i_x}$, and 
$\theta^{i_x}(x)=0$, otherwise, for $i_x=1,\ldots,N_{h_x}$. Similar basis is also constructed for $\chi^{i_v}(v)$, $i_v=1,\ldots,N_{h_v}$.
Then, taking $\psi^i(x,v)=\theta^{i_x}(x)\chi^{i_v}(v)$, for $i=1,\ldots,N_h$,  generates the
approximating space $D^{0,0}( \mathcal{T}_h ) = \mathrm{ span } \{ \psi^1,\ldots,\psi^{N_h} \}$.

The corresponding  upwind function $f^u$ defined in \eqref{eq:upwind} on $\partial K_i \setminus\Gamma$ 
is now written in the simpler form
\begin{equation} \label{warmup.upwind}
f^u\,(x,v,t;\alpha) = \left\{ \begin{array}
{r@{\quad,\quad}l}
f^-(x,v,t) & \mathrm{if\ }\alpha(x,v,t)\cdot\nu_{K_i}\ \ge\ 0\,,
\\
f^+(x,v,t) & \mathrm{if\ }\alpha(x,v,t)\cdot\nu_{K_i}\ <\ 0\,,
\end{array} \right.
\end{equation}
for  $f^{\pm}(x,v,t) =\lim_{s\to0^{\pm}} f\left( (x,v) + s \nu_{K_i},t\right)$  and 
the outward unit normal to $K_j$ denoted by $\nu_{K_i}(x,v)$ is simply defined by $(0,-1)$ for $v=v_{i_v-1}$, $(0,1)$ for $v=v_{i_v}$,
$(1,0)$ for $x=x_{i_x}$ and  $(-1,0)$ for $x=x_{i_x-1}$.

Therefore,  the corresponding lowest order UG scheme   is 
\begin{multline*} \label{eq:weak_vlasov4}
(\partial_t f,w)_{\Omega} +
\sum\limits_{k=1}^{P_{h}} \langle f^u [w],\alpha\cdot\nu_{k} \rangle_{F_k}
+ \sum\limits_{F_k\in\Gamma_{_O}} \langle fw,\alpha\cdot\nu_{k} \rangle_{F_k}
\\
= -\sum\limits_{F_k\in\Gamma_{_I}} \langle f_{_I}\,w,\alpha\cdot\nu_{k} \rangle_{F_k} \, .
\end{multline*}

In particular, for  the piecewise constant UG approximation, 
$f_h(x,v,t)\ =\ \sum\limits_{j=1}^{N_h} \beta^j(t)\,\psi^{j}(x,v)$ to $f(x,v,t)$, clearly 
one obtains  $(f_h)_{|K_j}=\beta^j(t)$, which implies
\begin{equation} \label{warmup.massterm}
\partial_t\Big( \int_{K_j} f_h\,dvdx \Big)\ =\ h_{j_x} h_{j_v}\,\dot{\beta}^{j}(t) \,, 
\end{equation}
and the corresponding  semi-discrete UG
approximation $f_h=\sum_{j=1}^{N_h} \beta^j(t)\,\psi^{j}$ is the unique function in
$C^1([0,T],D^{0,0}(\mathcal{T}_h))$ satisfying the initial condition $\int_{K_i}f_h(x,v,0)=\int_{K_i}f_0$, 
$\forall\,i\in\{1,\ldots,N_h\}$ and  $\forall\,t\in(0,T]$, and 
\begin{align} 
\label{warmup2.0}
h_{j_x} h_{j_v}\,\dot{\beta}^{i}(t) &+
\int_{\partial K_i/\partial\Omega} (f_h)^u(\beta(t))\,\alpha\cdot\nu_{K_i}\,dS +
\int_{\partial K_i\cap\Gamma_O} f_h(\beta(t))\,\alpha\cdot\nu_{K_i} \,dS
\notag\\
&+ \int_{\partial K_i\cap\Gamma_I} (\,f_h(\beta(t))\,)_I\,\alpha\cdot\nu_{K_i} \,dS
\ =\ 0\,, \qquad\mathrm{for}\ i=1,\ldots,N_h\,.
\end{align}
This last identity is a linear ODE system for any given electric field $E$,  where the  integration
along  the interior faces $\partial K_i\setminus\Gamma$ in (\ref{warmup2.0}) ( i.e., $\partial K_i\cap\partial\Omega=\emptyset$), 
is simply
\begin{align}
&\int_{\partial K_i/ \partial\Omega} f \alpha\cdot\nu_{K_i}\,dS
\ =\ \int_{x_{i_x-1}}^{x_{i_x}} E(x,t) \left( f(x,v_{{i_v}-1},t) -f(x,v_{i_v},t)\right) \,dx
\notag\\
&\hspace{ 1.4 in} +\int_{v_{i_v-1}}^{v_{i_v}} v\left( f(x_{{i_x}},v,t) - f(x_{i_x-1},v,t)\right) \,dv
\end{align}
and the integrations along $\partial K_i\cap\Gamma_O$ and $\partial K_i\cap\Gamma_I$ satisfy
\begin{equation}
\int_{\partial K_i\cap\Gamma_O} f_h\,\alpha\cdot\nu_{K_i} \,dS\ =\
\int_{\partial K_i\cap\Gamma_I} (f_h)_I\,\alpha\cdot\nu_{K_i} \,dS\ =\ 0\,.
\end{equation}

\section{Numerical Results}
\label{sect:NR}

In this section numerical results are presented for six examples chosen to  test the accuracy and convergence of the proposed DG method.
The examples chosen are typical for testing Vlasov-Poisson algorithms (see e.g.\   \cite{CheKno}),  but we have also included some  atypical,  more extensive comparison to theory.  Four of the examples test the linear dynamics and its  associated fine structure (filamentation)
in phase space, while two  examine the  nonlinear evolution.  The linear results are presented in  Section  \ref{sec:linear}:  in Section
  \ref{sec:Adv}  the ability of the DG method to solve the advection equation,  the Vlasov equation with the electric field  set to  zero,
is considered for both Maxwellian and Lorentzian equilibria, while in Section \ref{sec:Lin.Landau} the method is applied to the Landau problem
 and numerical results are extensively compared  with the theoretical results of linear Landau damping, also for both Maxwellian and Lorentzian
equilibria.    The nonlinear results are presented in  Section \ref{sec:nonlinear}:  in Section \ref{sec:nlLD} nonlinear Landau damping is
considered while in \ref{sec:Nonlin.two.stream} the nonlinear two-stream instability problem is computed.

For all examples,  piecewise constants are used to approximate the distribution $f$,  piecewise quadratic polynomials are used to approximate
the potential  $\Phi$, and  time is discretized using a conservative fourth-order Runge-Kutta method.  For all but  the  first two linear
advection examples,  the NIPG penalty  method is used  to approximate the Poisson system, and the linear system that results
from using the NIPG method is solved using an LU-decomposition algorithm.

Throughout this section, it is assumed that the distribution function $f$ has the form
\begin{equation}
\label{eq:f.form}
f(x,v,t)=f_{eq}(v)+\delta f(x,v,t) \,,
\end{equation}
and the initial and boundary conditions used in all of the examples are of the form
\begin{subequations}
\begin{eqnarray}
\label{Vlasov_IC}
\delta f(x,v,0) \ &=&\ A\,\mathrm{cos}(k x)\,f_{eq}(v) \,,
\\ \label{Vlasov_BC}
\delta f(0,v,t) \ &=&\ \delta f(L,v,t) \,,
\\ \label{Poisson_BC}
\Phi(0,t) \ &=&\ \Phi(L,t) \,,
\end{eqnarray}
\end{subequations}
for $(x,v,t)\in[0,L]\times[-V_c,V_c]\times(0,T),$ where $V_c>0,$ $L>0,$ and $T>0$ are given.  The constant $V_c$
is the cutoff velocity and is chosen large enough so that the values of $f$ are negligibly small when $|v|=V_c.$  It follows that each example
 is completely determined by specifying the governing equations
along with the parameters $f_{eq}(v),\ A,\ k,\ L,\ V_c,\ T$.  For both linear and nonlinear dynamics the initial condition is denoted by
$f_0(x,v) = f(x,v,0)=f_{eq} +\delta f(x,v,0)$.


\subsection{Linear Results}
\label{sec:linear}

 Both the linear advection example of  Section \ref{sec:Adv} with the initial condition $f_0=f_{eq} + \delta f(x,v,0)$, where $\delta f(x,v,0)$
is given by (\ref{Vlasov_IC}),  and the Landau damping example of Section \ref{sec:Lin.Landau},  governed by
 (\ref{eq:lin.vlasov.pb})-(\ref{eq:lin.poisson.pb}),  require the specification of $f_{eq}$.
For both examples,  the  two choices for $f_{eq}$  introduced in Section \ref{sect:VP} are
considered: the Maxwellian equilibrium $f_M$ of (\ref{Meq}) and  the Lorentz equilibrium  $f_L$ of (\ref{Leq}).

Because it is most common to consider the Maxwell equilibrium, we explicate here  several  reasons for considering the Lorentz equilibrium, which to our knowledge has not been numerically tested previously in the literature.
\begin{enumerate}
\item Naturally occurring plasmas are sometimes not  Maxwellian but posses  kappa distributions \cite{kiv,mere} that have  power law tails in $v$. The Lorentz equilibrium is in a sense an extreme case of these in that it has $v^{-2}$ decay  at infinity, with the existence of the particle density but not kinetic energy.    In any event, distributions with power law tails are of physical interest and thus worth studying in their own right. (See also e.g.\ \cite{Va,VaDa}.)

\item Because of the slow decay in $v$,   the effect of truncating the velocity domain is amplified and a greater velocity domain is needed.  This makes the Lorentz equilibrium  a more stringent test for a numerical algorithm.
\item The linear dynamics of Vlasov theory is dominated by  phase mixing, the mechanism   that underlies Landau damping (cf. Section \ref{sec:Lin.Landau}).    For the advection problem, the Lorentz equilibrium gives decay of the form $\exp{(-kt)}$,  as opposed to the Maxwell equilibrium that gives decay of the form   $\exp{(-k^2t^2)}$  (cf.~Section \ref{sec:Adv}),  and this suggests it might be a better test for getting Landau damping right.  In fact, the reason for this exponential decay is that  linear advection with the  Lorentz equilibrium shares the same analytic structure as that of the Landau damping problem (cf.~Section \ref{sec:Lin.Landau}), while linear advection with the Maxwell equilibrium does not.   The essence of Landau damping can be traced to the  Riemann-Lebesgue lemma \cite{Stein},  which states that charge density integrals of the form
\[
\lim_{t\rightarrow\infty}\int \!dv \, g(v) e^{ivt} =0
\]
provided $g\in L^1$, i.e.\ $\int dv |g(v)|<\infty$.  The rate of this temporal decay depends on the nature of the function $g(v)$.

The underlying reason for exponential decay in both the advection problem with the Lorentz equilibrium and the Landau damping problem with either equilibrium, is that the function $g(v)$ for these problems is analytic in a strip in the complex $v$-space, and the damping rate is determined by the pole closest to the real $v$ axis.  Thus,  the basic mechanism of Landau damping is tested in the simpler advection problem when $f$ is given by (\ref{eq:f.form})  with $\delta f$ given by (\ref{Vlasov_IC}) and $f_{eq}$ being  the Lorentz equilibrium.
\item With the  Lorentz equilibrium one can use residue calculus to explicitly obtain expressions for the damping rates (see Appendix~\ref{app:Landau}). Although for the advection problem this is also true for the Maxwellian, this is not the case for the   Landau damping problem of Section \ref{sec:Lin.Landau}.
\end{enumerate}


\subsubsection{Advection Results}
\label{sec:Adv}

The advection equation,
\begin{equation}
\label{eq:lin.advection}
f_{_t}\ +\ vf_{_x} \ =\ 0 \,,
\end{equation}
is a natural test for assessing Vlasov algorithms because the Poisson equation is removed from the calculation and the focus is placed on the resolution of phase space.  With a Maxwellian equilibrium this example has been treated in many works, for example in \cite{CheKno,FilSonBer,NakYab,PohShoKam}.  In \cite{PohShoKam}   four standard Vlasov solvers are compared.

After computing $f$,  the long time behavior of the solution can be checked by comparing the computational  results with the known theoretical damping behavior due to phase mixing.  To this end the net charge density $\rho_{tot}$ is given by
\begin{eqnarray}
 \label{eq:net.charge}
\rho_{tot}(x,t)  &=&   1 - \int_{-\infty}^{\infty} \!dv\, f(x,v,t) \nonumber\\
&=&
 \  1 - \int_{-\infty}^{\infty}\!dv\,  f_0(x-vt,v)
 \nonumber\\
&=&
 - A  \int_{-\infty}^{\infty}\!dv\,  \cos{[k( x-vt)]}\,f_{eq}(v) \,,
 \label{rhotot}
\end{eqnarray}
where the second equality follows  because the solution to (\ref{eq:lin.advection}) is given by $f(x,v,t)=f_{_0}(x-vt,v)$, and the
third upon substitution of (\ref{Vlasov_IC}).  According to the Riemann-Lebesgue lemma,   $\lim_{t\rightarrow \infty}\rho_{tot} =0$
under mild requirements on $f_{eq}$.  Below we give  explicit forms for the decay for the Maxwell and Lorentz equilibria.

Although it is common to consider the linear advection problem for testing numerical algorithms,   it is not so
well-known that there is an intimate relationship between the solution of the advection problem and the  actual
Landau damping problem.  In fact, there is a one-to-one correspondence between solutions of the two.   In
\cite{mp,m} an invertible  linear integral transform, a generalization of the Hilbert transform, called  the $G$-transform, was explicitly constructed that maps (\ref{eq:lin.vlasov.pb}) into the advection equation (\ref{eq:lin.advection}). Thus, given an initial condition for  the advection equation, there exists an initial condition for (\ref{eq:lin.vlasov.pb}) that transforms into the same solution.  For the Lorentz equilibrium one can use residue calculus to obtain explicit expressions. In particular, the $G$-transform of  the  Lorentz equilibrium $f_L$ of (\ref{Leq})  is
\begin{equation}
G[f_L]= \frac{1}{\pi}\frac1{1+ v^2}
\left[
1+ \frac1{k^2} \frac{(1 -3 v^2)}{(1 + v^2)^2}
\right]\,,
\end{equation}
where the procedure is done mode by  mode and $k$ is the mode number.  This means that a solution to the linear advection problem
with the initial condition
\begin{equation}
f_0= A\cos(kx) f_L
\end{equation}
is equivalent to the solution of the linear Vlasov-Poisson system with the initial condition
\begin{equation}
f_0= A\cos(kx)G[f_L]\,.
\label{g}
\end{equation}

The difference in long time decay between the advection problem and that of Landau damping can be
traced to poles that occur in the integral transform.   One could use the integral transform to further
test the veracity of a numerical algorithm by comparing the solutions of the advection and Landau problems,
but we will not do so here.  However,  given the understanding provided by the integral transform,  it is quite 
natural to examine the advection problem with initial conditions  that are meromorphic in velocity like the Lorentz equilibrium.

\medskip

\noindent\underline{Linear advection with a Maxwell equilibrium}

For our first example   (\ref{eq:lin.advection}) is solved for  $f_{eq}=f_M$ given by (\ref{Meq}).
We choose $A=0.1, k=0.5, L=4\pi, V_C=5$ and $T=40$. For  this particular case, it is easily shown by elementary methods
that the net charge density  of (\ref{rhotot})  is given by
\begin{equation}
 \label{eq:net.chargeM}
\rho_{tot}(x,t)\ = \  - A\cos{(k x)}\,e^{-k^2 t^2/2} \,.
\end{equation}
This implies that $\mathrm{max}_x\,|\rho_{tot}(x,t)|= 0.1\,e^{-t^2/8}$, since $k=0.5$.

\begin{figure}[ht]
\begin{center}
\includegraphics[angle=0,width=0.47\textwidth,height=0.35\textwidth]{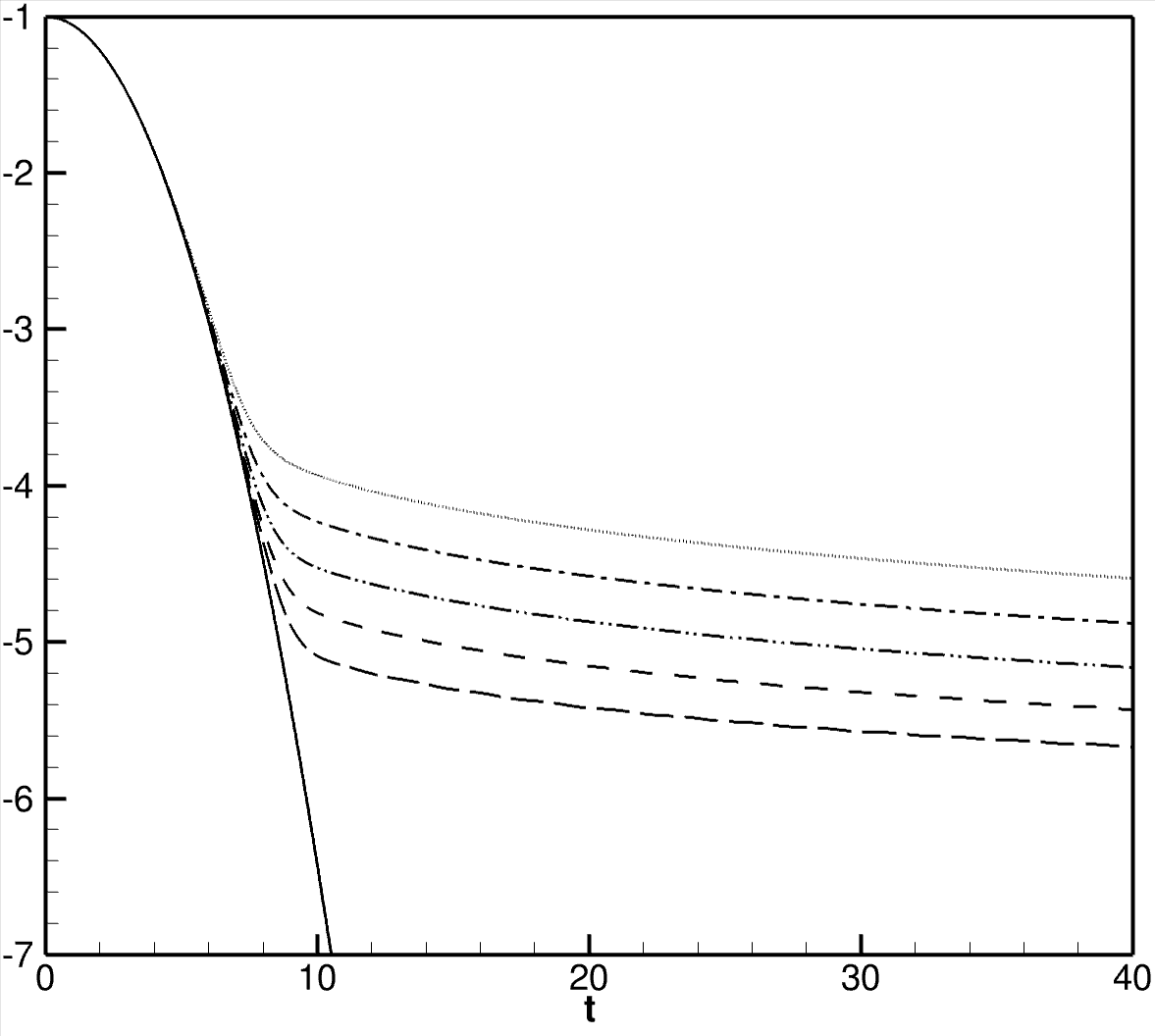}
\end{center}
\caption{(Linear advection with Maxwell equilibrium) Plot of $\rm{log}_{10}(\mathrm{max}_x\,|\rho_{tot}(x,t)|)=$
$-(\rm{log}_{10}e)t^2/8+1$ vs. $t$: analytic solution {\em (solid)}, $(N_{h_x}, N_{h_v})=$ (500,400){\em (dot)},
(1000,800){\em (dash-dot)}, (2000,1600){\em (dash-dot-dot)}, (4000, 400) {\em (short dash)}, (8000, 400)
{\em (long dash)}.}
\label{fig:Adv Maxwellian}
\end{figure}

To test the accuracy and convergence of the UG method, $\mathrm{max}_x\,|\rho_{tot}(x,t)|$ is computed numerically and the results
are plotted in Fig.~\ref{fig:Adv Maxwellian}.  The numerical results were generated using the five uniform meshes
$(N_{h_x},N_{h_v})=(500,400)$, $(1000,800)$, $(2000,1600)$, $(4000,400)$, $(8000,400)$, where $N_{h_x}$ and $N_{h_v}$
denote the number of partitions of the $x$-axis and the $v$-axis, respectively.

The first three meshes are such that $h_x\approx h_v$, whereas the fourth and fifth meshes are such that
$h_x\approx h_v/8$ and $h_x\approx h_v/16$, respectively.  The motivation for using the last two meshes
comes from the fact that the problem
being approximated involves only advection in the $x$-direction.  Hence,  it is reasonable to assume that mesh
refinements in $x$ will improve the numerical accuracy as much as performing refinements in both $x$ and $v$,
provided the refinement
level in $v$ is sufficiently small so that the error is almost entirely due to the refinement level in $x$.
Figure~\ref{fig:Adv Maxwellian} clearly shows that the UG method is both accurate and numerically convergent
under mesh refinements.

Our results compare with those of  \cite{CheKno,FilSonBer,NakYab,PohShoKam} for early times where solutions are accurate, but unlike the others we do not obtain the later time recurrence that arises  from periodicity of $f_{_0}(x-vt,v)$ in its first argument and the velocity mesh size used in evaluating the integral of (\ref{rhotot}).  This is because of the fineness of our mesh and because of  the specific dissipative properties of the DG method give monotonic error.  With a mesh of $N_{h_v}=400$ the recurrence time is $T_R=2\pi k/\Delta v= \pi N_{h_v}/(2V_c)\approx 126$.

\medskip

\noindent\underline{Linear advection with a Lorentz equilibrium}

In this second example, we consider the advection problem with the Lorentz equilibrium  (\ref{Leq}).  Numerical results
 are given for four different values of the  wavenumber $k$.   We will revisit these cases in  Section \ref{sec:Lin.Landau},
 where  we consider the actual Landau damping problem with the electric field  not taken to be zero.
Specifically, here Eq.~(\ref{eq:lin.advection}) is solved for $f_{eq}=f_L=\pi^{-1}/(v^2+1)$, $A=0.01$,
 and the large value $V_c = 30$ for each
of the wavenumbers $k=1/8$, 1/6, 1/4, and 1/2.  The corresponding values for $L$ and $T$ for $k=$1/8, 1/6, 1/4,
and 1/2 are $L=16\pi$, $12\pi$, $8\pi$ and $4\pi$ and $T=75,$ 75, 50, and 50, respectively.  The uniform mesh
$(N_{_x},N_{_v})=(1000,2000)$ was employed in each of the four cases.  For this case, it is easily  shown using
 residue calculus  that the   charge density of (\ref{rhotot})  is given by
\begin{equation}
\label{eq:net.chargeL}
\rho_{tot}(x,t)\ =\  -A\cos{(kx)}\,e^{-kt} \,.
\end{equation}
This implies that $\mathrm{max}_x\,|\rho_{tot}(x,t)|= 0.01\,e^{-kt} $.

\begin{figure}[ht]
\includegraphics[angle=0,width=0.47\textwidth,height=0.35\textwidth]{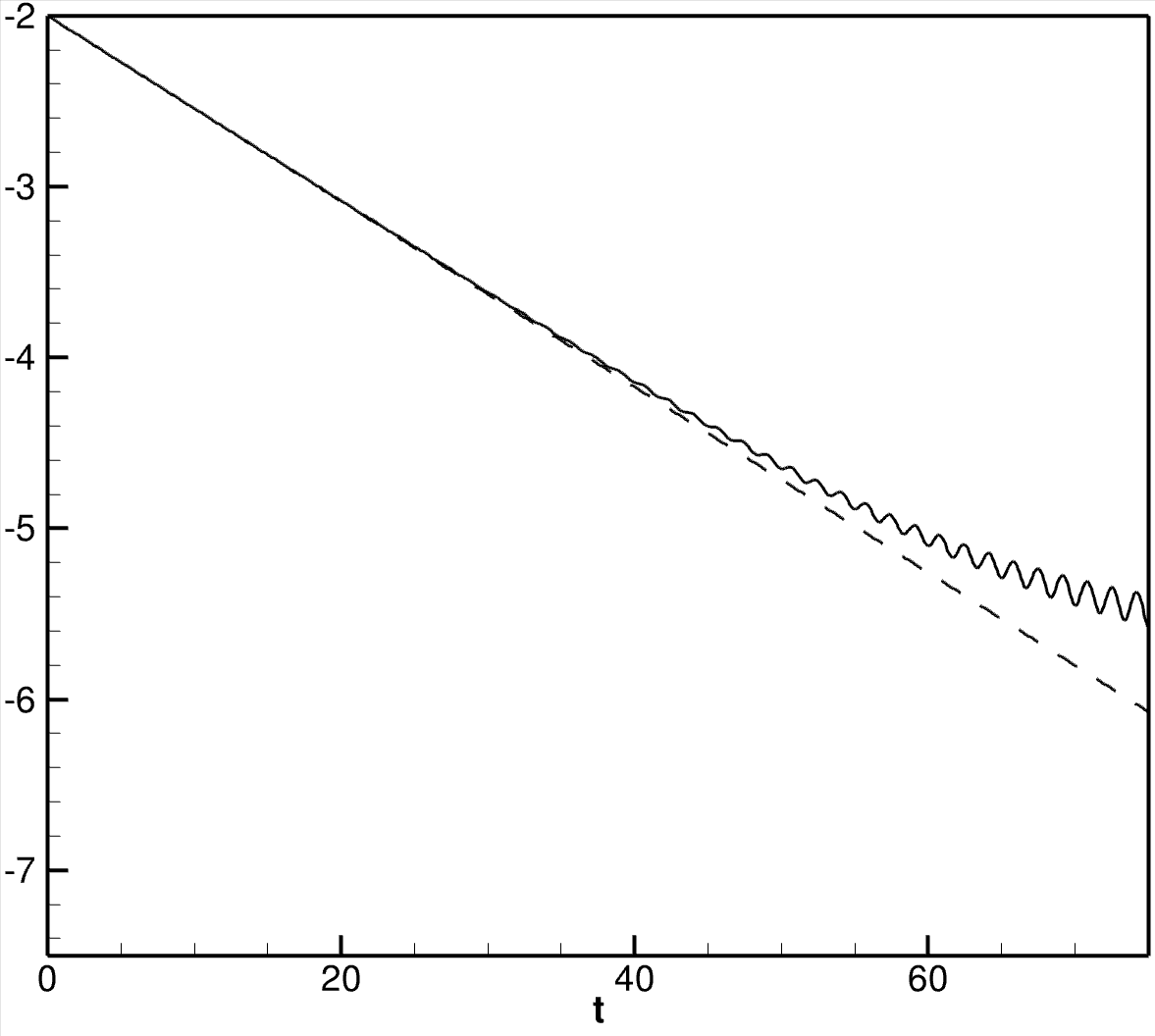}
\hspace{0.02\textwidth}
\includegraphics[angle=0,width=0.47\textwidth,height=0.35\textwidth]{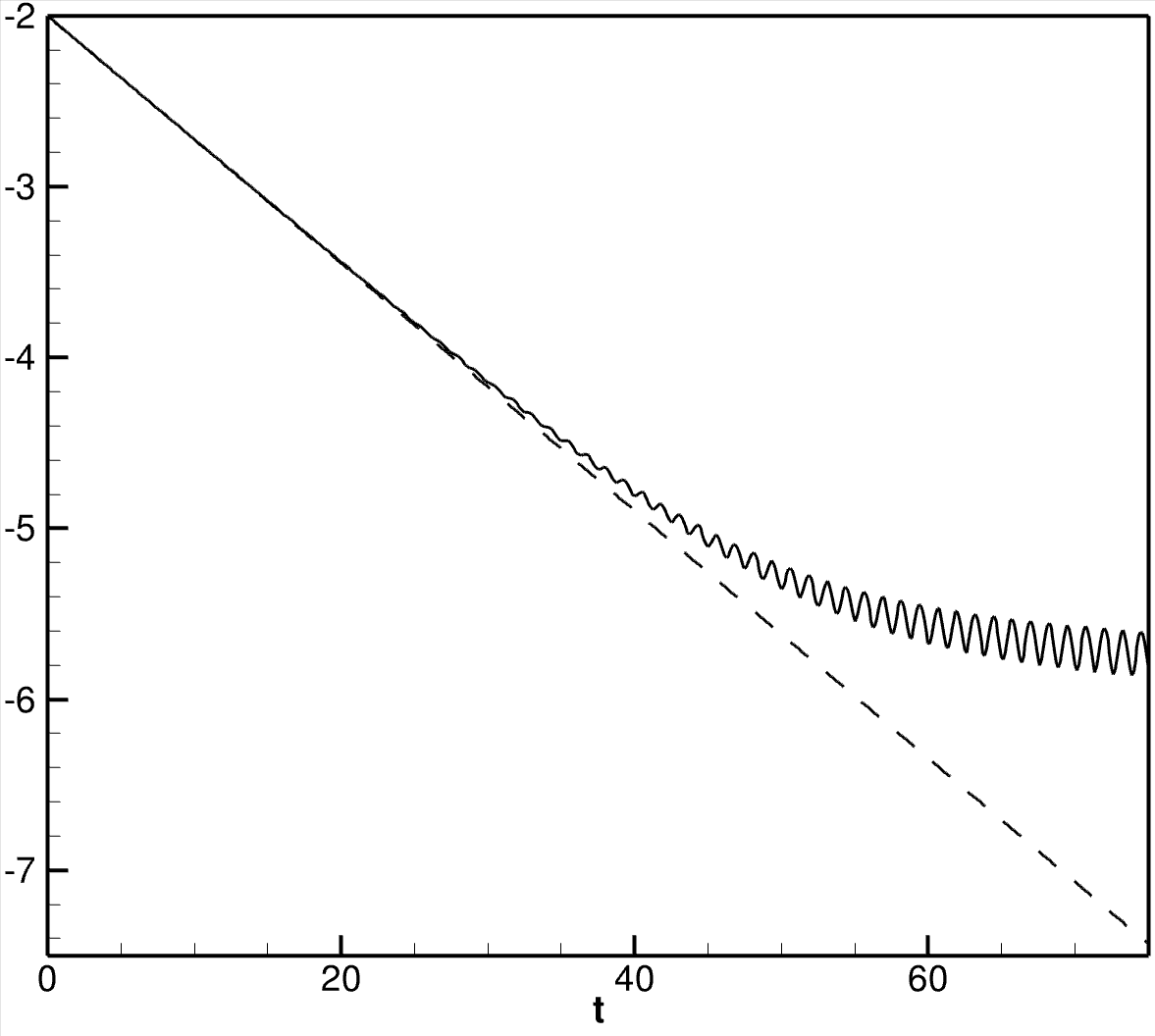}
\vspace{0.5cm} \\
\includegraphics[angle=0,width=0.47\textwidth,height=0.35\textwidth]{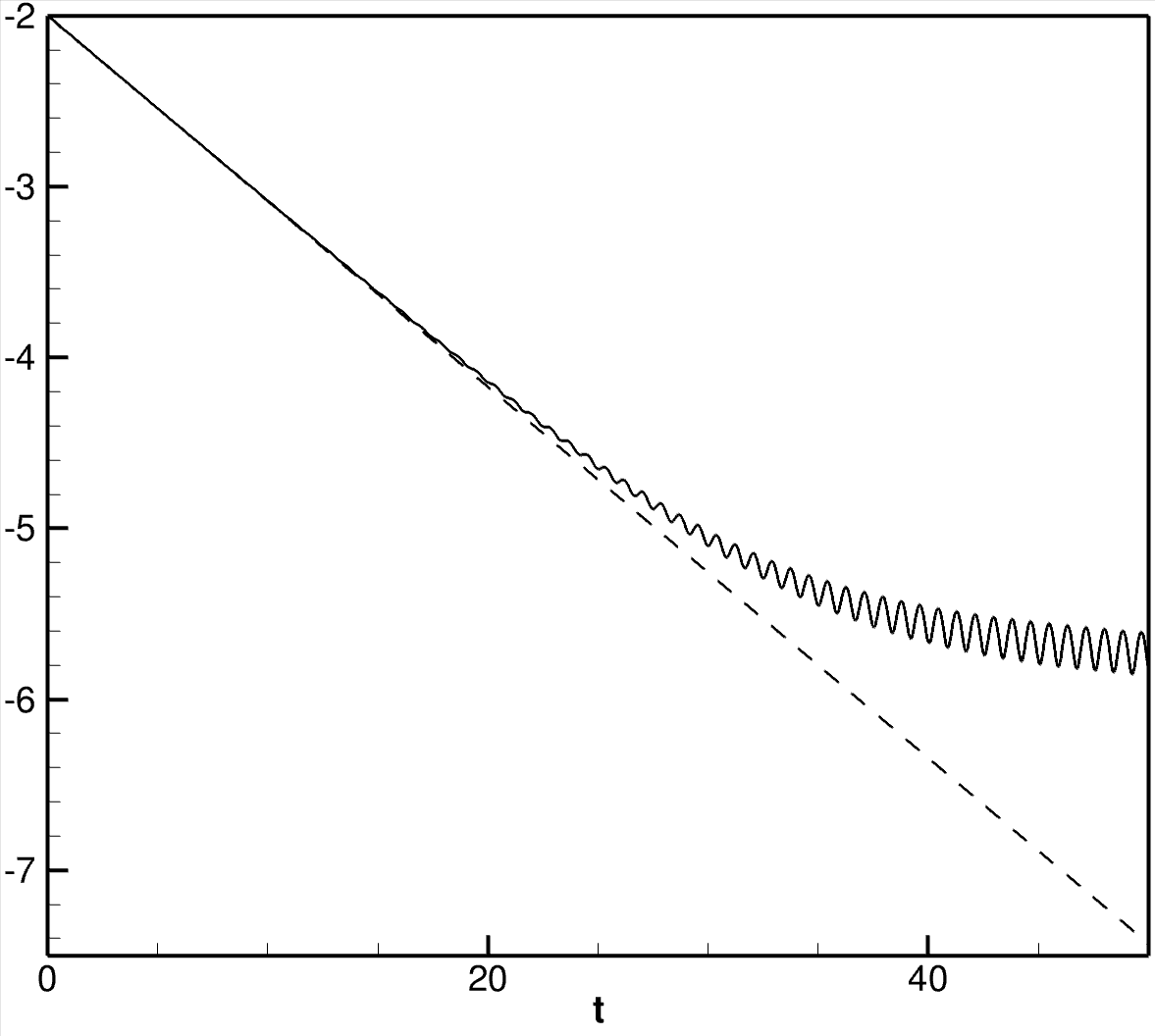}
\hspace{0.02\textwidth}
\includegraphics[angle=0,width=0.47\textwidth,height=0.35\textwidth]{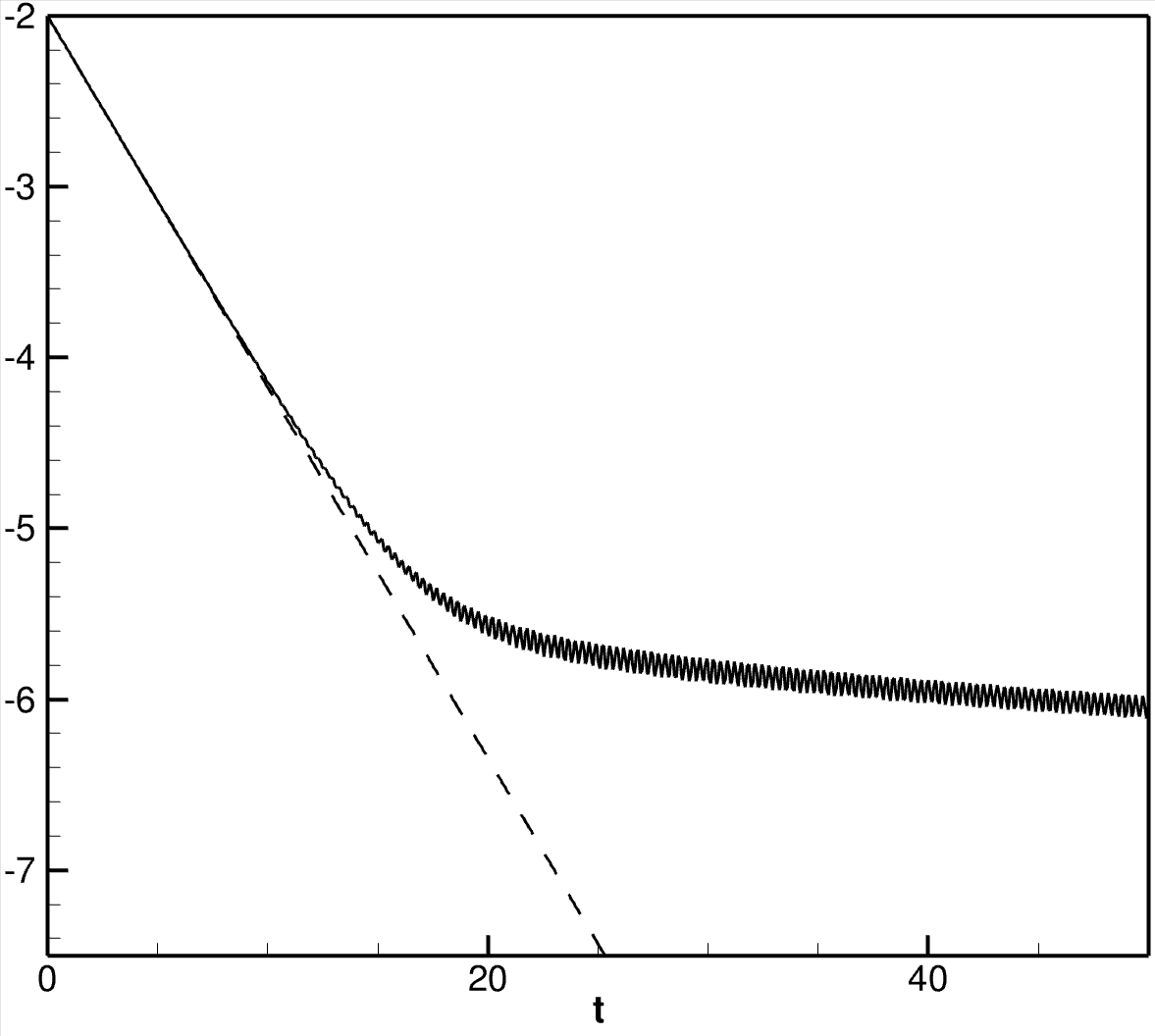} \\
\caption{(Linear advection with Lorentz equilibrium) Plot of $\rm{log}_{10}(\mathrm{max}_x\,|\rho_{tot}(x,t)|)=$
$-(\rm{log}_{10}e)kt+2$ vs. $t$: $k$=1/8 {\em (top left)}, $k$=1/6 {\em (top right)}, $k$=1/4 {\em (bottom left)} and
$k$=1/2 {\em (bottom right)}; analytic solution {\em (dash)}, numerical solutions {\em (solid)}.}
\label{fig:Adv Lorentzian}
\end{figure}

The computed results for each of the four wavenumbers $k$ are shown in Fig.~\ref{fig:Adv Lorentzian}
along  with the exact result of (\ref{eq:net.chargeL}).  From the figure it is seen that for early times
the computations match the theoretical result (\ref{eq:net.chargeL}). At late times the computations
diverge and, as anticipated, it is more difficult to resolve cases with larger $k$, i.e.\ with finer spatial structure.


\subsubsection{Linear Landau Damping Results}
\label{sec:Lin.Landau}

The next two examples test the ability of the method to reproduce results consistent with the theoretical
results of linear Landau damping.  We emphasize that it is not enough to merely show exponential decay,
but to believe an algorithm one must compare the decay rates and the parametric dependence of the theoretical rates.   As above, both Maxwellian and Lorentzian equilibria are considered.  For the Maxwellian equilibrium, which were previously treated in \cite{CheKno,FilSonBer,GraFei,NakYab,ZakGarBoy2,ZhoGuoShu}, results are computed using
four successively refined uniform meshes in order to demonstrate that the numerical decay rates converge
to the theoretical decay rate under mesh refinement.  For the Lorentz equilibrium, we will see that the
UG method is robust in the sense that it produces the correct decay rates for  different wavenumbers $k$.

As noted in Section \ref{sect:VP},   Landau showed  the electric field decays exponentially in the time-asymptotic limit (for more rigor see  \cite{MasFed} for the linear case and the recent nonlinear results of  \cite{MouVi}, of which Ref.~37 is an early version of the present work).   More specifically, if  we write the frequency as $\omega(k)=\omega_R(k)+i\gamma(k)$, where $\omega_R(k)$ and $\gamma(k)$ are real-valued, then in the time-asymptotic limit $E_k(\omega,t)$ decays at a rate $\gamma(k)$ and oscillates at a frequency $\omega_R(k)$.

Besides the  Maxwellian equilibrium of (\ref{Meq}),  the Lorentzian equilibrium $f_L$ of (\ref{Leq})  gives rise to exponential damping of the electric field.    In this case, the decay rate $\gamma(k)$ for the
$n$-th electric field mode with  $k=2\pi n/L$ is given by
\be
\gamma(k) = -k\,,
\label{eq:lorentz.damp}
\ee
where $\gamma <0$ implies damping, and the corresponding frequency of the electric field oscillations satisfies
\be
\omega_R(k) = 1\,.
\label{eq:lorentz.freq}
\ee
The derivation of (\ref{eq:lorentz.damp})-(\ref{eq:lorentz.freq}) is described  in Appendix~\ref{app:Landau}.  It is important to note that
formulae (\ref{eq:lorentz.damp}) and (\ref{eq:lorentz.freq}) are explicit, which directly results from using the Lorentz equilibrium,
whereas, as noted above,  the formula for $\gamma(k)$ and $\omega_R(k)$ when a Maxwell equilibrium is used are implicitly defined \cite{Landau}.

\medskip


\noindent\underline{Linear Landau damping with a Maxwell equilibrium}

We solved the  linear system~(\ref{eq:lin.vlasov.pb})-(\ref{eq:lin.poisson.pb})  with
$f_{eq}=f_M=(2\pi)^{-1/2}e^{-v^2/2}$, $A=0.01$, $k=0.5$, $L = 4\pi$, $V_c = 4.5$ and $T = 80$.
For this problem, the theoretical decay rate and frequency of oscillations to the third decimal
digit are respectively equal to $\gamma =-0.153$ and $\omega_R=1.415$ \cite{CheKno}. Approximate
solutions are computed using the four uniform meshes $(N_{h_x},N_{h_v})=(250,400)$, $(500,800)$, $(1000,1600)$ and $(2000,1600)$.

\begin{figure}[ht]
\includegraphics[angle=0,width=0.48\textwidth,height=0.36\textwidth]{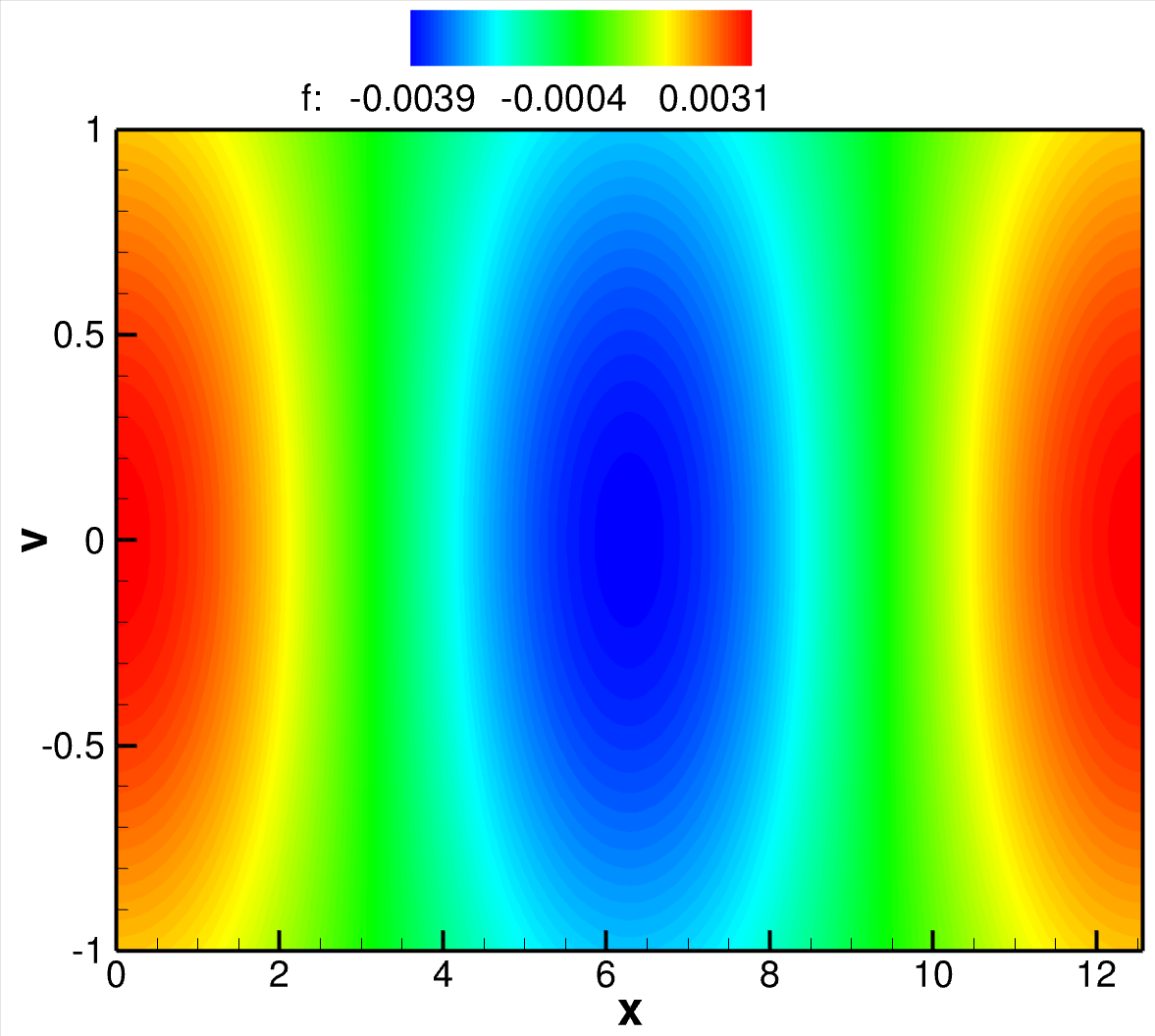}
\hspace{0.02\textwidth}
\includegraphics[angle=0,width=0.48\textwidth,height=0.36\textwidth]{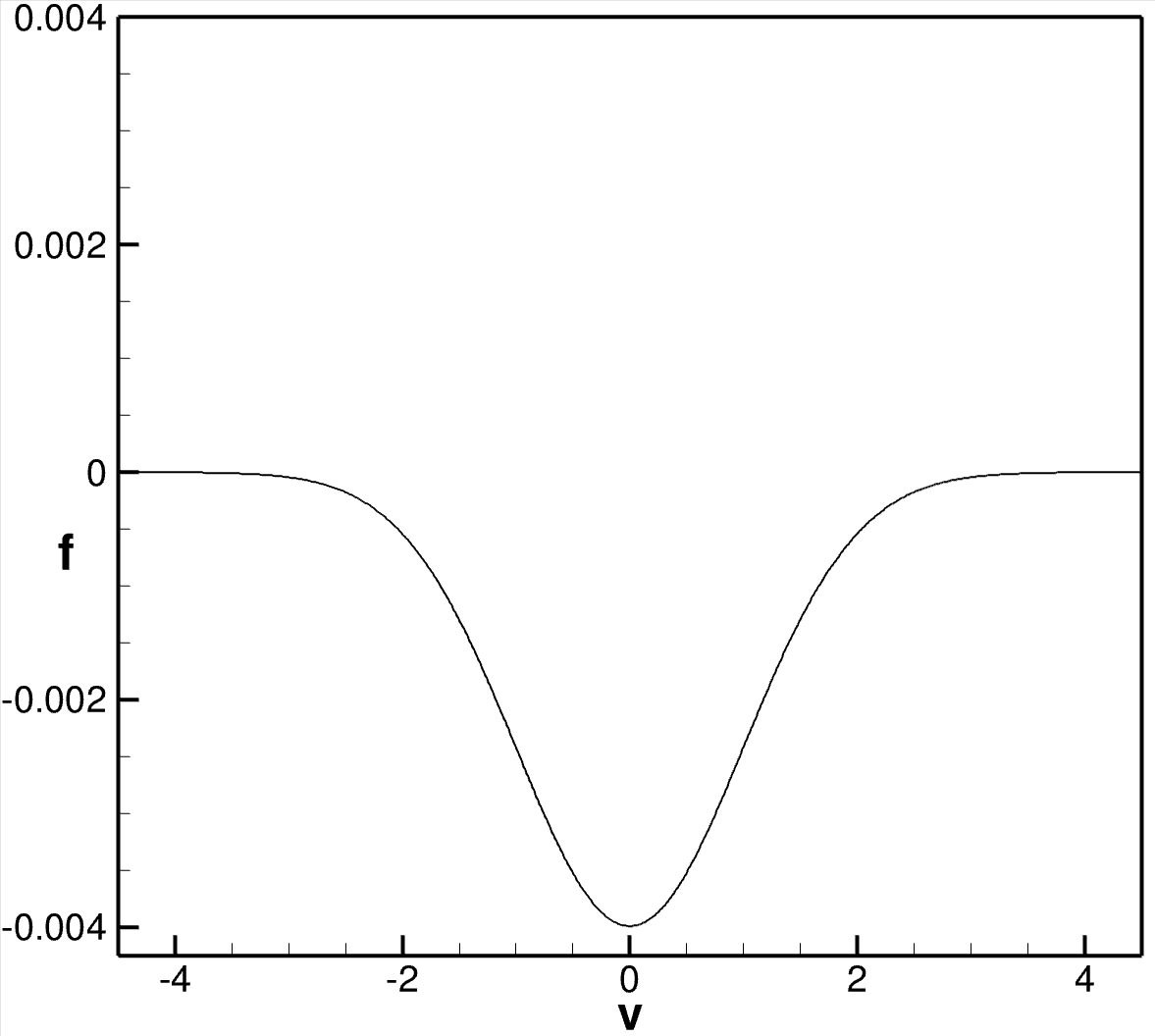}
\vspace{0.08cm} \\
\includegraphics[angle=0,width=0.48\textwidth,height=0.36\textwidth]{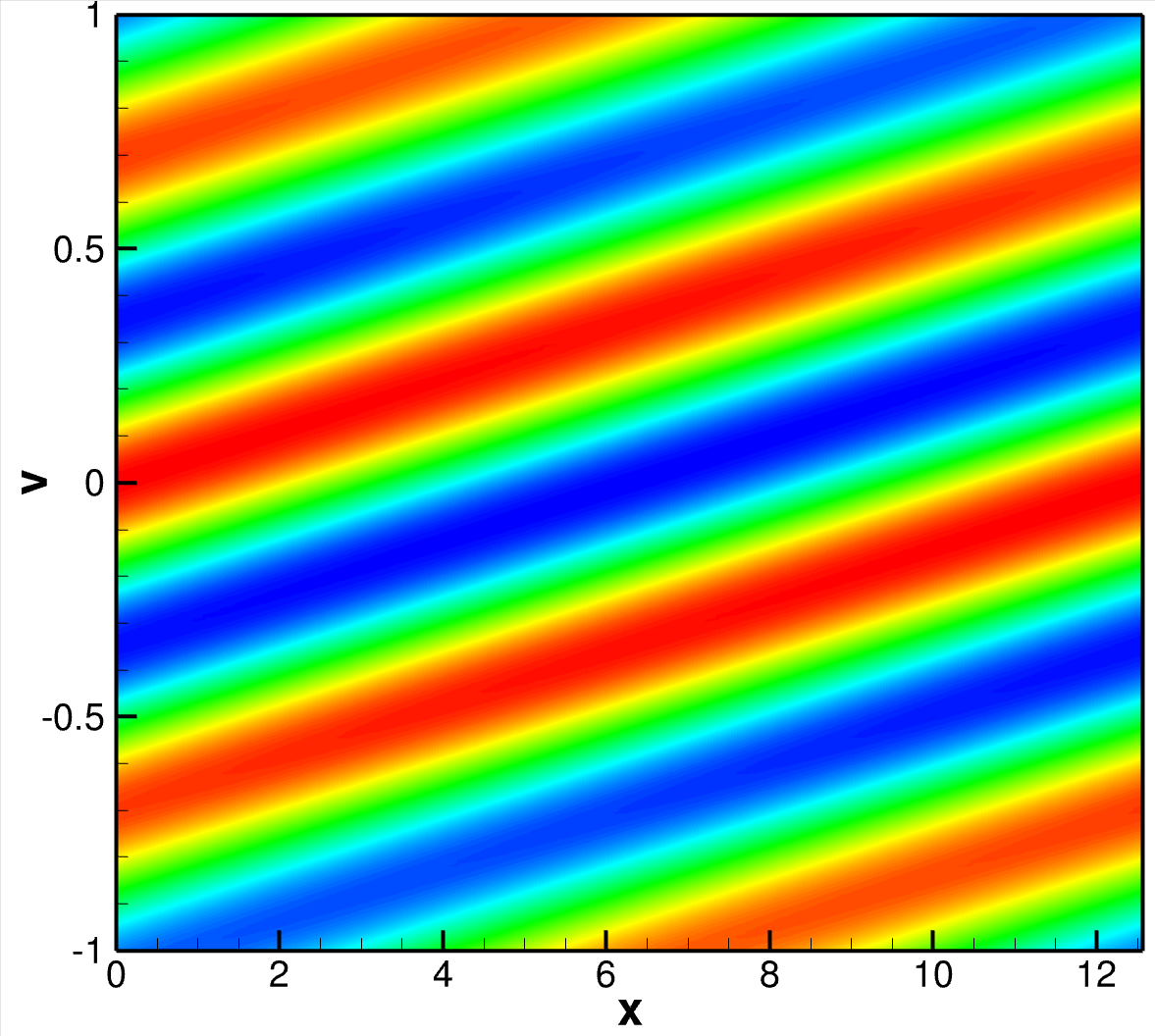}
\hspace{0.02\textwidth}
\includegraphics[angle=0,width=0.48\textwidth,height=0.36\textwidth]{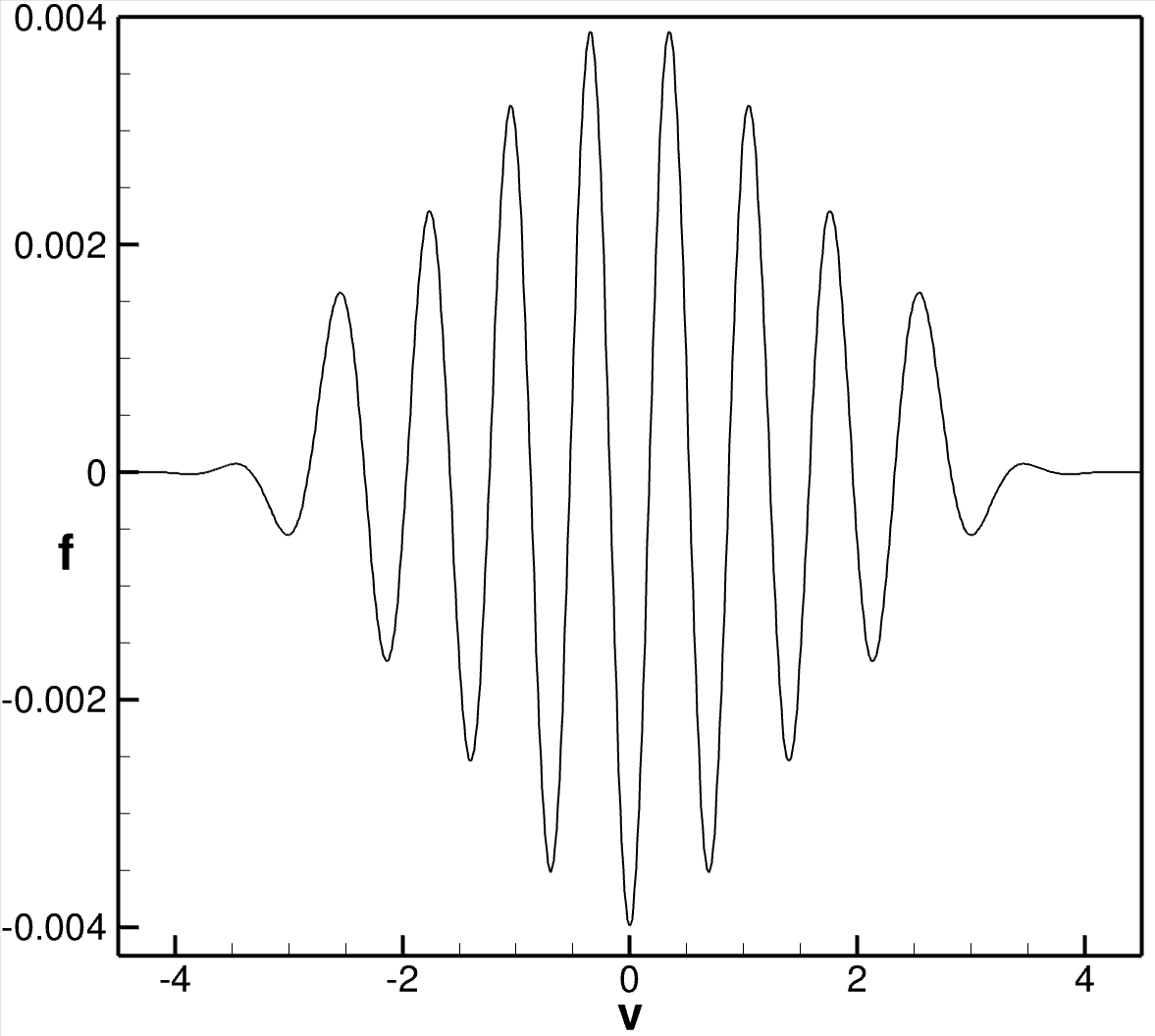}
\vspace{0.08cm} \\
\includegraphics[angle=0,width=0.48\textwidth,height=0.36\textwidth]{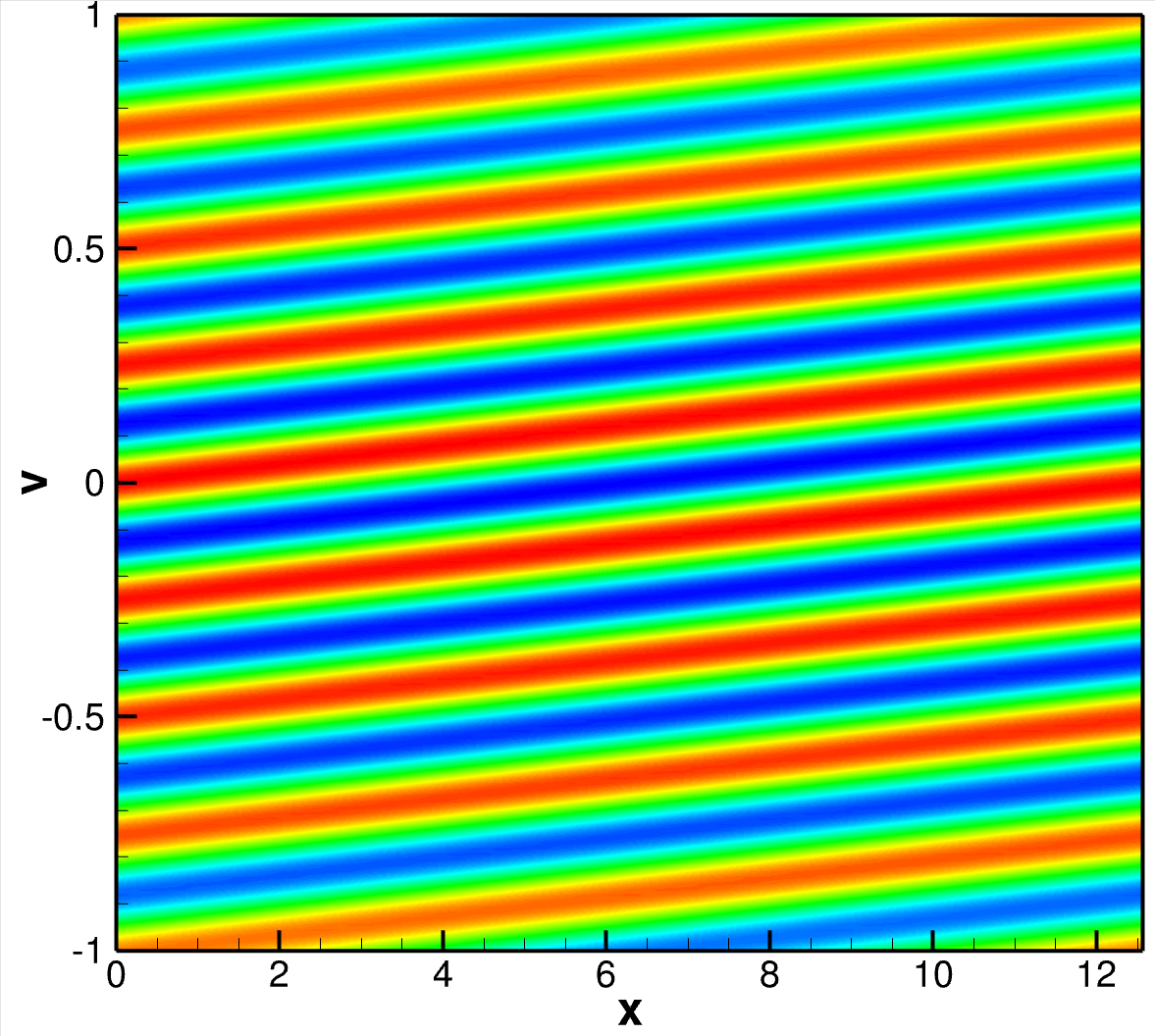}
\hspace{0.02\textwidth}
\includegraphics[angle=0,width=0.48\textwidth,height=0.36\textwidth]{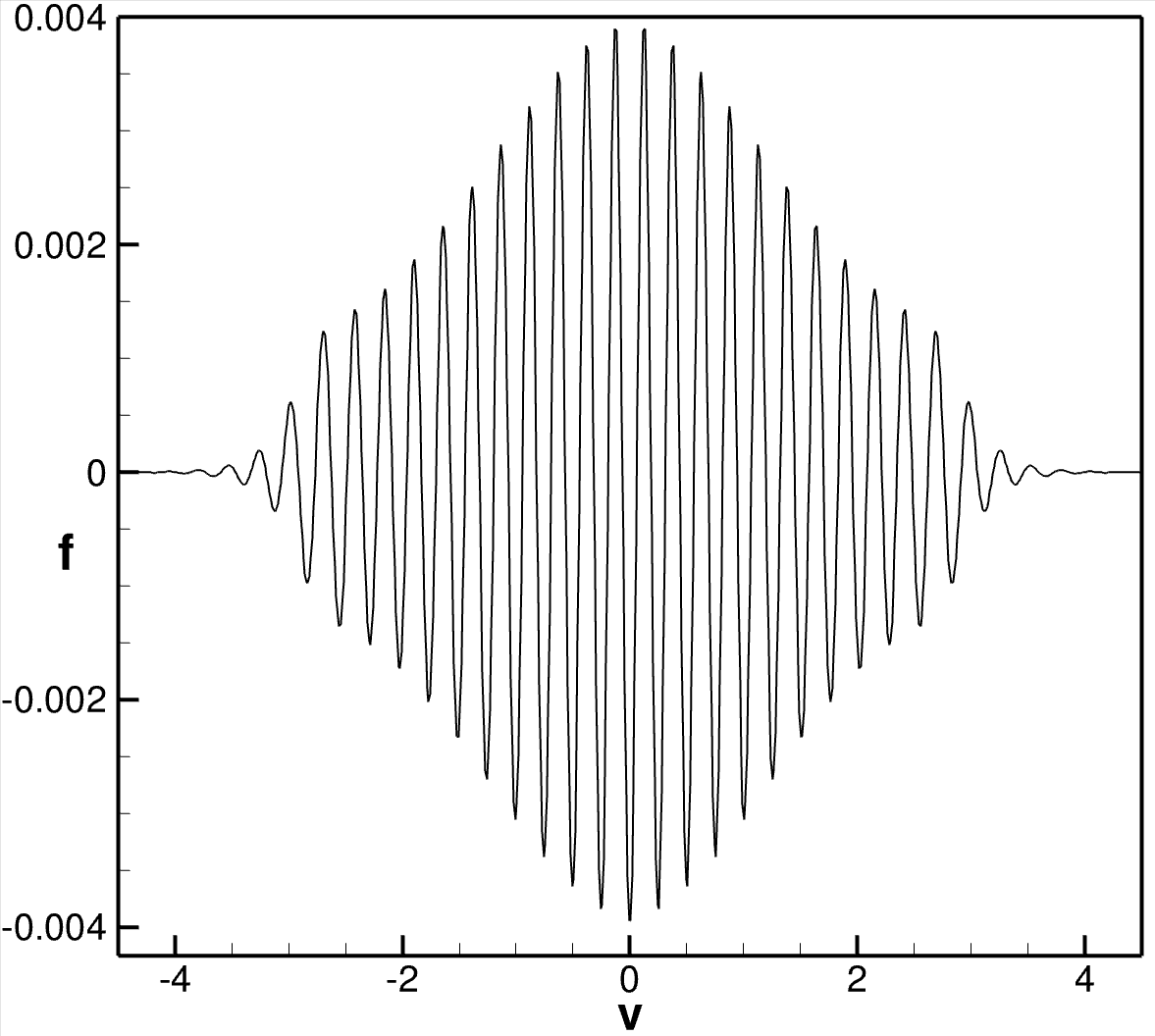}
\vspace{0.08cm} \\
\includegraphics[angle=0,width=0.48\textwidth,height=0.36\textwidth]{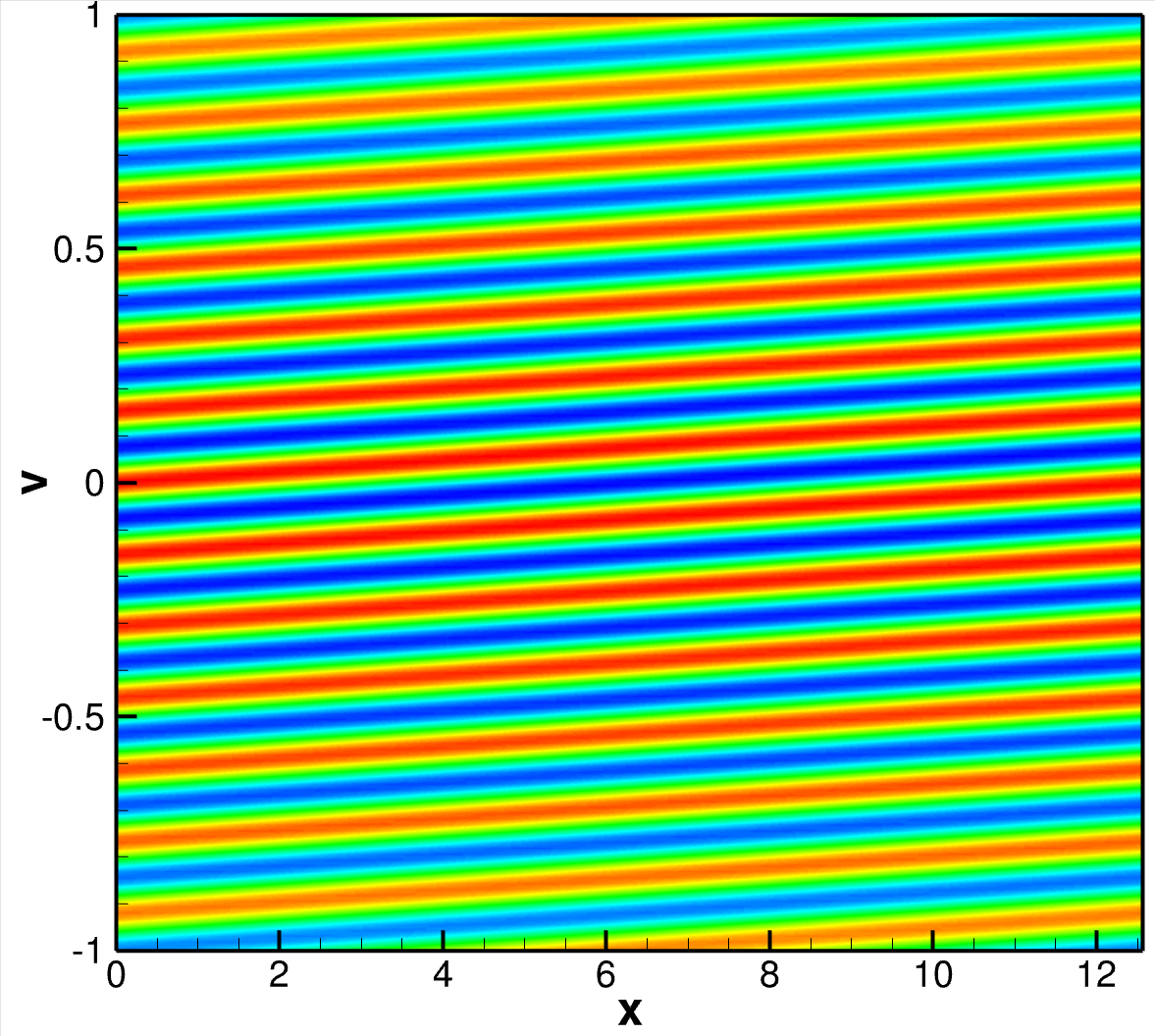}
\hspace{0.02\textwidth}
\includegraphics[angle=0,width=0.48\textwidth,height=0.36\textwidth]{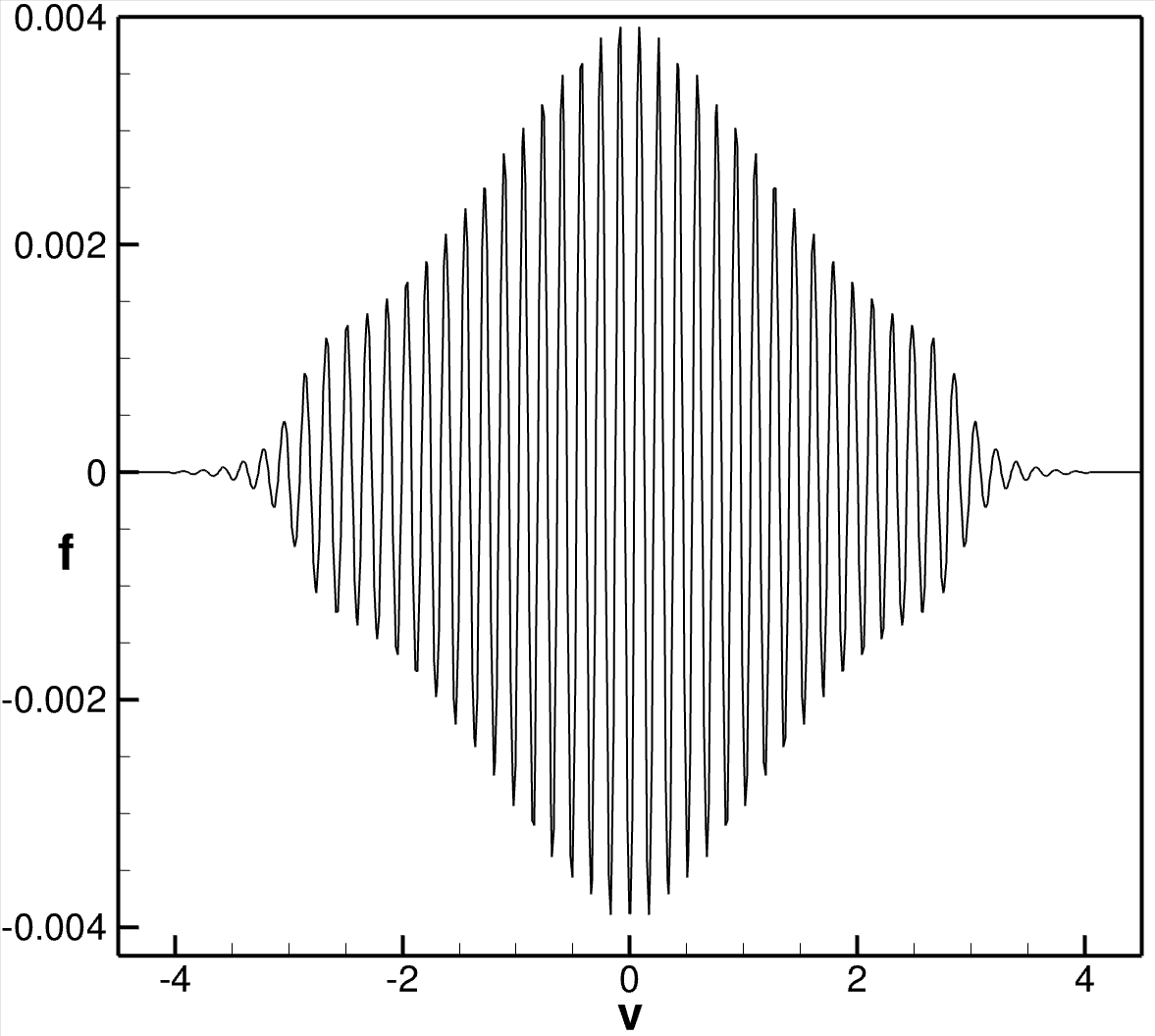} \\
\caption{(Linear Landau damping with Maxwell equilibrium) Contour plots ({\em left}) and cross-sectional
plots ({\em right}), $x=2\pi$, for $\delta f$  at $t=0$, $t=25$, $t=50$, $t=75$ ({\em descending order}).}
\label{fig:Maxwellian sol}
\end{figure}

Phase-space contour plots and cross-sectional plots in $v$ of  the approximate solution $f_h$
for $(N_{h_x},N_{h_v})=(500,800)$ and $t=0,$ $t=25$, $t=50$, and $t=75$ are displayed in
Fig.~\ref{fig:Maxwellian sol}.  These sequential plots show the increase
in filamentation of $f_h$ as time elapses.

\begin{figure}[ht]
\includegraphics[angle=0,width=0.47\textwidth,height=0.35\textwidth]{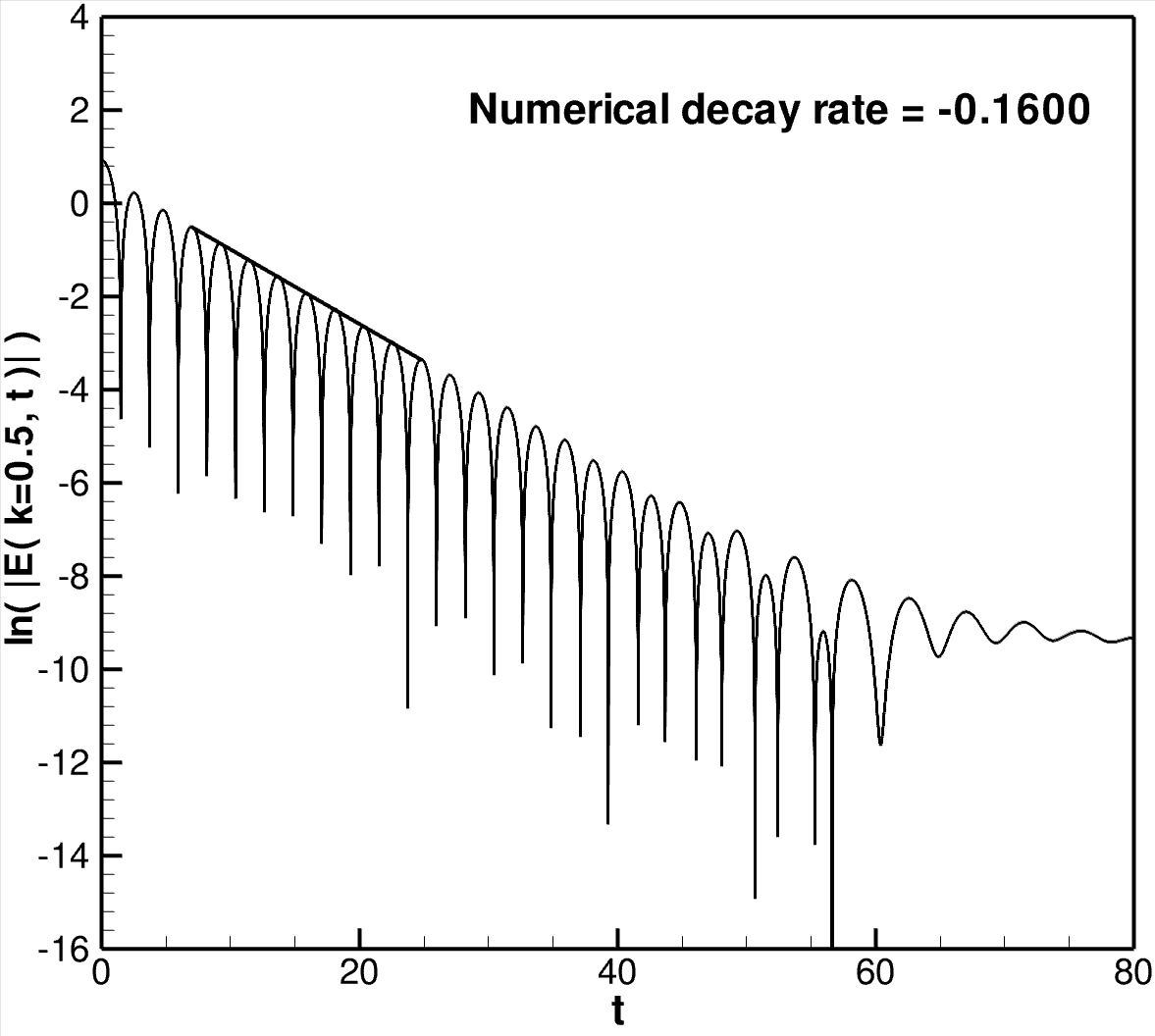}
\hspace{0.02\textwidth}
\includegraphics[angle=0,width=0.47\textwidth,height=0.35\textwidth]{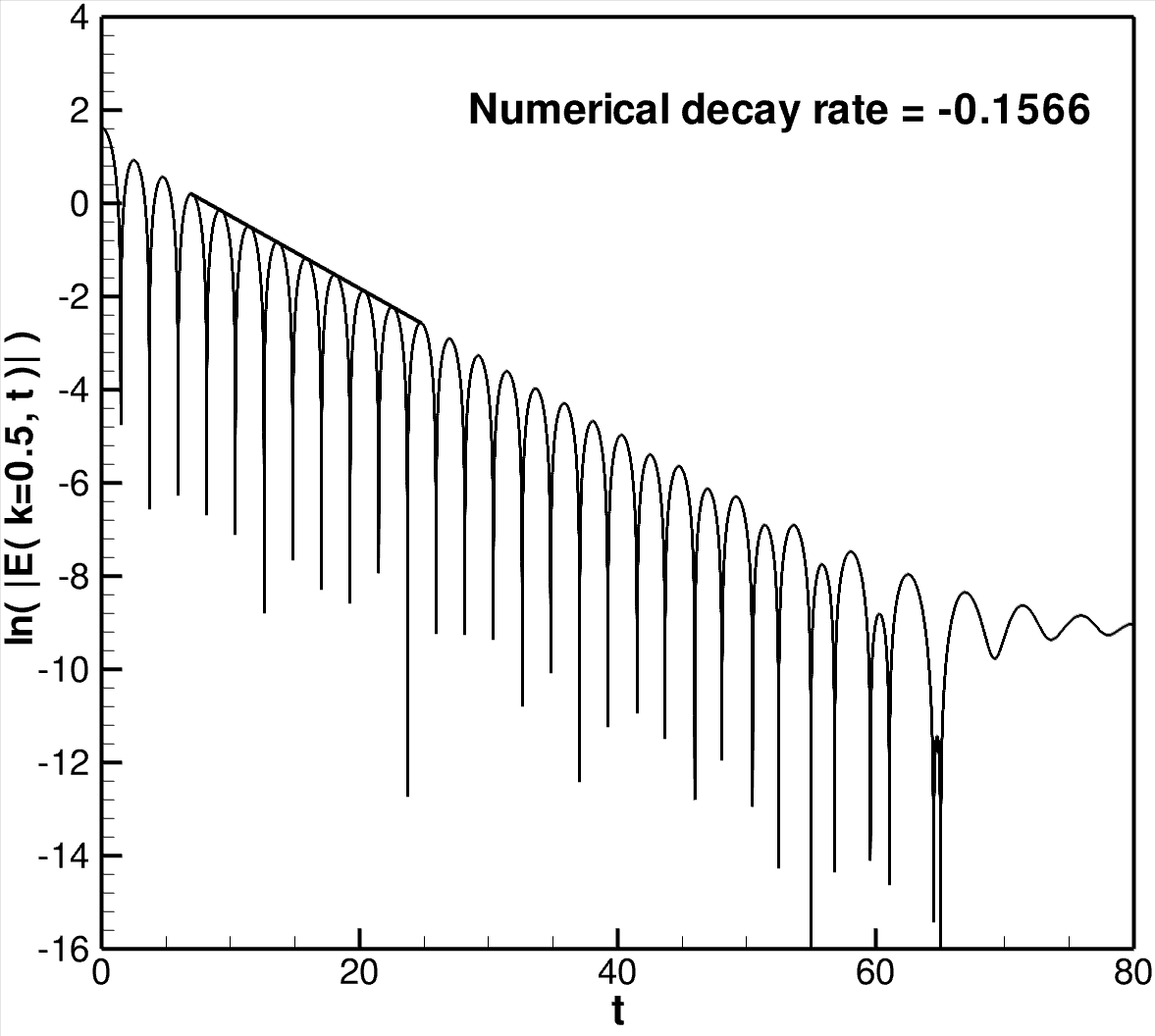}
\vspace{0.5cm} \\
\includegraphics[angle=0,width=0.47\textwidth,height=0.35\textwidth]{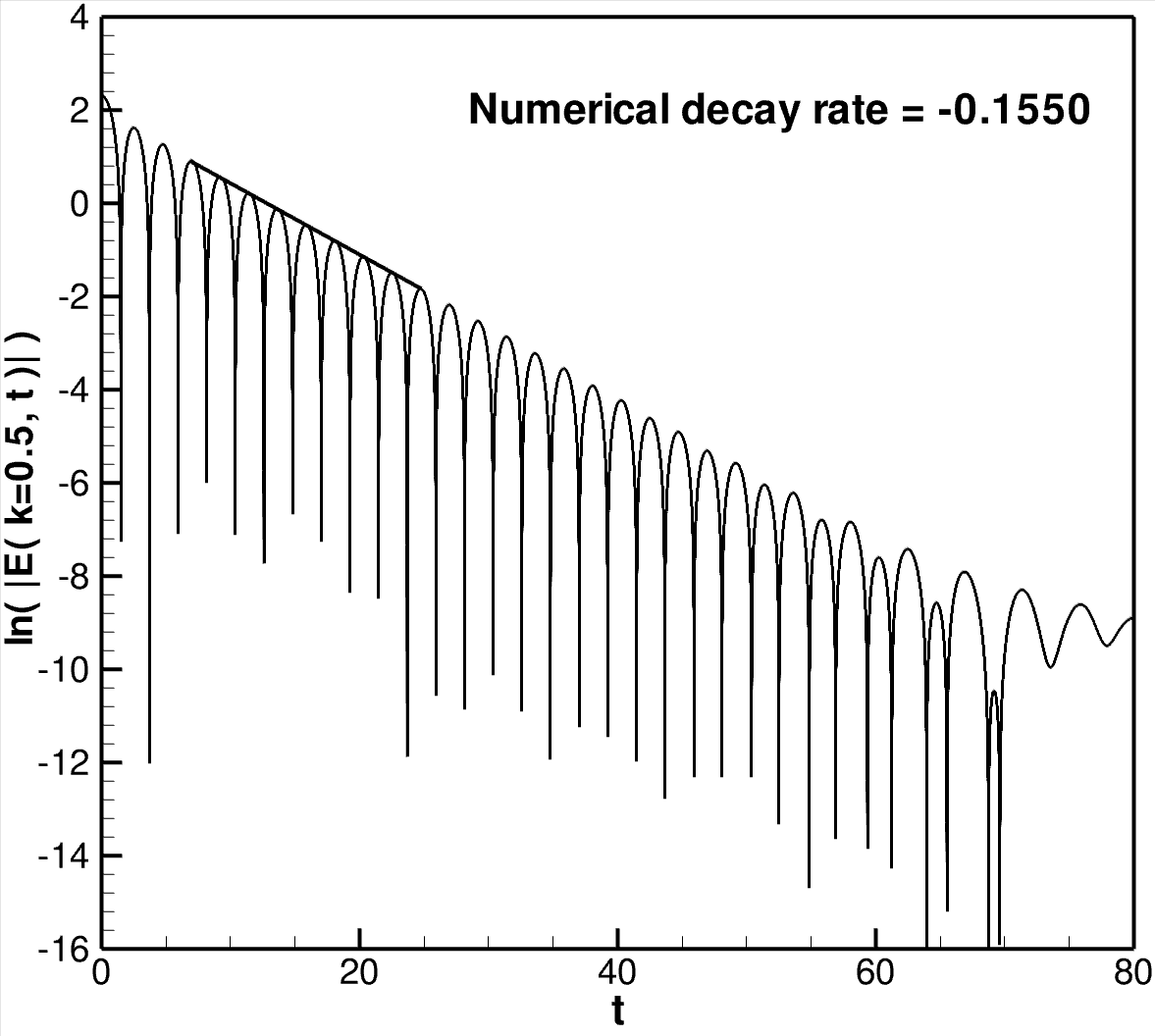}
\hspace{0.02\textwidth}
\includegraphics[angle=0,width=0.47\textwidth,height=0.35\textwidth]{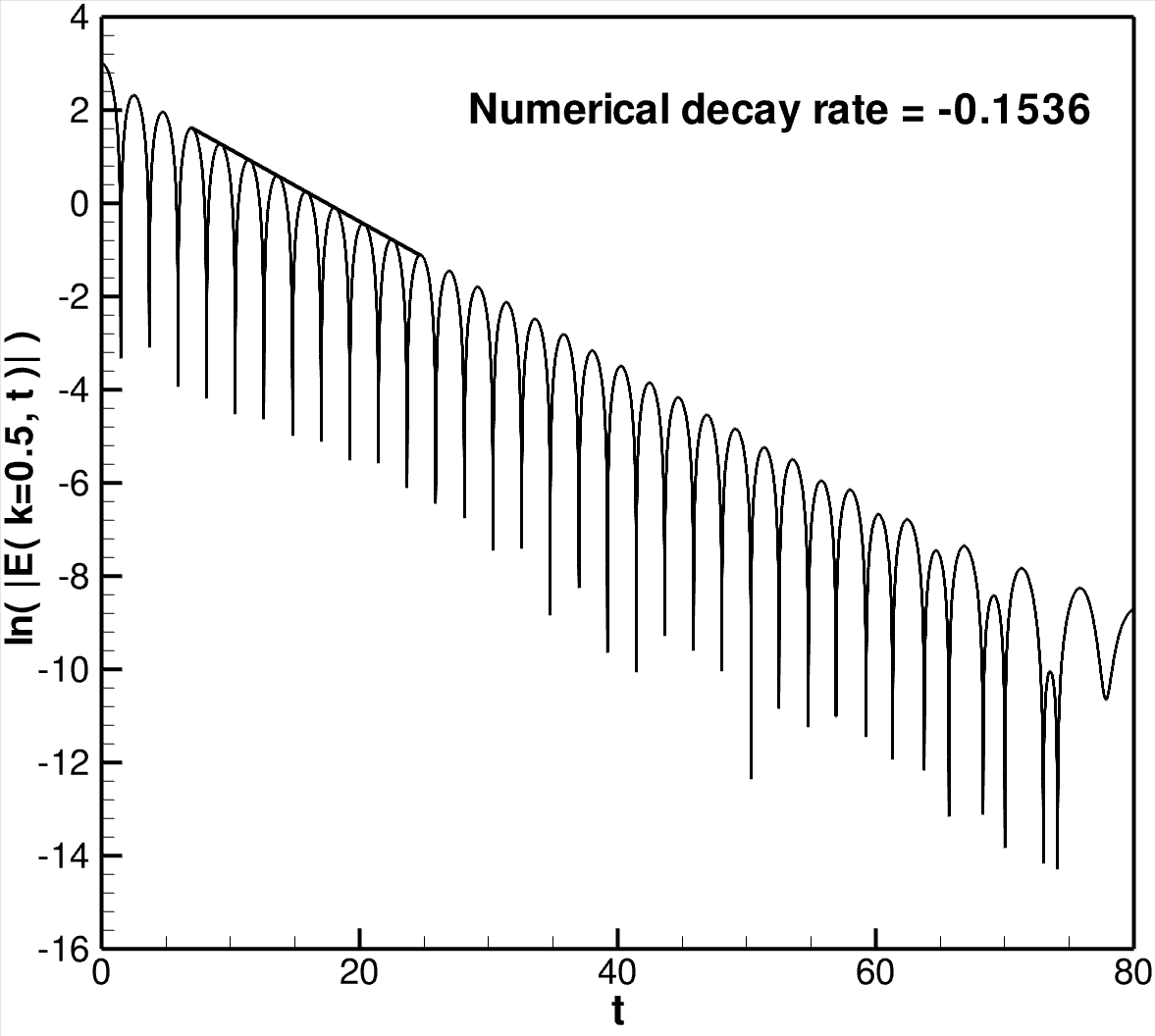} \\
\caption{(Linear Landau damping with Maxwell equilibrium) Time decay plots of fundamental mode
(with arbitrary ordinate origin)  under mesh refinement: $(N_{h_x}, N_{h_v})=$(250, 200) {\em (top left)}, (500, 400) {\em (top right)},
(1000, 800) {\em (bottom left)} and (2000, 1600) {\em (bottom right)}. The numerical decay rate converges to the theoretical value of  -0.153  to within three decimal-digit accuracy; similarly, the numerical oscillation frequency agrees with the theoretical frequency of $\omega_R = 1.415$ to within three decimal digits.}
\label{fig:Maxwellian refinement}
\end{figure}

The convergence of numerical decay rates of the dominant electric field Fourier mode is demonstrated in
Fig.~\ref{fig:Maxwellian refinement}.  The decay rate resulting for each mesh was computed by calculating
the slope of the straight line plotted in the figure.  In each case, the line was defined by the point occurring at the peak of the
third oscillation and the point occurring at the peak of the ninth oscillation.  A time step of $\Delta t=0.001$
was used in order to ensure that the points defining each of the lines were actual computed data points rather
than points that were determined using some interpolation of the data.  Under mesh refinement, the numerical decay
rate is seen to converge, up to three decimal-digit accuracy, to the theoretical decay rate of -0.153.
In all four cases, the numerical frequency is observed to correspond to the theoretical frequency up to three
decimal-digit accuracy.  We also note that, upon refining the mesh, the decay of the dominant mode is sustained
for longer times before leveling off.  Our results compare favorably with  those of  previous works  \cite{CheKno,FilSonBer,GraFei,NakYab,ZakGarBoy2,ZhoGuoShu}, which were  obtained by various other methods.  Because of the fineness of our mesh we were able to proceed to longer times than all but \cite{ZhoGuoShu} which achieved machine zero for small perturbations.

As with the advection results above, in contrast to all other works,   we do not see  recurrence.  It is important to note that recurrence is in fact an indication of numerical error.   Recurrence is a general phenomenon in finite-dimensional dynamical systems with time advancement maps that are measure preserving, one-one, onto,   bicontinuous, and have a bounded phase space. Poincar\'e proved recurrence for finite-dimensional Hamiltonian systems, although it can hold for other systems as well. The Vlasov-Poisson system is an infinite-dimensional Hamiltonian system \cite{m80}, and there is no general recurrence theorem for such systems.

Numerical truncation procedures generally are not Hamiltonian, and indeed we have shown this to be the case for  the DG method used here.  However, it has been shown that the DG method, although not Hamiltonian, does give a finte-dimensional system that preserves phase space volume \cite{cheng-gamba-morrison11} and has a bounded energy.   Thus our semi-discrete system has all the ingredients for making a  general estimate for the recurrence time for the distribution function in  terms of the number of degrees of freedom,  like that done by  Boltzmann for gas dynamics.    This results is a very long time for  meshes with any significant resolution.   Note, this approach differs from a result that is often quoted in numerical works  that follows for the  method of  \cite{CheKno}; viz.\  for a given Fourier mode and a mesh of size $\Delta v$ the advection equation was shown to have a recurrence time for the electric field of $T_R= 2\pi k/ \Delta v$.  Many Vlasov algorithms see recurrence at a value that is close to this advection value  $T_R$.  It is important to note that this does not mean that the distribution function is recurring in phase space on this time scale.  In  \cite{cheng-gamba-morrison11}  we present detailed recurrence calculations for our DG algorithm with piecewise constant and higher order polynomials.

In any event, recurrence is not physical and the same can be said for the non-recurrent flattening of the electric field decay rate that is seen in our results with the DG algorithm at late times. The lack of recurrence in our results is due to both the very fine mesh size as well as the dissipative nature of DG algorithms, which is monotonic in nature. However, from Fig.~\ref{fig:Maxwellian refinement} (and also Fig.~\ref{fig:Lorentzian damping}  below) we would argue that DG can do a good job.

\medskip


\noindent\underline{Linear Landau damping with a Lorentz equilibrium}

The ability of the UPG method to achieve accurate damping results across different wavenumbers is now investigated.
As noted above, the  Lorentz equilibrium is used in this example because explicit formula for the electric field damping rates and
the frequencies of the damped oscillations are easily obtained (see
Appendix~\ref{app:Landau}).  Moreover, this example also tests the ability of the method to produce accurate results
for an equilibrium that has a much heavier tail in $v$ than does the Maxwell equilibrium.  The heavy (algebraic)
tail of the Lorentz equilibrium leads to a faster rate of filamentation because there are more resonant electrons than for the Maxwell equilibrium.

The linear system (\ref{eq:lin.vlasov.pb})-(\ref{eq:lin.poisson.pb}) is solved with ${f}_{eq}=f_L=\pi^{-1}/(v^2+1)$,
$A=0.01$, and $V_c = 30.$ for each of the wavenumbers $k=1/8$, 1/6, 1/4, and 1/2.  Note, $V_c$ needs to be  large because of the algebraic tail of $f_L$. The corresponding values for
$L$ and $T$ for $k=$1/8, 1/6, 1/4, and 1/2, are $L=16\pi$, $12\pi$, $8\pi$ and $4\pi$ and $T=75,$ 75, 50, and 50,
respectively.  Uniform meshes were employed in all four cases, where in each case $(N_{_x},N_{_v})=(1000,2000)$.

In Fig.~\ref{fig:Lorentzian damping}, log plots of $\rho_{tot}$ for the four cases are given. The observed damping for
$k=1/2$ lasts for a shorter time duration than does the damping for the other three wavenumbers, even though the mesh
for $k=1/2$ has the finest resolution in $x$.  This result is due to the fact that the filamentation in the velocity
direction develops more rapidly than it does in the other three cases. From (\ref{eq:lorentz.damp}),  obtained in Appendix~\ref{app:Landau}, it follows that
for a given wavenumber $k$ the fundamental mode of the electric field damps
exponentially in time at a rate equal to $\gamma(k)=-k$ and the frequency for all of the damped oscillations
is $\omega(k)=1$.  Therefore, the theoretical damping rates corresponding to $k=1/8$, 1/6, 1/4, and 1/2
are $\gamma(k)=-1/8$, -1/6, -1/4, and -1/2.  In all of the four cases shown in Fig.~\ref{fig:Lorentzian damping}, the
numerical damping rates and frequencies of oscillation are equal
to the theoretical values up to the first two decimal digits.

\begin{figure}[ht]
\includegraphics[angle=0,width=0.47\textwidth,height=0.35\textwidth]{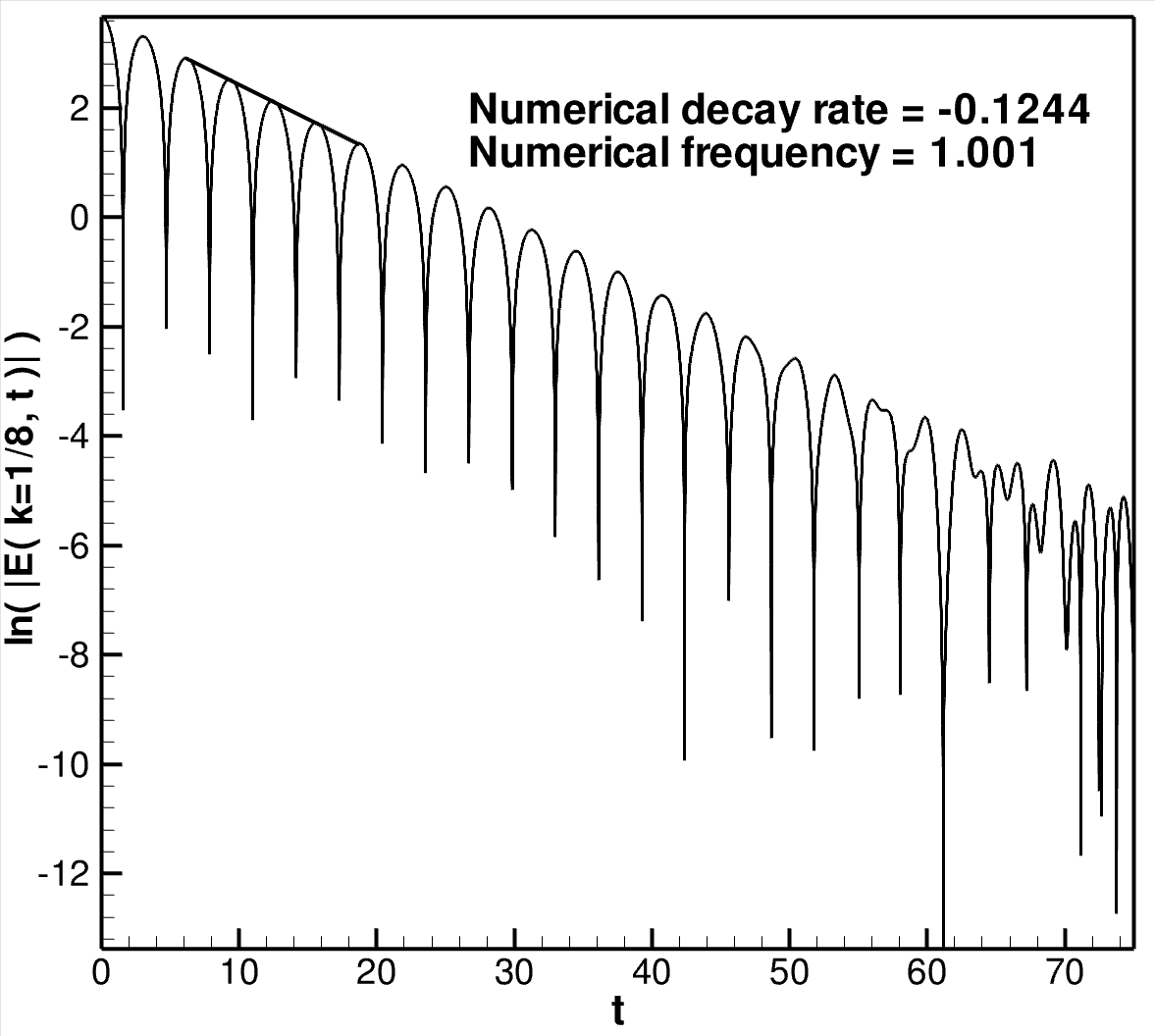}
\hspace{0.02\textwidth}
\includegraphics[angle=0,width=0.47\textwidth,height=0.35\textwidth]{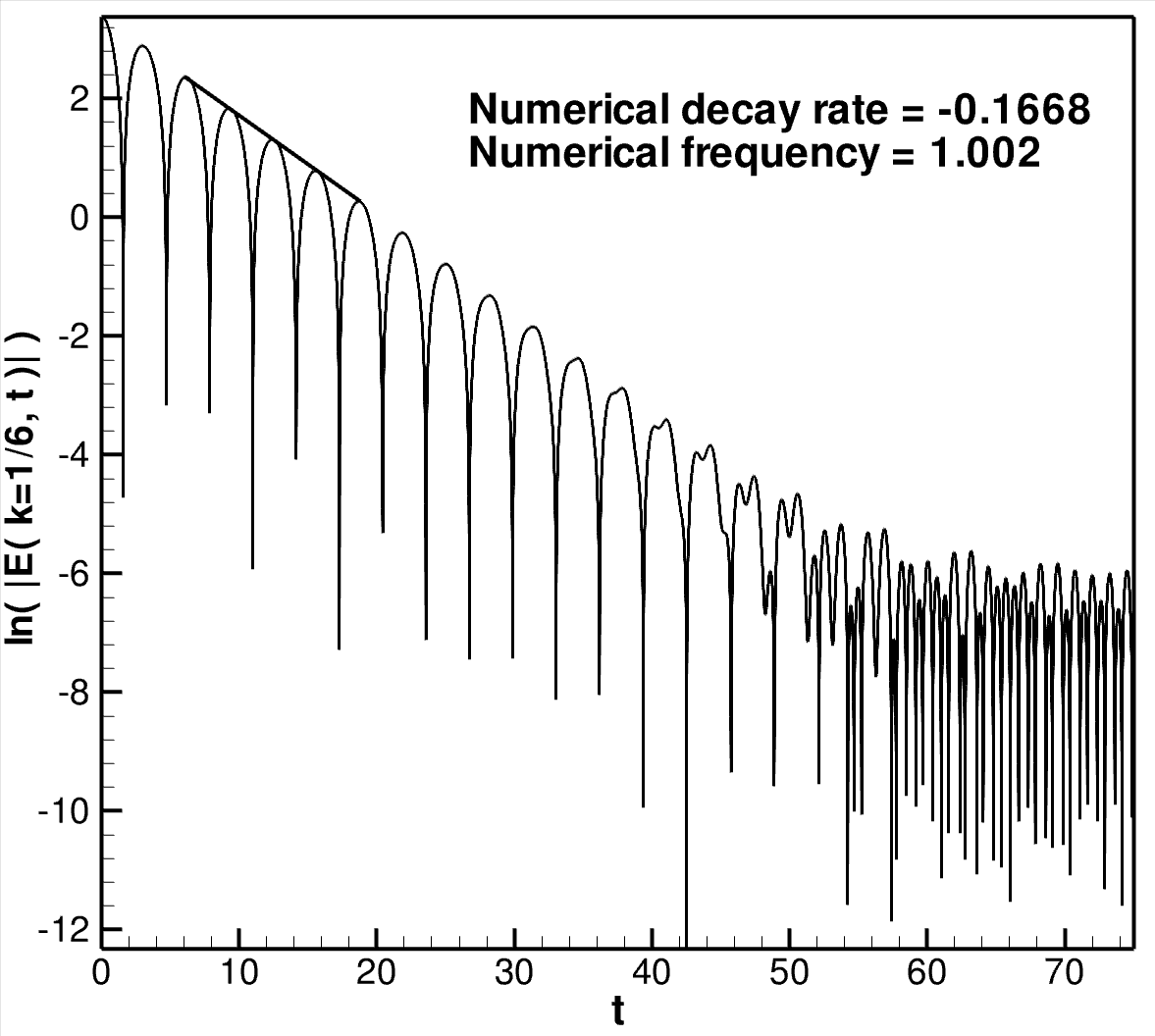}
\vspace{0.5cm} \\
\includegraphics[angle=0,width=0.47\textwidth,height=0.35\textwidth]{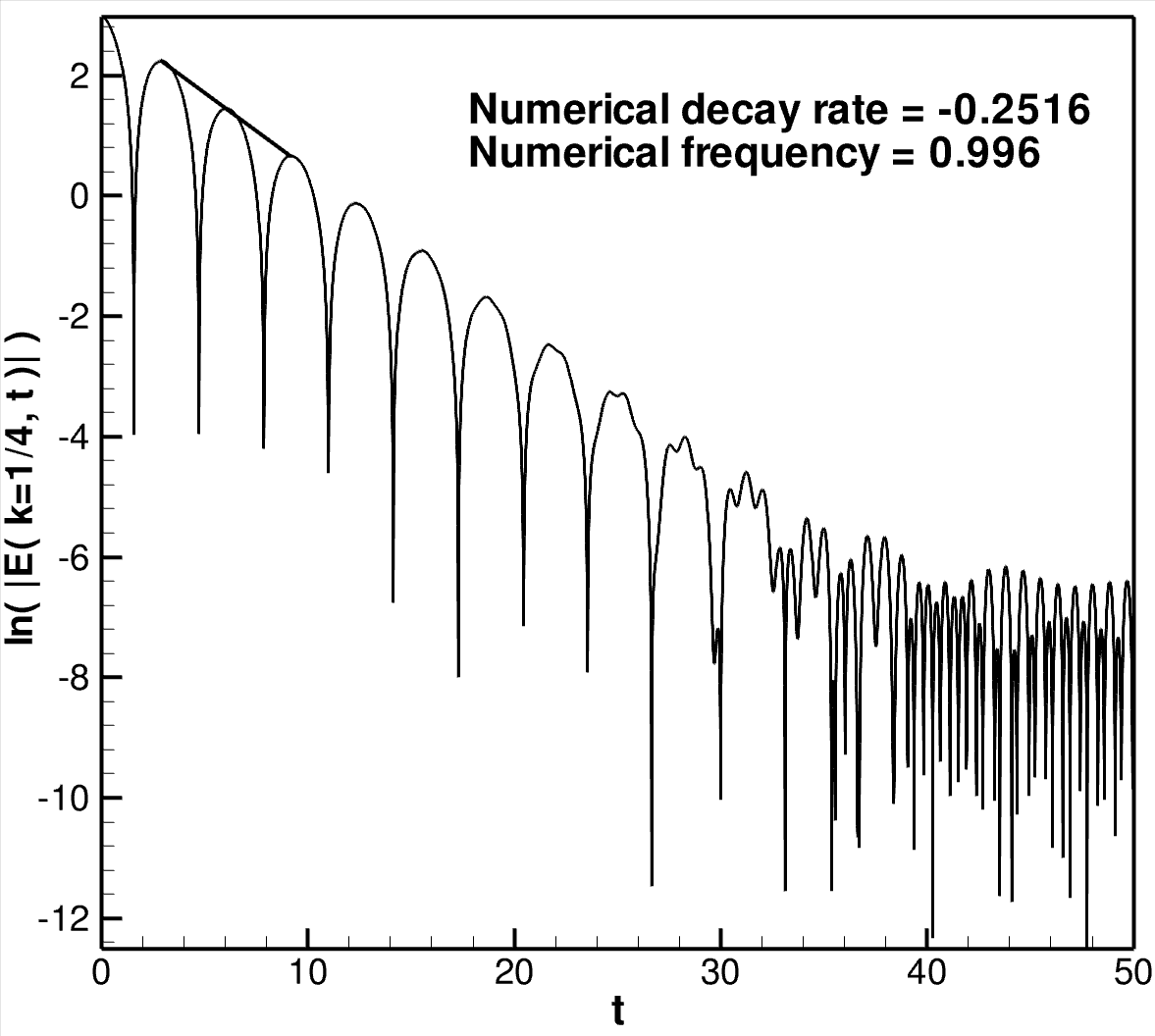}
\hspace{0.02\textwidth}
\includegraphics[angle=0,width=0.47\textwidth,height=0.35\textwidth]{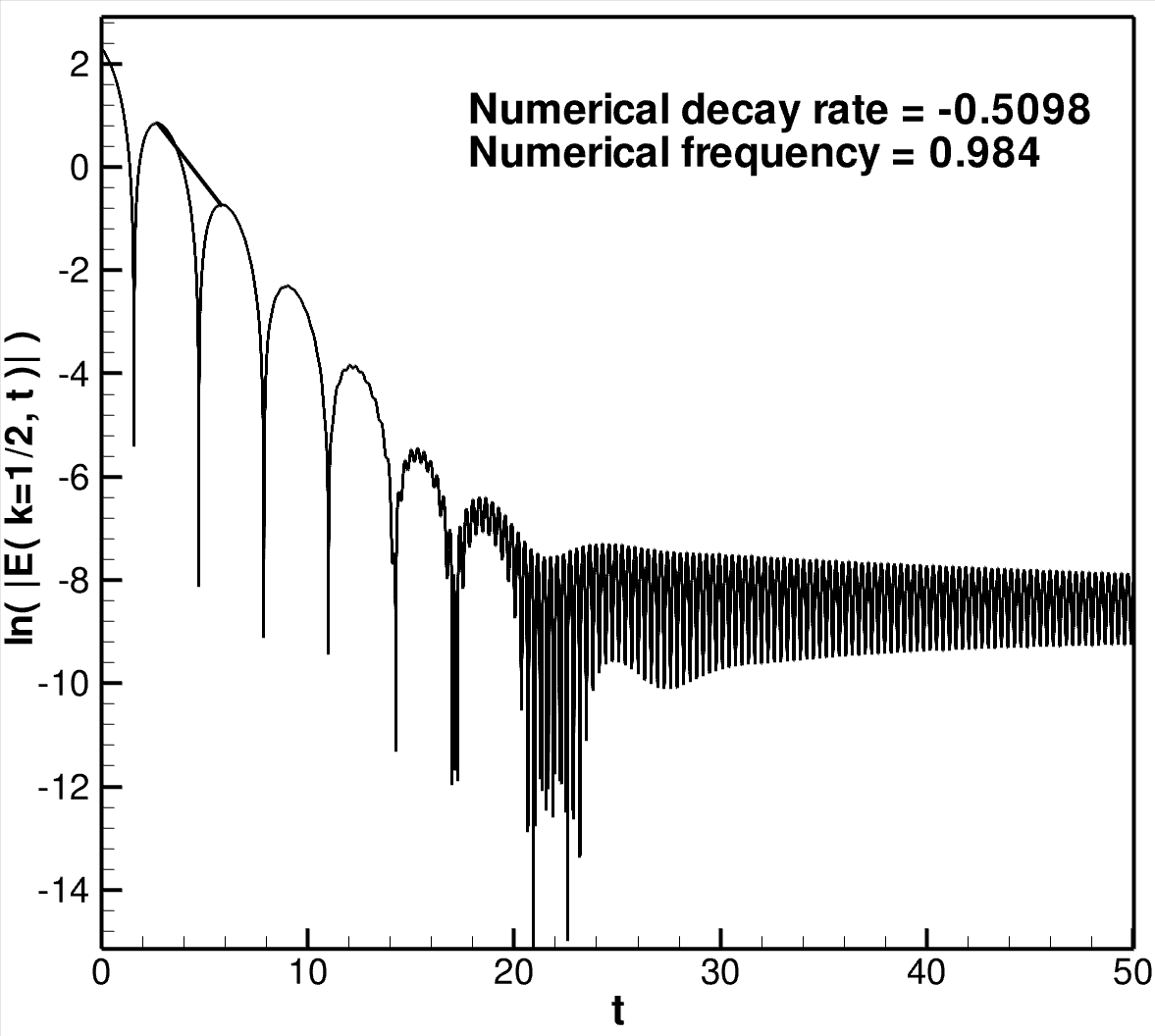} \\
\caption{(Linear Landau damping with Lorentz equilibrium) Time decay plots of fundamental modes (with arbitrary ordinate origin):
$k$=1/8 {\em (top left)}, $k$=1/6 {\em (top right)}, $k$=1/4 {\em (bottom left)} and
$k$=1/2 {\em (bottom right)}.}
\label{fig:Lorentzian damping}
\end{figure}

\clearpage


\subsection{Nonlinear Results}
\label{sec:nonlinear}

Two nonlinear calculations are performed,  an example of strong nonlinear Landau damping  and a version of  the two-stream instability that we examine in greater detail.  In the former case, in addition to early time damping,  we expect to see motions on the bounce time scale due to particle trapping, while in the latter we expect to see trapping and an asymptotic approach to a BGK state.

\subsubsection{Nonlinear Landau Damping}
\label{sec:nlLD}

Following  \cite{CheKno,CanGazFro,Eli,FilSonBer,KliFar,NakYab,ZakGarBoy2,ZhoGuoShu} we consider an initial condition
near a Maxwellian and of the form  given by (\ref{eq:f.form}) and (\ref{Vlasov_IC})--(\ref{Poisson_BC}) with  $f_{eq}=f_M=(2\pi)^{-1/2}e^{-v^2/2}$,
$A=0.5$, $k=0.5$, $L = 4\pi$, $V_c = 5.0$ and a run time of $T = 120$.
Unlike the case considered in Section~\ref{sec:Lin.Landau} we evolve under the full nonlinear Vlasov equation.
Solutions are computed using a  uniform mesh with $(N_{h_x},N_{h_v})=(1000,750)$.

\begin{figure}[htb]
\centering
\subfigure[{\footnotesize $k=0.5$.}]{
\includegraphics[width=0.47\textwidth]{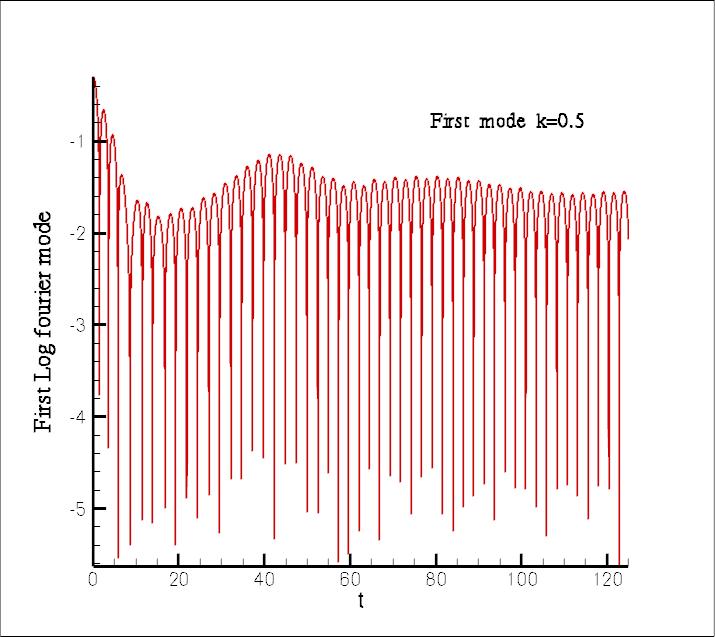}
\label{fig:nlLDmodes_k=.5}
}
\subfigure[{\footnotesize  $k=1.0$.}]{
\includegraphics[width=0.47\textwidth]{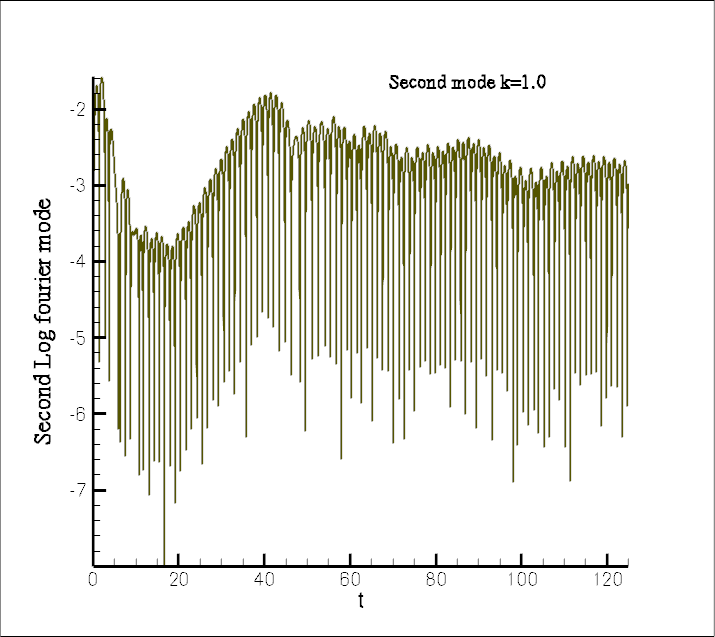}
\label{fig:nlLDmodes_k=1}
}
\subfigure[{\footnotesize  $k=1.5$.}]{
\includegraphics[width=0.47\textwidth]{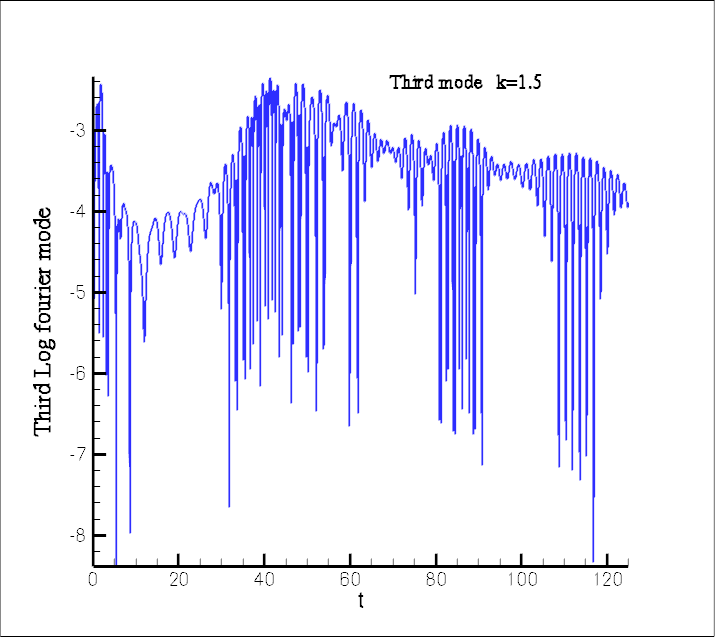}
\label{fig:nlLDmodes_k=1.5}
}
\subfigure[{\footnotesize  $k=2.0$.}]{
\includegraphics[width=0.47\textwidth]{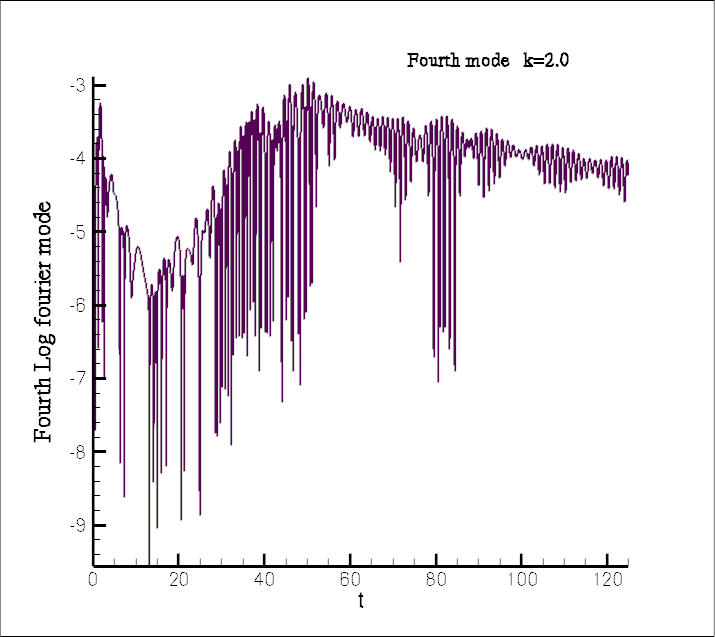}
\label{fig:nlLDmodes_k=2}
}
\caption[modes of nonlinear Landau damping]
{(Nonlinear Landau damping  with Maxwell equilibrium) Amplitudes of the first four modes versus time.  Mesh size  $(N_x,N_v)=(1000,750)$.
}
\label{fig:nlLDmodes}
\end{figure}

Because $A$ has been increased to $0.5$,  higher modes are excited at very early times and the damping rate significantly exceeds the linear rate of $\gamma =-0.153$.    This is  because energy leaves the first mode as the higher modes get excited; i.e.,   in addition to the phase mixing process there is an energy transfer process caused by the nonlinear interaction of the modes.  Figure \ref{fig:nlLDmodes} shows the amplitudes of the first four modes as a function of time, with a form similar to previous calculations.   These plots are obtained from our mesh data by using the following `log Fourier Mode' function:
\bq
\log FM_k(t)  =
\log_{10}
 \left(
             \sqrt{\left|E_{s,k}(t) \right|^2 + \left|E_{c,k}(t)\right|^2}/L\,,
 \right)
\label{logfour}
\eq
with
\bq
E_{s,k}(t):=\int_0^L\!dx\,  E(x,t) \sin\left(  \frac{2\pi k x}{L}\right)
\eq
and
\bq
\  \  E_{c,k}(t):=\int_0^L\!dx\, E(x,t) \cos\left( \frac{2\pi k x}{L}\right)\,,
\eq
where $k$ is the mode number sought.  From Fig.~\ref{fig:nlLDmodes} it is seen that mode-one reaches its minimum value at around $t\approx15$ and then all modes grow until they reach their  maxima at $t\approx40$, consistent with previous calculations.  Using the maximum amplitude of mode-one, the bounce time is calculated to be $T_B\approx 20$ and this is also in agreement with previous results.

\begin{figure}[ht]
\begin{center}
\includegraphics[angle=0,width=.75\textwidth,height=0.65\textwidth]{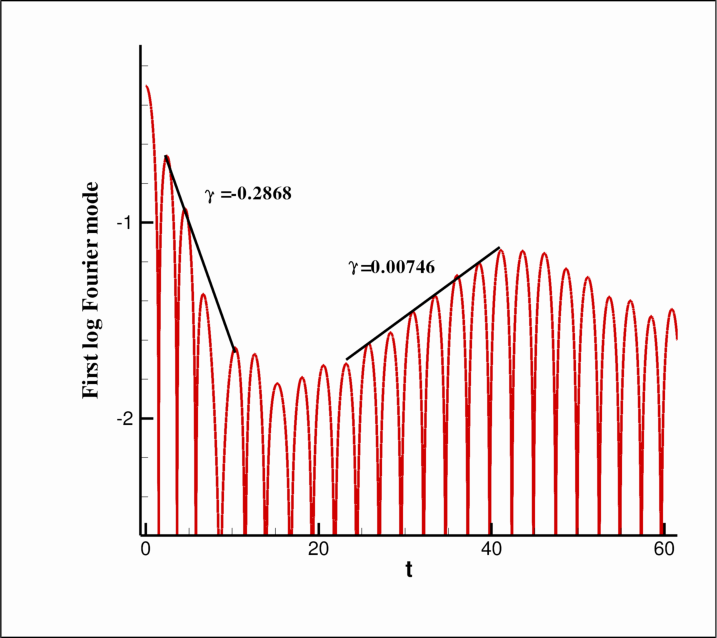}
\end{center}
\caption{(Nonlinear Landau damping  with Maxwell equilibrium)  Initial damping of dominant mode followed by growth due to particle trapping.}
\label{fig:NLDdamprate}
\end{figure}

\begin{figure}[htb]
\centering
\subfigure[{\footnotesize Energy.}]{
\includegraphics[width=0.47\textwidth]{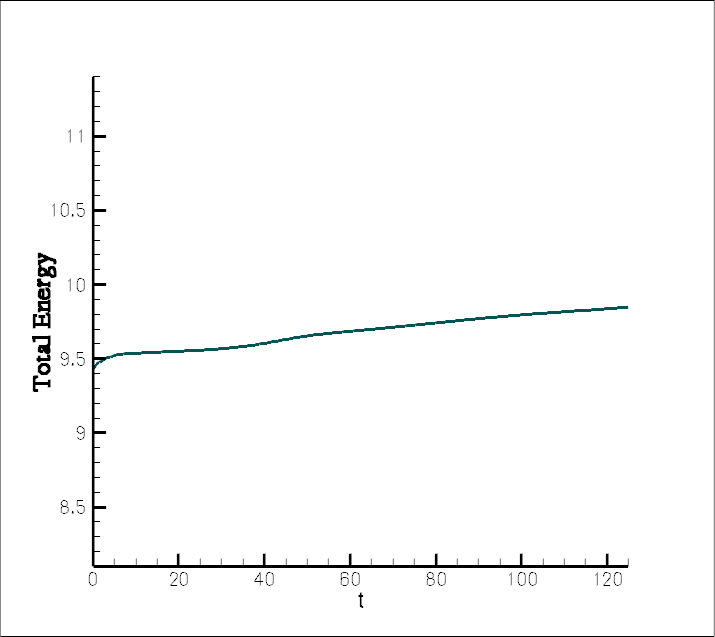}
\label{fig:nlLDenergy}
}
\subfigure[{\footnotesize  Entropy.}]{
\includegraphics[width=0.47\textwidth]{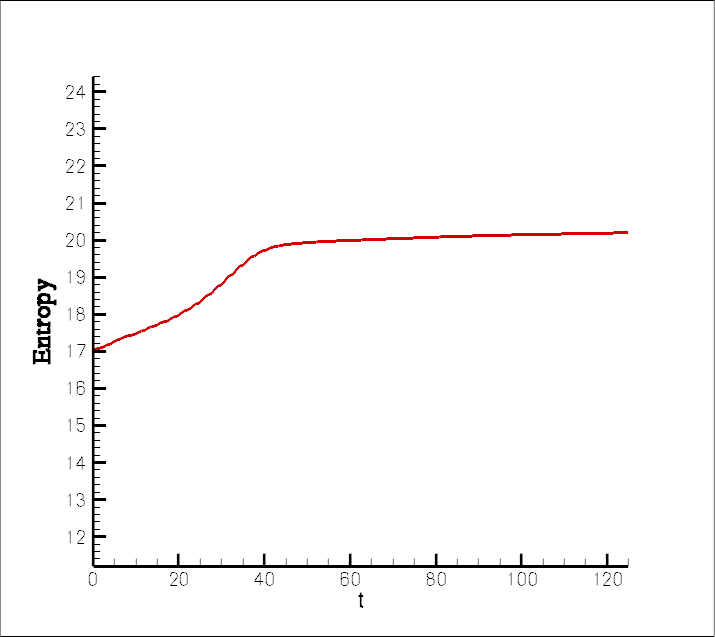}
\label{fig:nlLDentropy}
}
\caption[Energy and entropy for  nonlinear Landau damping]
{(Nonlinear Landau damping  with Maxwell equilibrium) The total energy $H$ of  (\ref{eq:energy}), kinetic plus electrostatic,  and entropy,  (\ref{eq:casimir}) with $\mathcal{C}=-f\ln f$,   versus time.
}
\label{fig:nlLDenergy-enstrophy}
\end{figure}

 Examination of  Fig.~\ref{fig:NLDdamprate}  reveals that we obtain a damping rate for the first mode of about $\gamma=-0.287$, a  value consistent with  the $-0.281$ obtained by  \cite{CheKno} and $-0.243$  by  \cite{ZakGarBoy2},  given the different ways authors have used to make this kind of fit.   Also, given that we have a finer mesh,  some  deviation could be due to our more precise coupling to the higher modes.  Examination of Fig.~\ref{fig:nlLDenergy-enstrophy}  shows there is significant entropy [(\ref{eq:casimir}) with $\mathcal{C}=-f\ln f$)] dissipation at short times,  as also seen by \cite{FilSonBer},  and this introduces some error.     Also  note, we have run to $T=120$,  which is significantly longer than previous calculations and a small decay in all four of the amplitudes is seen.  This could be due to transfer to higher modes, cascading,  or due to dissipation in the algorithm.  Over the full length of the run the total energy $H$ of (\ref{eq:energy}) is  c
 onserved to within a few percent, and in the later part of the run entropy is well conserved.  This is improved in the next section, where we treat the two-stream instability,  by decreasing the mesh size.

\begin{figure}[htb]
\centering
\subfigure[{\footnotesize $t=0$.}]{
\includegraphics[width=0.47\textwidth]{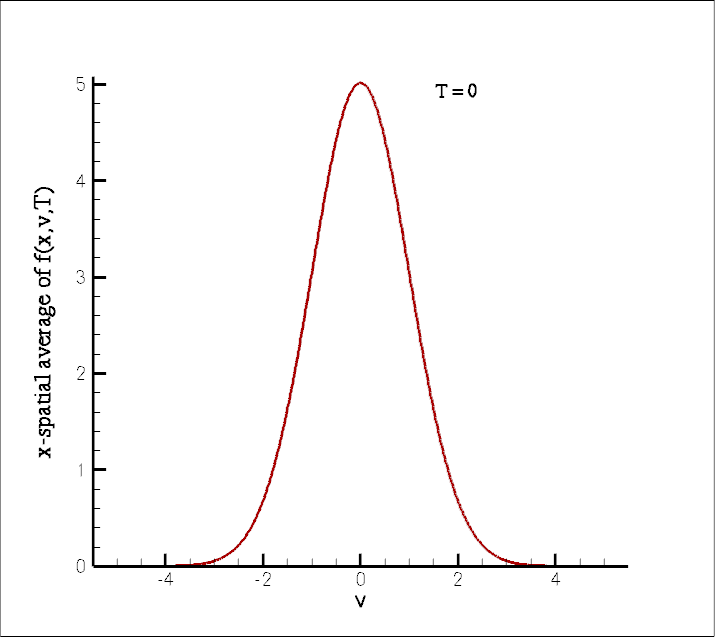}
\label{fig:nlLDavDist_t=0}
}
\subfigure[{\footnotesize  $t=12.5$.}]{
\includegraphics[width=0.47\textwidth]{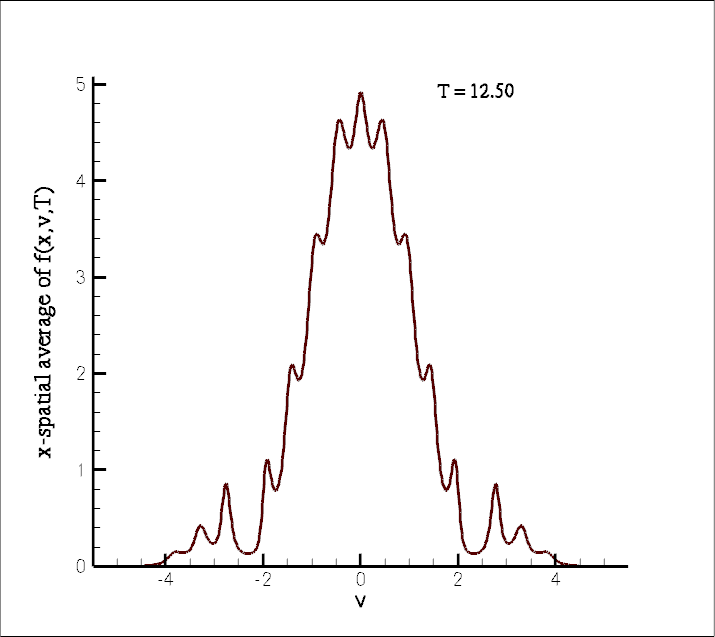}
\label{fig:nlLDavDist_t=12}
}
\subfigure[{\footnotesize  $t=31.5$.}]{
\includegraphics[width=0.47\textwidth]{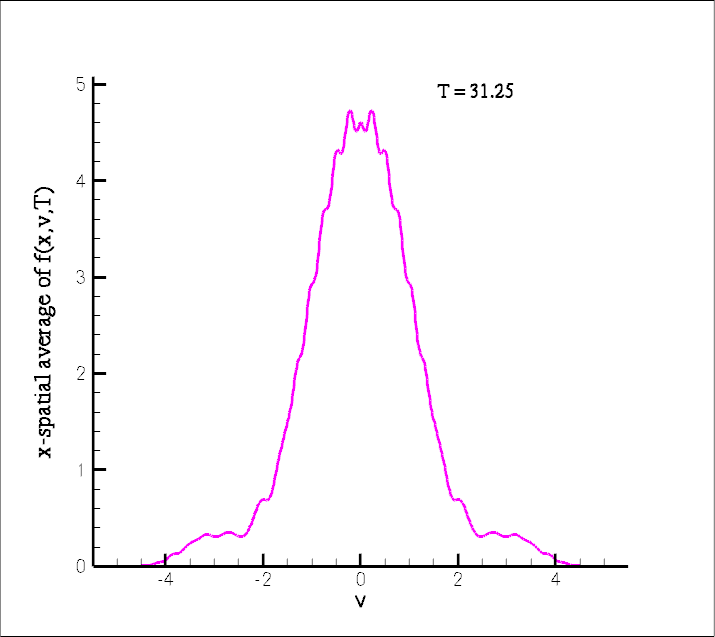}
\label{fig:nlLDavDist_t=31}
}
\subfigure[{\footnotesize  $t=43.75$.}]{
\includegraphics[width=0.47\textwidth]{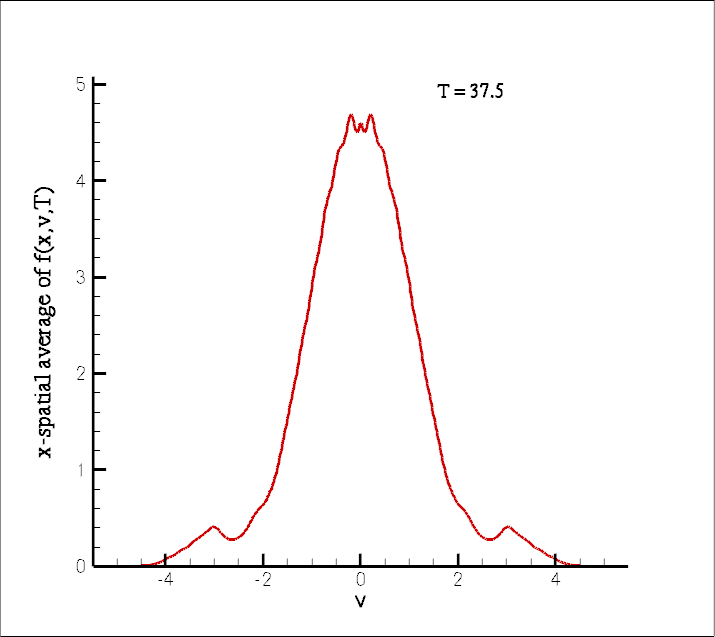}
\label{fig:nlLDavDist_t=37}
}
\subfigure[{\footnotesize  $t=75$.}]{
\includegraphics[width=0.47\textwidth]{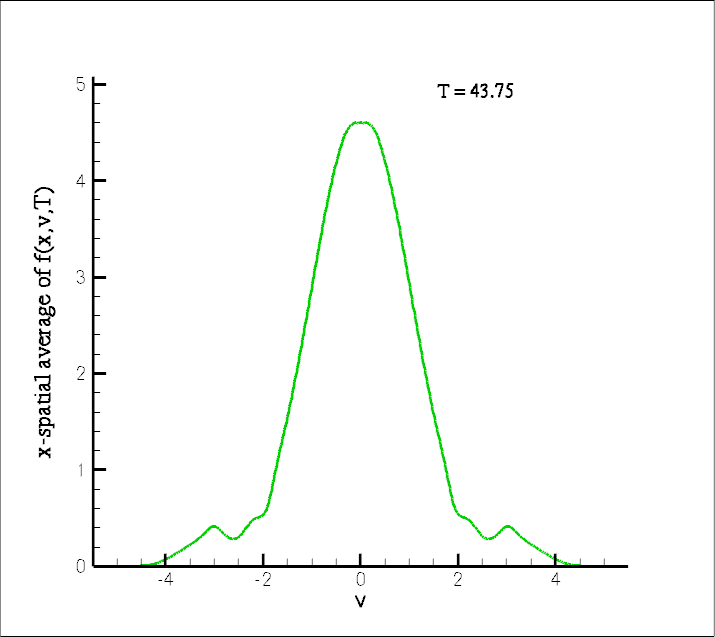}
\label{fig:nlLDavDist_t=43}
}
\subfigure[{\footnotesize  $t=112.5$.}]{
\includegraphics[width=0.47\textwidth]{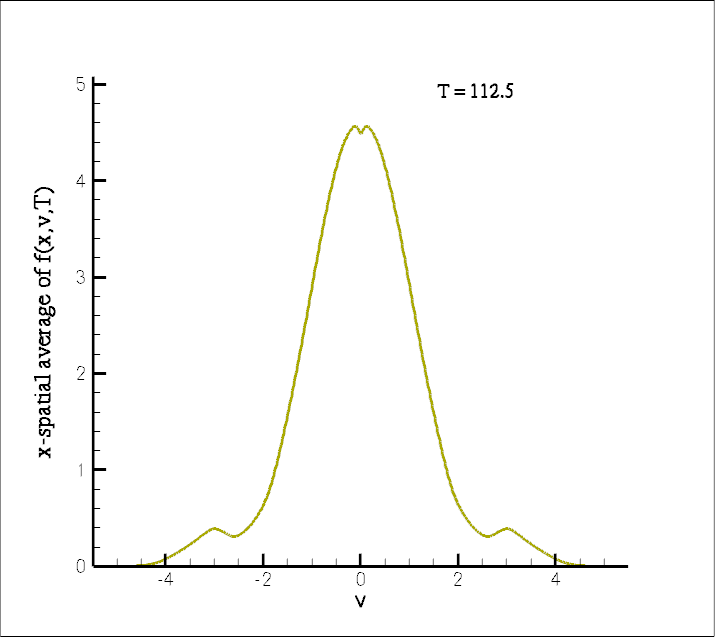}
\label{fig:nlLDavDist_t=112}
}
\caption[modes of nonlinear Landau damping]
{(Nonlinear Landau damping  with Maxwell equilibrium) Spatial average of the distribution function at the times indicated.
}
\label{fig:nlLDavDist}
\end{figure}

In Fig.~\ref{fig:nlLDavDist} we plot the spatial average of the distribution function.  Like other authors, we obtain early plateau formation in the vicinity of the phase velocity of the wave, seen in panel (b), which broadens as the higher order modes are excited.  At around $t=40$, approximately the time of the first bounce maximum,  significant smoothing takes place and the system settles into a nearly constant average state with a persistent  electron hole.  The smoothing at $t\approx 40$ can also be seen in the results of \cite{CheKno,FilSonBer}. Finally,  we note the presence a small periodic dimpling behavior at the maximum that persists for late times.



\clearpage

\subsubsection{Nonlinear Two-Stream Instability}
\label{sec:Nonlin.two.stream}

In this subsection, we numerically study the long-time nonlinear evolution of the  two-stream instability, a  standard example that has been used for demonstrating  the efficacy of a variety of numerical algorithms  \cite{CheKno,DenKru,GraFei,KliFar,NakYab,PohShoKam,ZakGarBoy2}.  Like these previous calculations, we consider  the equilibrium state
\begin{equation}
f_{TS}(v)\ =\ \frac1{\sqrt{2\pi}}\, v^2\,e^{- v^2/2}\,
\label{TSequil}
\end{equation}
and  apply our algorithm to the nonlinear system (\ref{eq:vlasov.pb})-(\ref{eq:poisson.pb}) with data of the form
given by (\ref{eq:f.form}) and (\ref{Vlasov_IC})--(\ref{Poisson_BC}). We choose parameters values  to be $A=0.05$, $k=0.5$, $L=4\pi$, $V_c=5$, and $T=100$,   a uniform mesh defined by $(N_x,N_v)=(1800,1400)$,  and a time step $\Delta t=0.00125$.

\noindent\underline{Qualitative Behavior}

\begin{figure}[ht]
\centering
\includegraphics[angle=0,width=0.48\textwidth,height=0.52\textwidth]{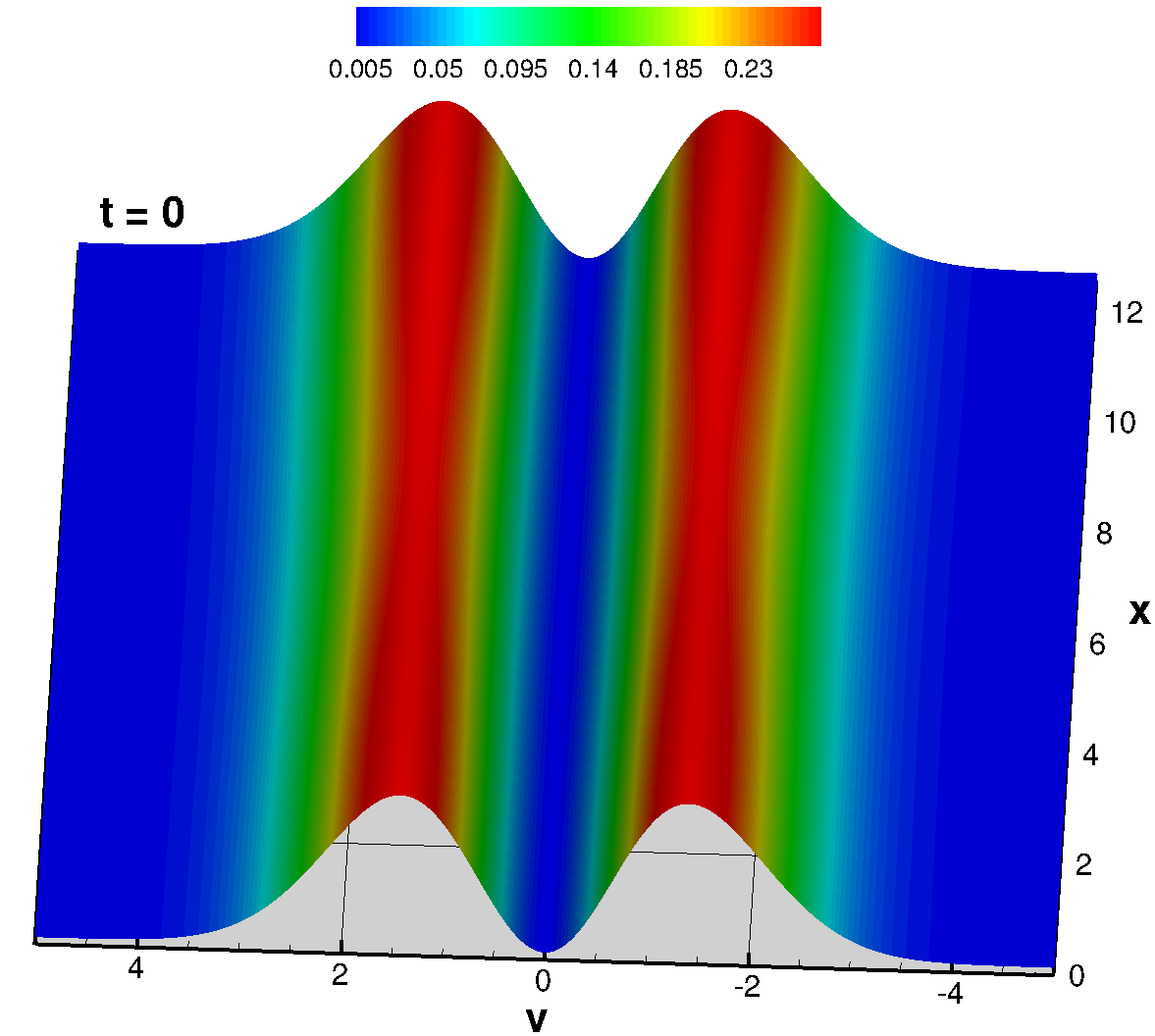}
\hspace{0.02\textwidth}
\includegraphics[angle=0,width=0.48\textwidth,height=0.52\textwidth]{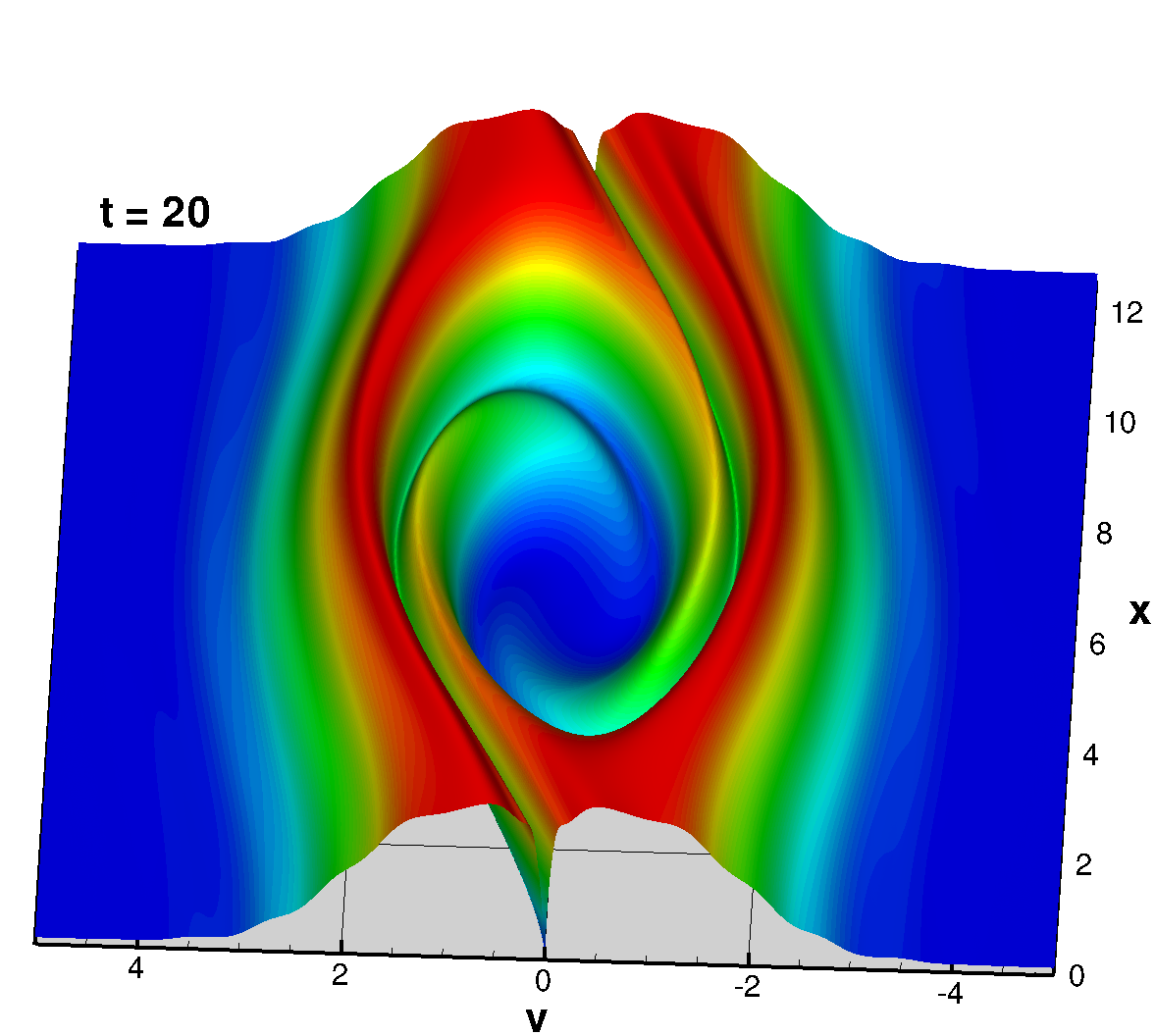} \\
\includegraphics[angle=0,width=0.48\textwidth,height=0.52\textwidth]{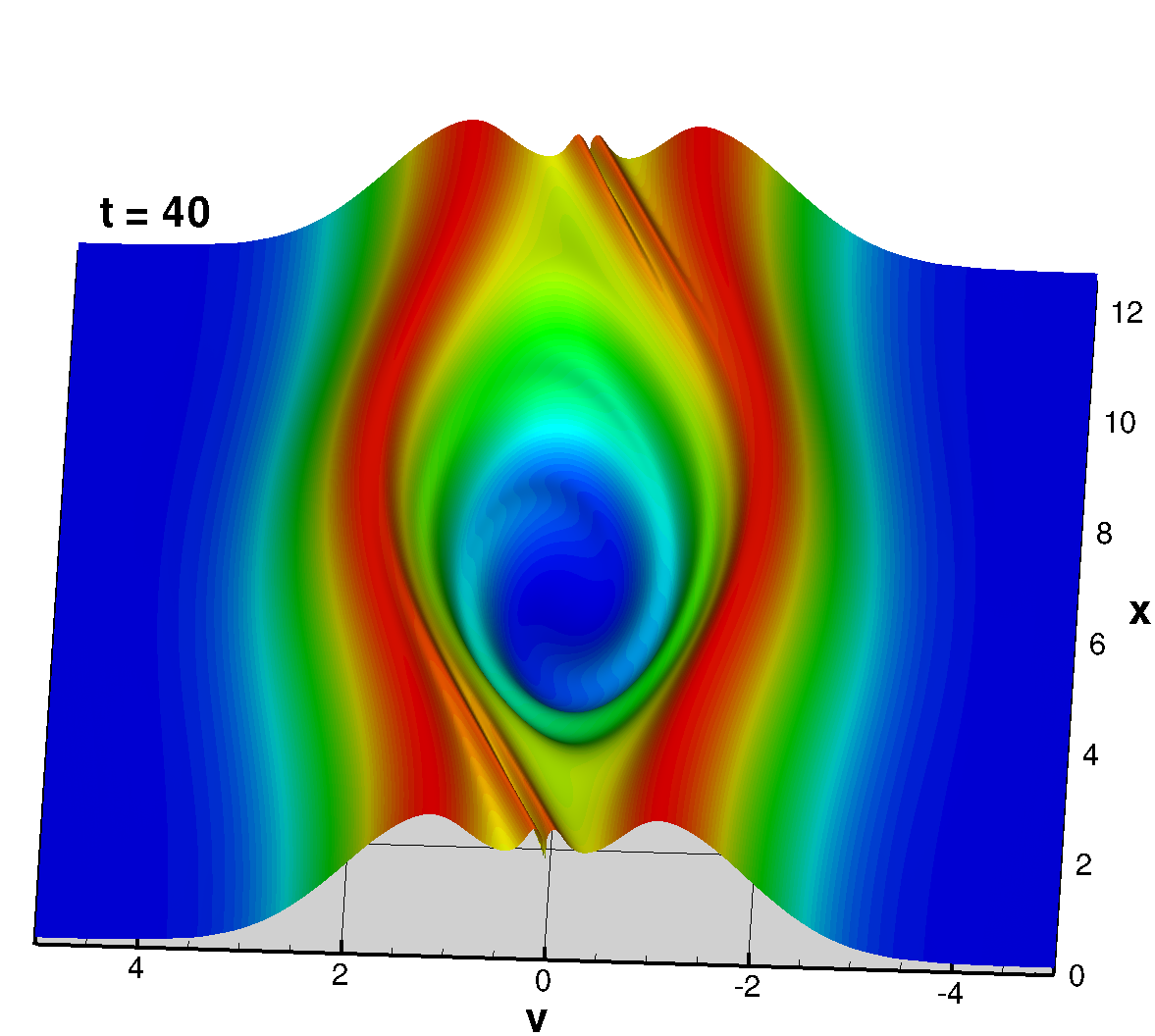}
\hspace{0.02\textwidth}
\includegraphics[angle=0,width=0.48\textwidth,height=0.52\textwidth]{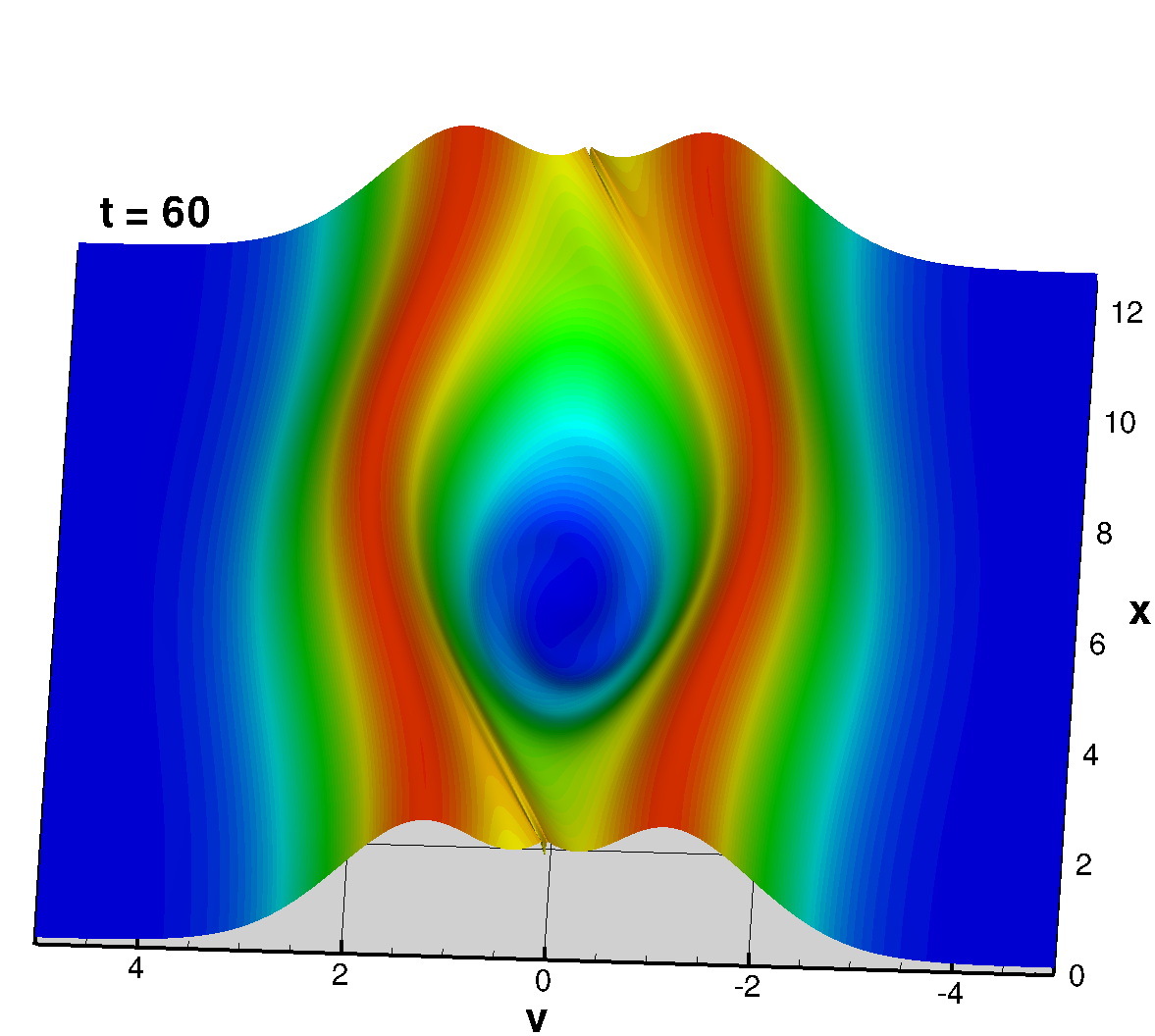} \\
\includegraphics[angle=0,width=0.48\textwidth,height=0.52\textwidth]{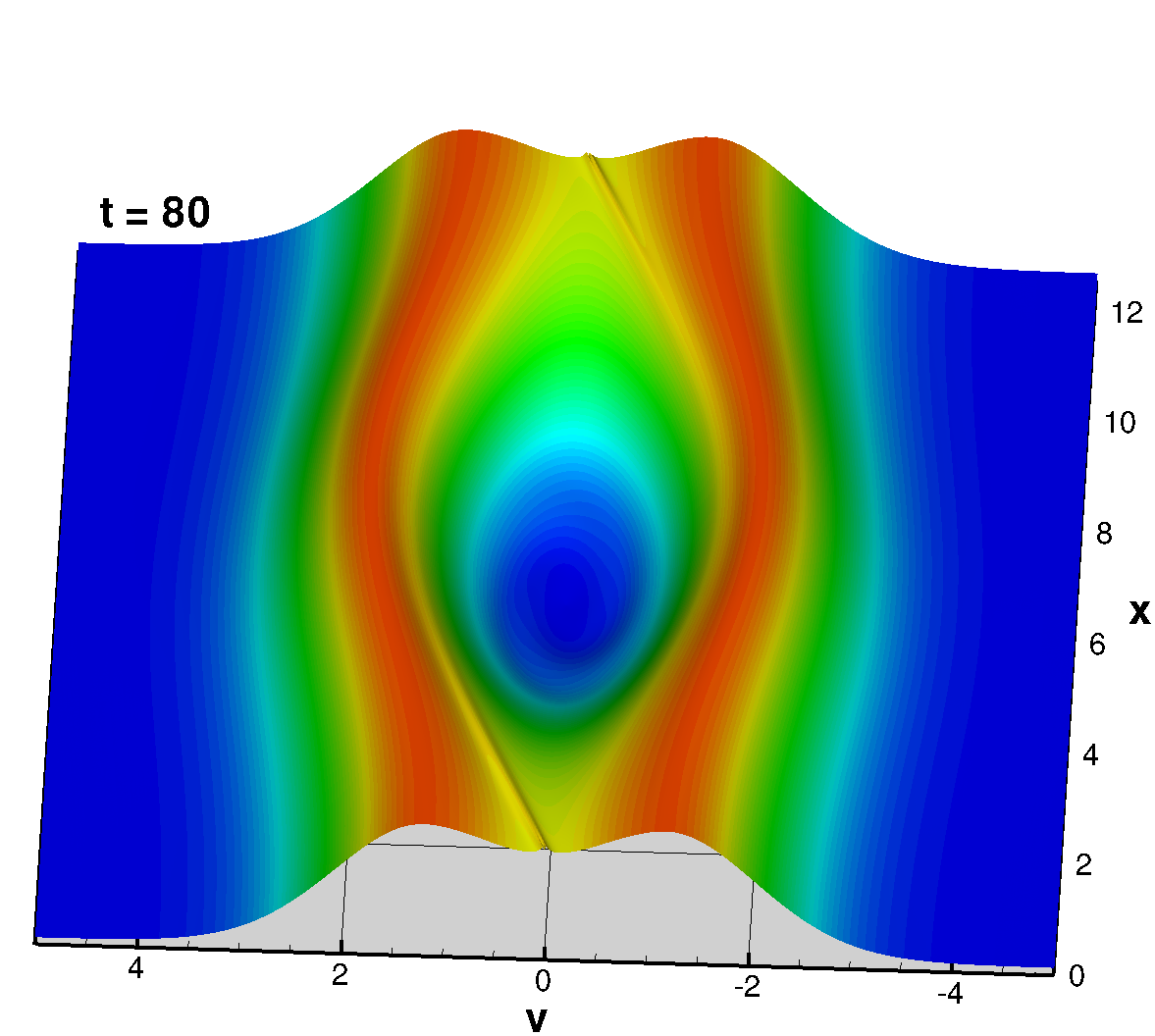}
\hspace{0.02\textwidth}
\includegraphics[angle=0,width=0.48\textwidth,height=0.52\textwidth]{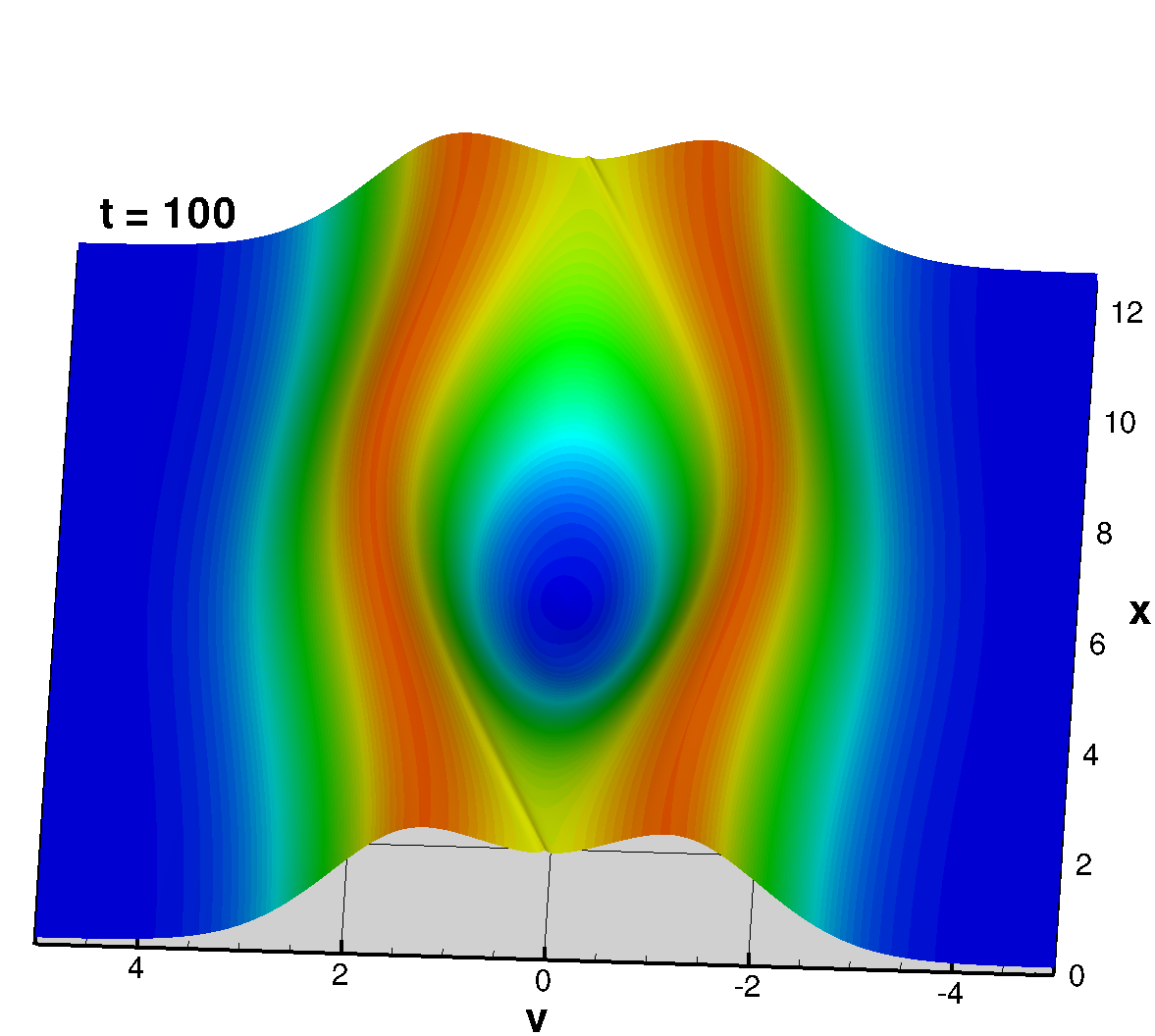} \\
\caption{(Nonlinear two-stream instability) 3D plots of the solution $f$ at $t=0$, $t=20$, $t=40$,
$t=60$, $t=80$ and $t=100$.}
\label{fig:Two-stream 3D}
\end{figure}

\begin{figure}[ht]
\centering
\includegraphics[angle=0,width=0.48\textwidth,height=0.5\textwidth]{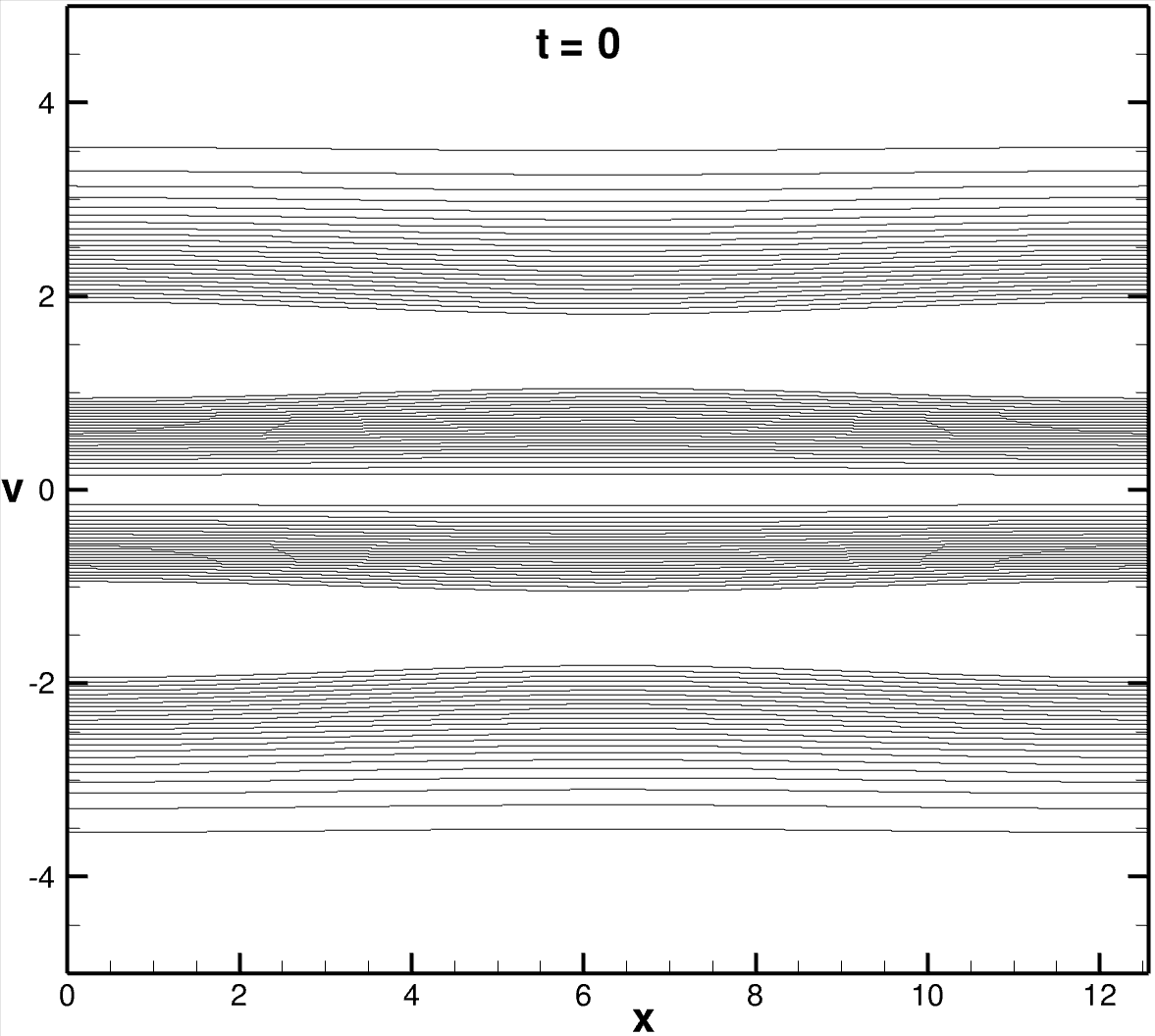}
\hspace{0.02\textwidth}
\includegraphics[angle=0,width=0.48\textwidth,height=0.5\textwidth]{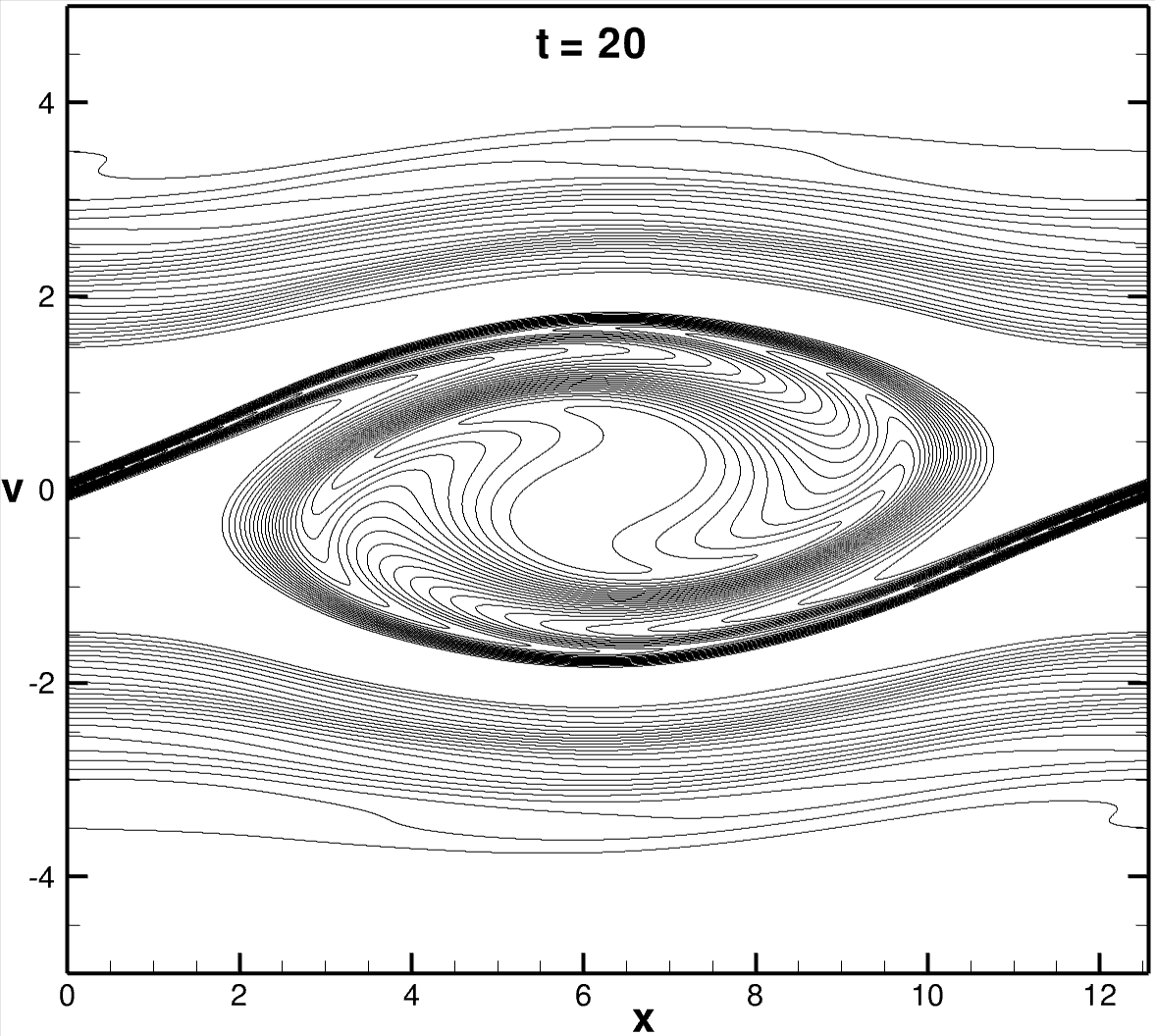} \\ \vspace{0.4cm}
\includegraphics[angle=0,width=0.48\textwidth,height=0.5\textwidth]{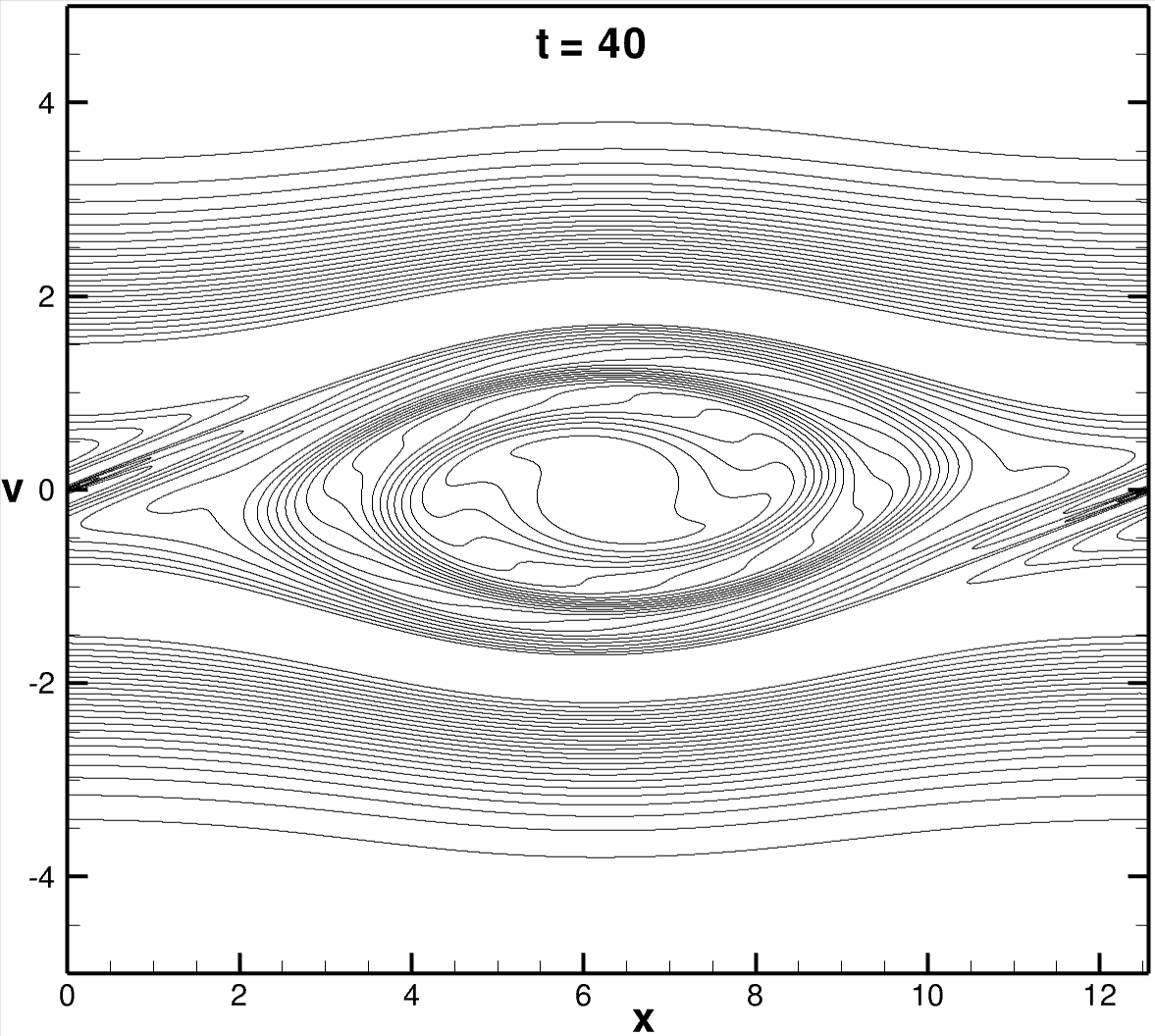}
\hspace{0.02\textwidth}
\includegraphics[angle=0,width=0.48\textwidth,height=0.5\textwidth]{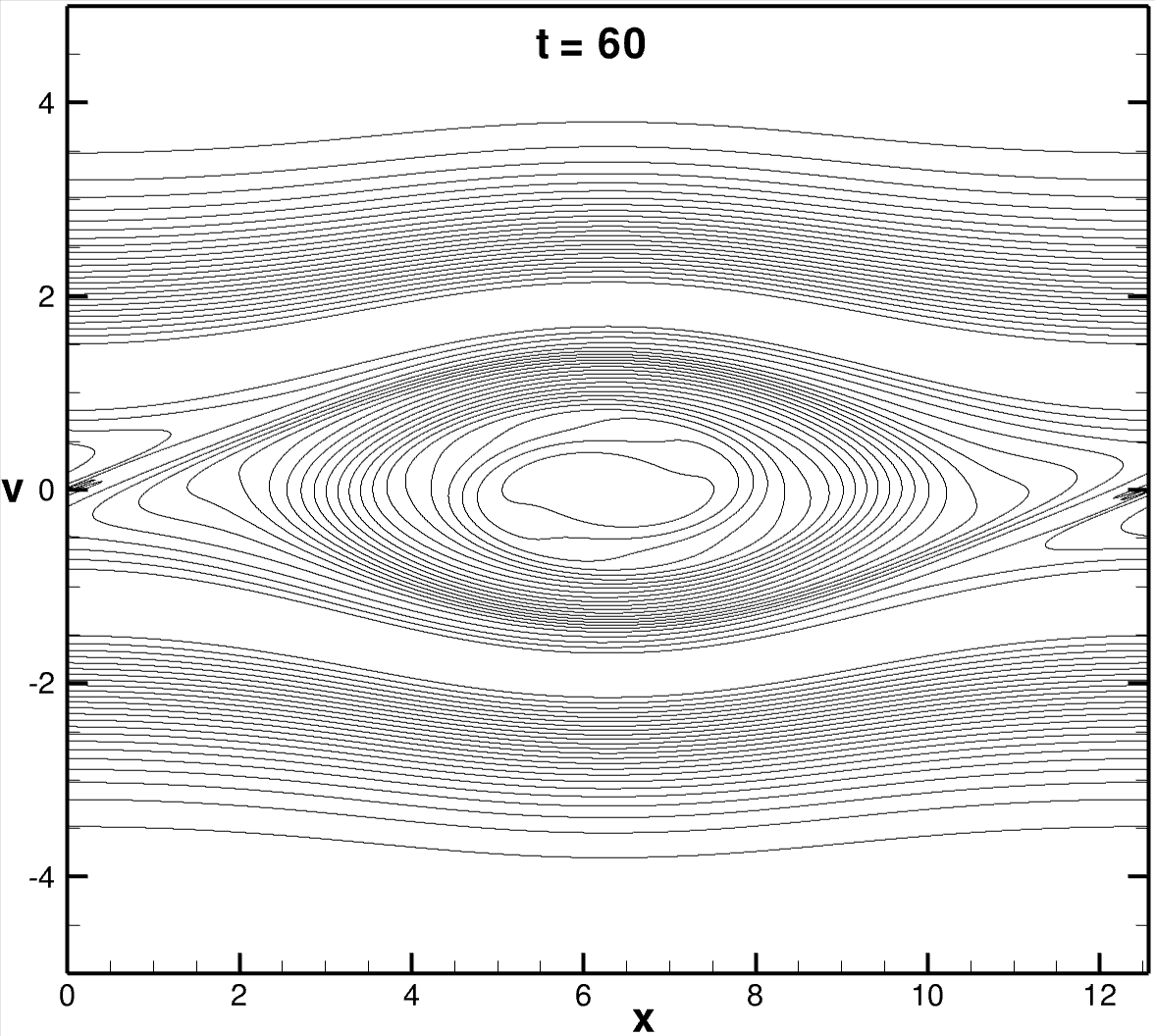} \\ \vspace{0.4cm}
\includegraphics[angle=0,width=0.48\textwidth,height=0.5\textwidth]{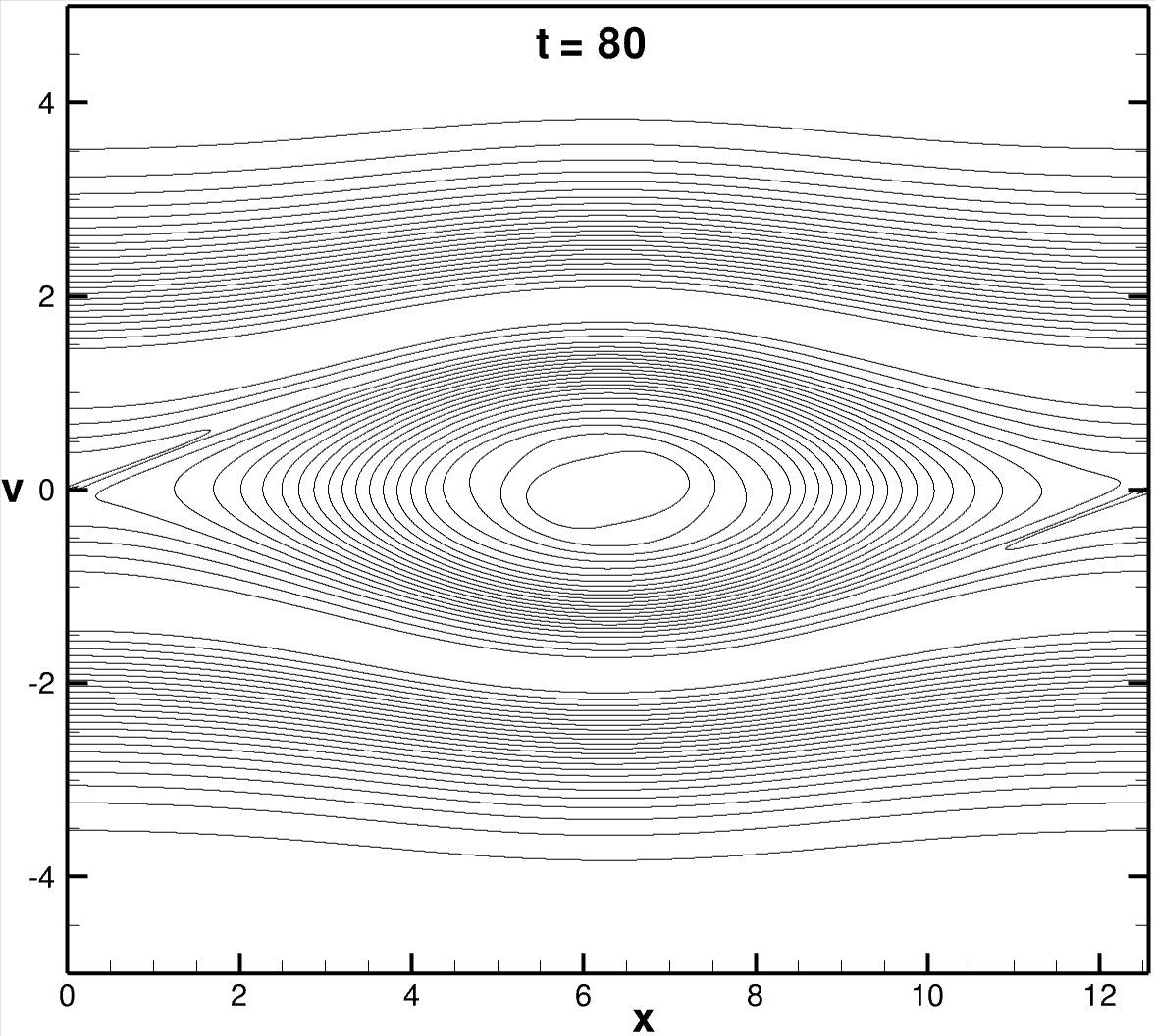}
\hspace{0.02\textwidth}
\includegraphics[angle=0,width=0.48\textwidth,height=0.5\textwidth]{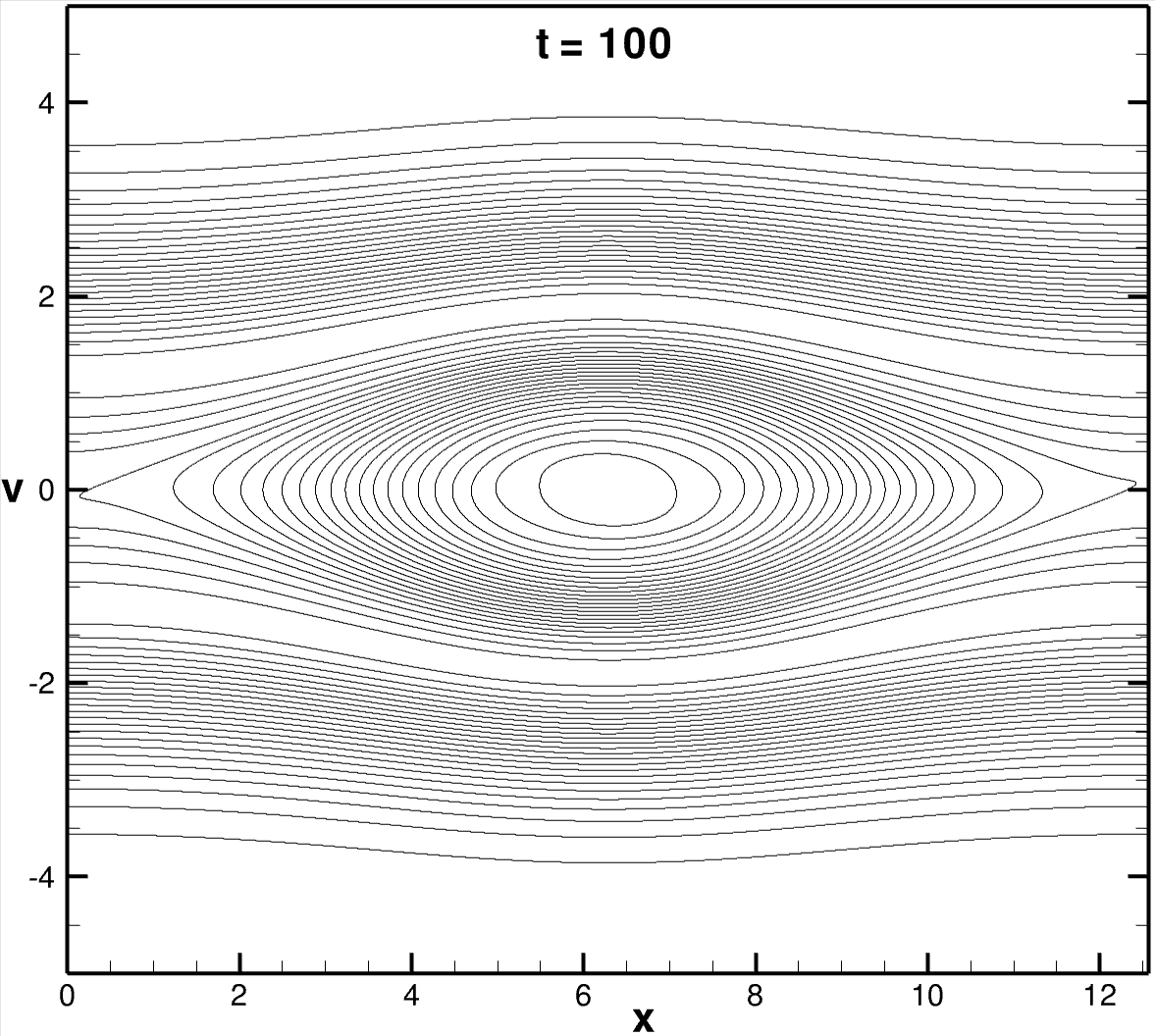} \\
\caption{(Nonlinear two-stream instability) Contour plots of the solution $f$ at $t=0$, $t=20$, $t=40$, $t=60$,
$t=80$ and $t=100$.}
\label{fig:Two-stream 2D}
\end{figure}

\begin{figure}[ht]
\includegraphics[angle=0,width=0.48\textwidth]{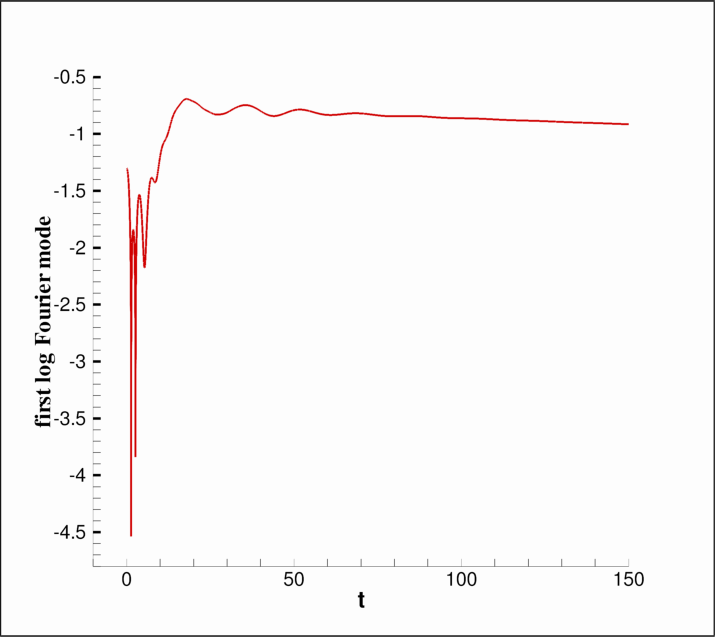}
\includegraphics[angle=0,width=0.48\textwidth]{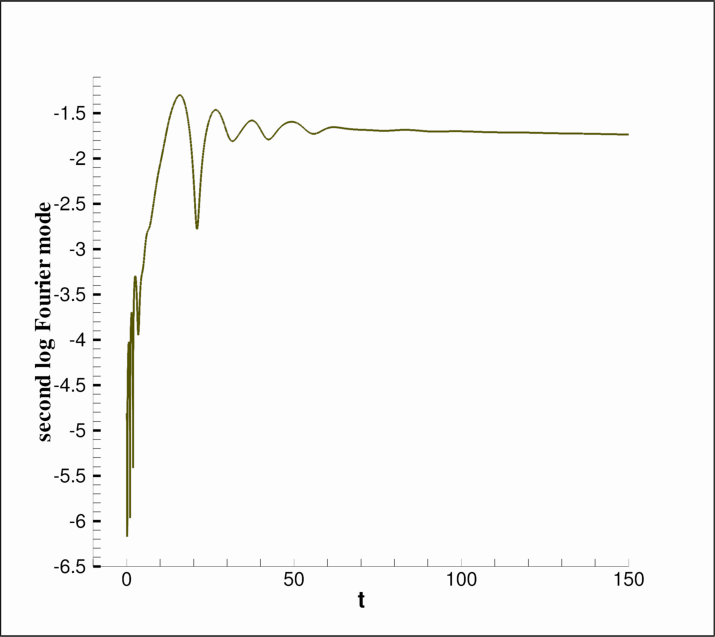}
\vspace{0.5cm}\\
\includegraphics[angle=0,width=0.48\textwidth]{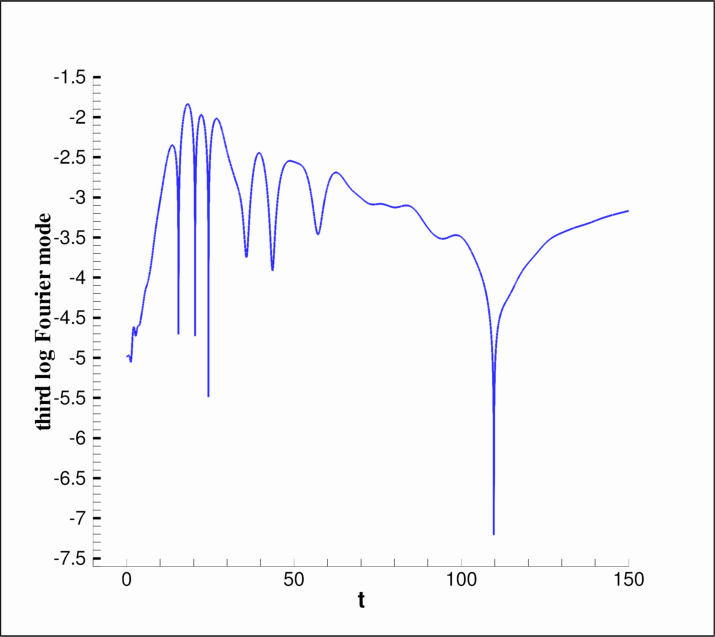}
\includegraphics[angle=0,width=0.48\textwidth]{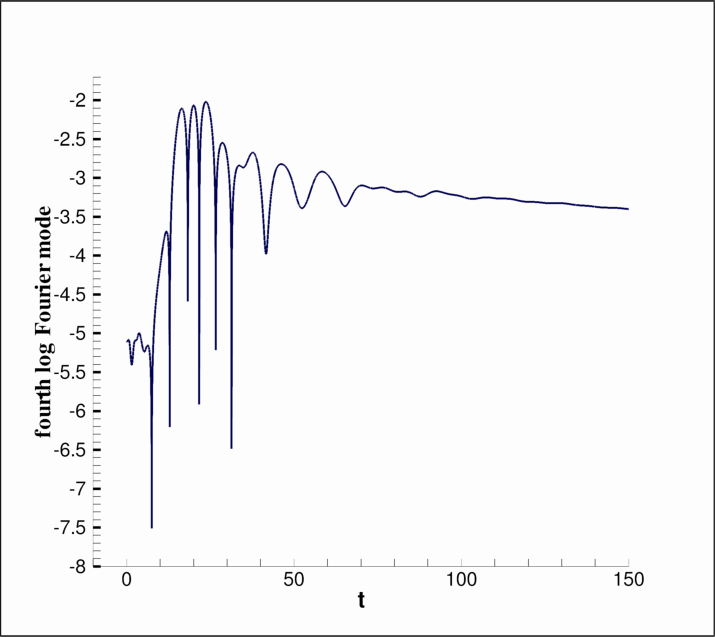} \\
\caption{(Nonlinear two-stream instability) Time series plots of the first four modes:
$k=1/2$ {\em (top left)}, $k=1$ {\em (top right)}, $k=3/2$ {\em (bottom left)}, and $k=2$ {\em (bottom right)}.}
\label{fig:2streamNewModes}
\end{figure}

Figures~\ref{fig:Two-stream 3D} and \ref{fig:Two-stream 2D} show 3D and 2D contour plots of the distribution function $f$ in
phase space at different instances of  time.  The initial data consists of two symmetric, counter-streaming, warm beams with  the small sinusoidal perturbation  superimposed, as  described above.   We obtain the qualitative behavior expected: the linear two-stream instability grows exponentially until nonlinearity becomes important and trapping occurs, with an eventual long time asymptotic Bernstein-Greene-Kruskal (BGK) state  \cite{BerGreKru} with  an apparent electron hole-like structure (cf.\ e.g.\  \cite{Sch}).   As the nonlinear system evolves, linear modes grow and saturate as shown in Fig.~\ref{fig:2streamNewModes}, the  phase space hole forms as  a portion of the  electron distribution becomes trapped and exhibits   filamentation.  Over time, the sharp contour variation associated with filamentation is smoothed out,  a consequence of  nonlinearity and the numerical dissipation of the UPG method.  (We note, here for aesthetic reasons some smoothing was also done by 
 the  plotting routine.)
Our numerical results are indicative of  a very stable computational scheme. This is partly due to the fact that our
DG method  is monotone and mass conservative. The particle number error accumulation in time is due to the nonlinear coupling.


\noindent\underline{Early Growth}

As noted above, Fig.~\ref{fig:2streamNewModes}
shows the time evolution of the first four modes, which have wavenumber values $k=0.5$, $k=1$, $k=1.5$, and $k=2$, from $t=0$ to $t= T=100$.   (Recall, our domain is of size $4\pi$.)
At early  times,  $0\le t <20$, the  behavior illustrated in  Fig.~\ref{fig:2streamNewModes}   is consistent with the results of previous authors  where the initialized mode-one with $k=0.5$ dominates   the other modes that are not initially excited.  All modes reach a maximum at   $t\approx 18$ with growth rates on the order of a couple of tenths of $\omega_p$ which is typical of linear theory for this problem.   During early times,  noise in the system and the growth of mode-one excites the other modes as was the case for the nonlinear Landau damping of Section \ref{sec:nlLD}.    A comparison was made in the calculations of \cite{DenKru} using results from \cite{GraFei}, but we give some further details here.

\begin{figure}[ht]
\includegraphics[width=.4\textwidth]{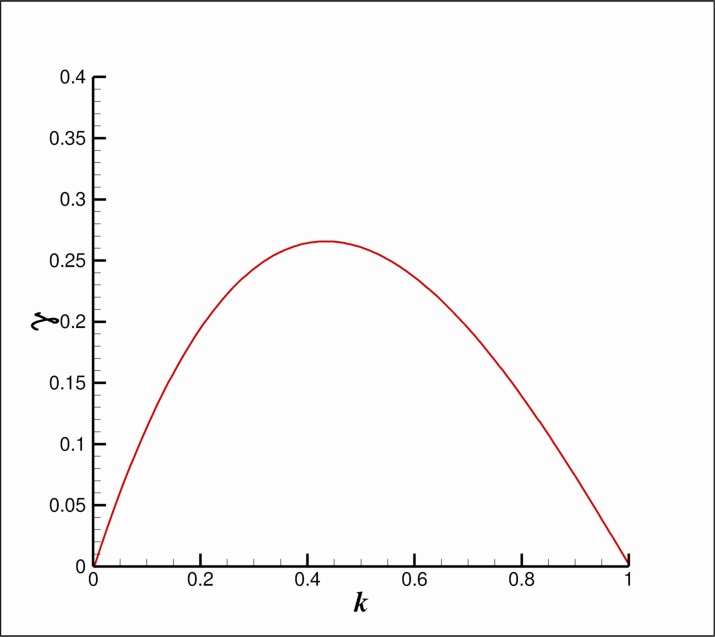}
\caption{Plot of growth rate $\gamma$ vs.\  $k$ for  the  equilibrium distribution function of (\ref{TSequil}).}
\label{df}
 \end{figure}

The plasma dispersion function from Vlasov linear theory is given by
 \bq
 \epsilon=1-\frac1{k^2}\int_{-\infty}^{\infty} f_0'(v) \frac{dv}{v -\omega/k}\,,
 \label{pdf}
 \eq
where $\omega$ will be off  the real axis as is known for the two-stream instability.  We  evaluate this $\epsilon$  with the two-stream equilibrium of (\ref{TSequil}) in order to  obtain the linear growth rates.   In Appendix \ref{app:Landau} we show  that $\epsilon$  can be rewritten in the form
\bq
\epsilon(z, k) =
1 - \frac1{k^2} \left[1 - 2 z^2  + 2 Z(z)(z -z^3)\right]\,,
 \eq
where $z=\omega/(k\sqrt{2})$ and $Z$ is the usual plasma dispersion function as defined in \cite{FriCon}.   Figure~\ref{df}  is a  plot of the growth rate obtained from $\epsilon=0$ adapted to our domain with $L=4\pi$ (cf.~\cite{GraFei}  Fig.~3) confirming that our early growth is about right.


\noindent\underline{Conservation Properties}

\begin{figure}[ht]
\includegraphics[angle=0,width=0.48\textwidth]{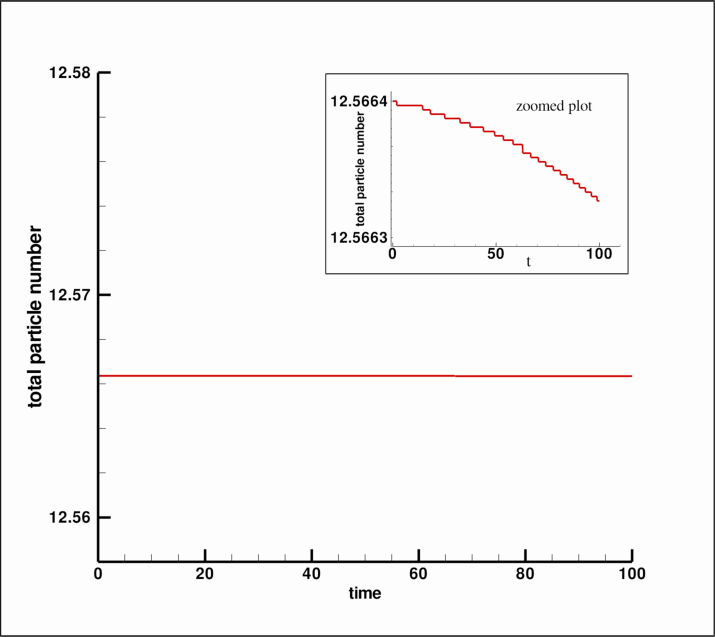}
\includegraphics[angle=0,width=0.48\textwidth]{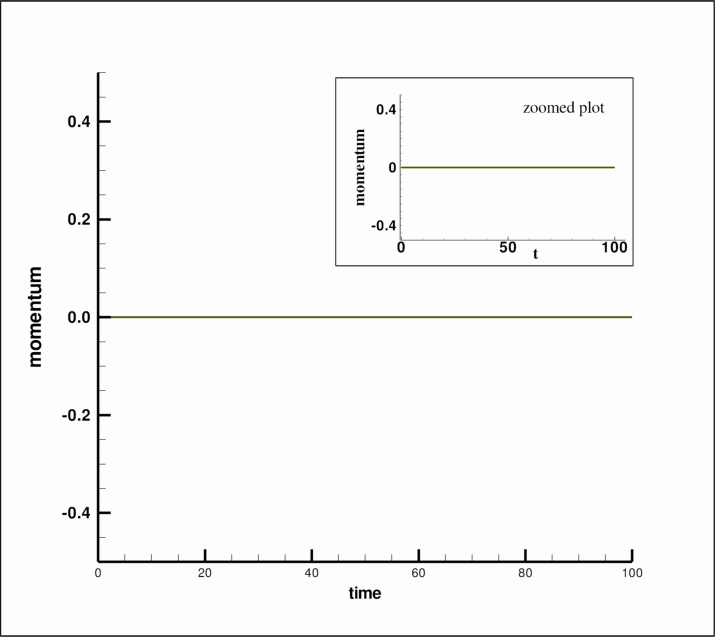}
\vspace{0.5cm}\\
\includegraphics[angle=0,width=0.48\textwidth]{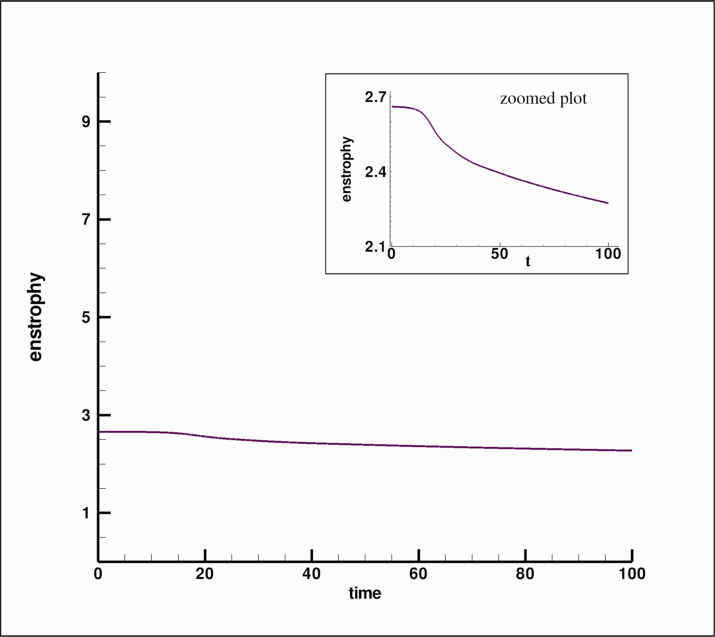}
\includegraphics[angle=0,width=0.48\textwidth]{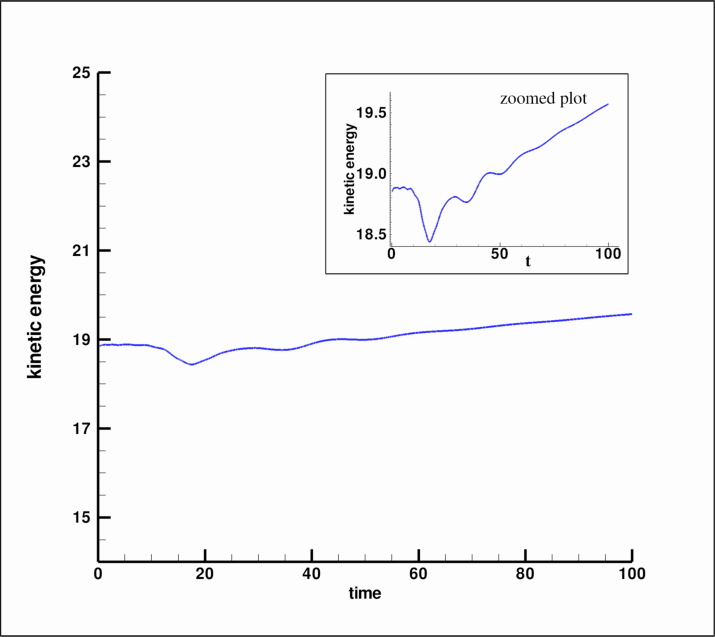}
\vspace{0.5cm} \\
\includegraphics[angle=0,width=0.48\textwidth]{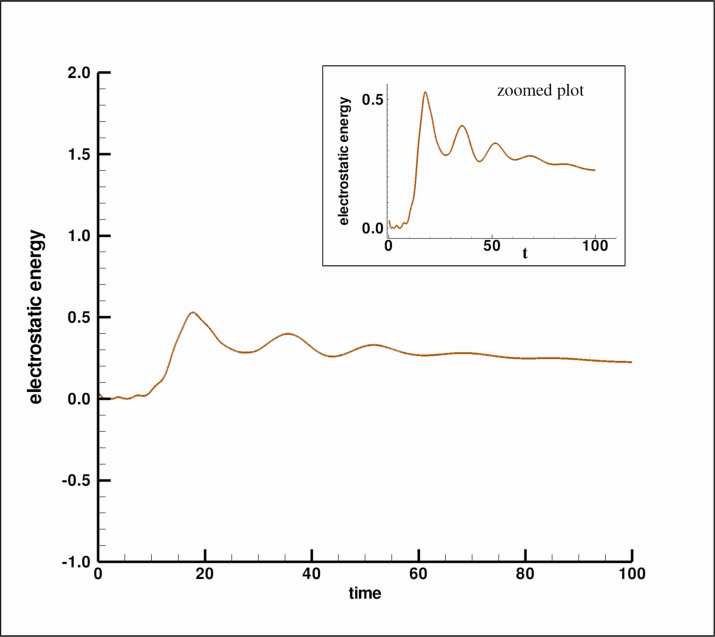}
\includegraphics[angle=0,width=0.48\textwidth]{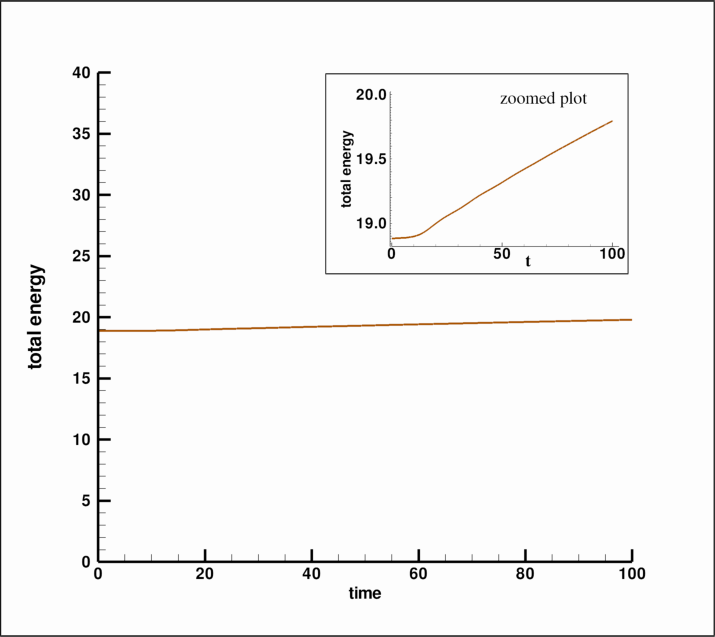} \\
\caption{(Nonlinear two-stream instability) Temporal evolution of the total particle number $N$ {\em (top left)},
momentum $P$ {\em (top right)}, enstrophy {\em (middle left)}, and kinetic energy {\em (middle right)},
electrostatic energy  {\em (bottom left)}, and  the total energy  $H$ {\em (bottom right)}. See Section \ref{sect:VP} for definitions of these quantities.}
\label{fig:Two-stream-macroscopic-quantities}
\end{figure}

\begin{figure}[htb]
\centering
\subfigure[{\footnotesize}]{
\includegraphics[width=0.47\textwidth]{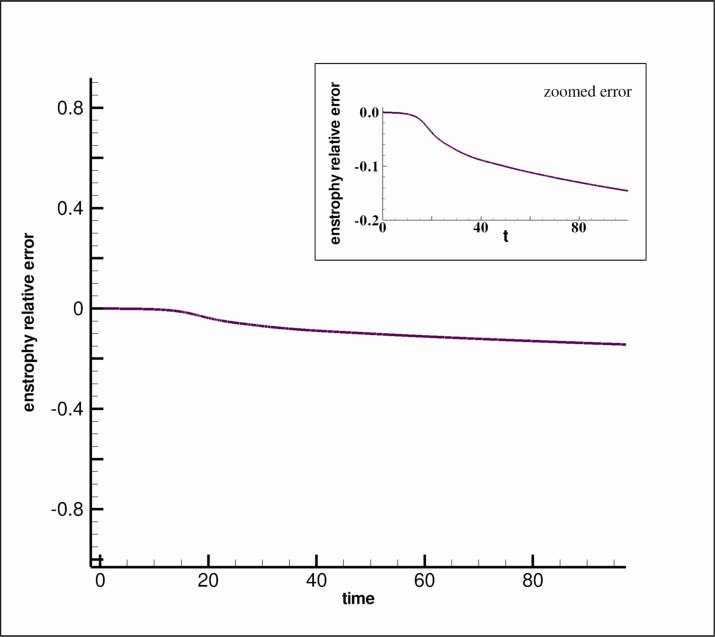}
\label{fig:enstrophy-rel-err}
}
\subfigure[{\footnotesize }]{
\includegraphics[width=0.47\textwidth]{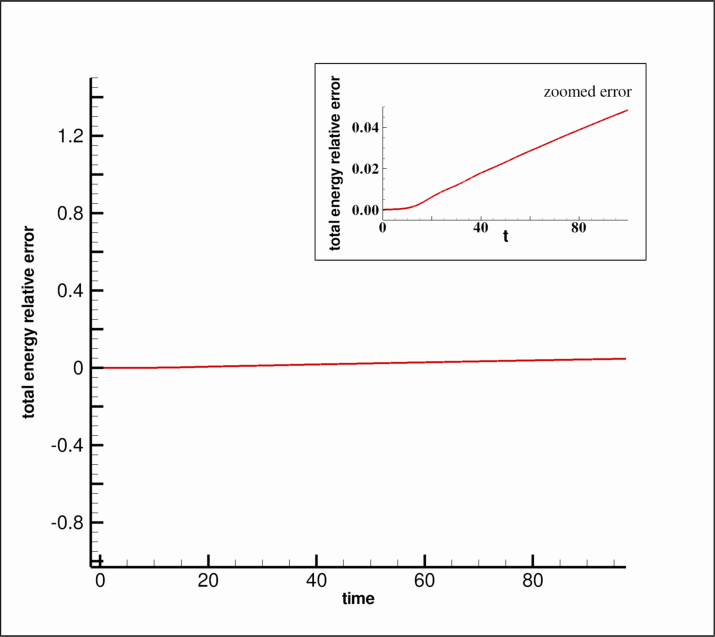}
\label{fig:ene-rel-err}
}
\caption[relative error]
{(Nonlinear two-stream instability) Relative error  of the
enstrophy  {\em (left)} and the total energy $H$ {\em (right)}.
}
\label{fig:Two-stream modes-relative-error}
\end{figure}

Figure~\ref{fig:Two-stream-macroscopic-quantities}  shows plots of the invariants of Section~\ref{sect:VP} as functions of time, while Fig.~\ref{fig:Two-stream modes-relative-error}  depicts the relative error  of the total energy, $(H-H_0)/H_0$, and similarly the enstrophy.  The top left panel of Fig.~\ref{fig:Two-stream-macroscopic-quantities}   shows that   the total particle number is conserved quite well, with an  error no larger than   $0.01\%$ over the full 100 units of time of our simulations.
Actually, the DG discretization and  Runge-Kutta (RK) method used for time advancement perfectly conserves this quantity for the transport 
equation.  However the nonlinear iteration generates a monotone in time error for the total particle number of order of $10^{-4}$ in 100 time units.    The top right panel of Fig.~\ref{fig:Two-stream-macroscopic-quantities}   shows the  evolution of  the total momentum, $P$.  
This quantity is exactly conserved because the DG method cannot break symmetry, and the same is true for the time advancement algorithm.  In fact the method perfectly conserves all odd velocity moments of $f$.   The middle left panel of Fig.~\ref{fig:Two-stream-macroscopic-quantities}  depicts the evolution of  the enstrophy Casimir invariant, $\int dx dv f^2$, a q
 uantity that appears to be seldom-monitored in Vlasov codes (\cite{FilSonBer} being an exception).      From the inset of Fig.~\ref{fig:Two-stream modes-relative-error},   it is seen to be conserved to within about 10\%, which we suspect is comparatively  good, and can be considerably improved by refinement and an increase in polynomial order.  Conservation of this quantity arises because the solution of the Vlasov equation is a rearrangement, a property that we discuss below in more detail.  It is  violated because of small scale error and the diffusive nature of numerical algorithms.  Note, like total particle number $N$, the error in the enstrophy is monotonic in time.  The middle right and bottom left panels of Fig.~\ref{fig:Two-stream-macroscopic-quantities}   show the evolution and error of the kinetic and electrostatic energies  (\ref{eq:kin.energy}) and (\ref{eq:pot.energy}), the first and second terms of (\ref{eq:energy}), respectively.  Individually these quantitie
 s are not conserved by the Vlasov equation, as is clearly evident from the figures,  but their sum,  $H$, shown in the bottom left panel, is conserved with a relative error less than $4\%$
over the entire course of the run.  The oscillations in the kinetic and electrostatic energies, indicative of the trapping process, cancel upon addition and give an error that increases monotonically in time.  In  Fig.~\ref{fig:Two-stream modes-relative-error}  the relative error of the enstrophy and total energy are depicted.   In other algorithms the error in $H$ is oscillatory in nature, typically with small temporal mean.   Even though this mean error could be small,  it is important to realize that the oscillations amount, in a sense, to successively reinitializing the code, and the cumulative error of the solution for such a process is hard to assess.

In addition to the conserved quantities of (\ref{eq:number}), (\ref{eq:momentum}), (\ref{eq:energy}), and (\ref{eq:casimir}), the Vlasov equation possesses {\em topological  conservation}.  As mentioned above, it is well-known that the solution of the Vlasov equation is a rearrangement \cite{gardner}.  This means that the solution at any time $t$ can be obtained formally  from its initial value as follows:
\begin{equation}
f(x,v,t)=f_0(x_0(x,v,t),v_0(x,v,t))=f_0\circ z_0\,,
\label{rearrange}
\end{equation}
where $z_0=(x_0(x,v,t),v_0(x,v,t))$ is obtained upon inverting the solution of the characteristic equations $z=z(z_0,t)$.   Because the characteristic equations have Hamiltonian form they preserve volume (here area), and
\begin{equation}
\frac{\partial(x_0,v_0)}{\partial(x,v)}=1\,.
\label{jac}
\end{equation}
  A consequence of this is the  family of Casimir invariants of  (\ref{eq:casimir}), whose invariance  follows directly upon effecting the coordinate change $z_0\leftrightarrow z$ and making use of  (\ref{rearrange}) and (\ref{jac}):
\begin{equation}
\int dx_0\int dv_0\,  \mathcal{C}(f_0(x_0,v_0)=\int dx \int dv \,  \mathcal{C}(f(x,v,t) \,,
\end{equation}
where we have suppressed limits of integration.

Under  continuity assumptions,  the level sets of  $f_0$ are topologically equivalent to the composition $f_0\circ z_0=f$.  This is what is meant by  topological  conservation.  It follows that the number and nature of extrema,  the values of $f$ at these extrema, and the kinds of separatrices   connecting saddle points must correspond to those of $f_0$.   In addition, although not a topological property,   the area between any two contours of $f$ must also be conserved,  a consequence  of the area preservation property of the characteristics.  Since  DG is a weak formulation with lack of continuity, the extent to which level sets are actually topologically conserved remains to be seen. Casimir invariants such as enstrophy and entropy may be conserved well  and be consistent with  rearrangement inequalities such as Jensen's inequality  without the continuity assumption.

Because the initial  condition has the form  $f_0(x,v)= f_{TS}(v)[1 +0.05\cos(x/2)]$, the initial phase space has nearly imperceptible initial `tears'  in the first panels of Figs.~\ref{fig:Two-stream 3D} and \ref{fig:Two-stream 2D}, that should be  consistent with asymptotic state at $t=T=100$.  Extrema of $f_0(x,v)$ occur at points $(x,v)$:  $(x,0)$, for all $x\in\{0,4\pi\}$, $(0,\pm\sqrt{2})$, and $(2\pi,\pm\sqrt{2})$.  Thus there is a trough along the line $v=0$. The points $(0,\pm\sqrt{2})=(4\pi,\pm\sqrt{2})$ atop the beams are maxima straddled by  thin separatrices entering and emerging from the saddle points located at $(2\pi,\pm\sqrt{2})$.  The following extremal values of $f_0$ remain fixed in time under Vlasov dynamics, although their locations can change:  $f_0(0,0)=0$, $f_0(2\pi,0)=0$, $f_0(0,\pm\sqrt{2})=0.308$, and $f_0(2\pi,\pm\sqrt{2})= 0.279$. A glance at Fig.~\ref{fig:Two-stream 3D} shows some degradation of the value of $f$ atop  the ridge-like structures, 
 while the vanishing value of $f$ at the points  $(0,0)$ and  $(2\pi,0)$ is rigorously maintained.   A  measure of error in an  algorithm is the extent to which it  fills in the hole, i.e.\ returns   values of $f_0(2\pi,0)\neq0$.

In our simulations the value of $f$ remains zero at the minimum $(2\pi,0)$  and it remains  zero to high accuracy at $(0,0)$.  The values of $f$ at the extrema and the associated separatrices atop the beams are never very discernible, their existence due to only a few percent in the variation of $f$,  and are washed out by numerical dissipation as the contours of the distribution function wrap around and `trap particles' over the course of the simulation.   However, the trough at $v=0$ is structurally unstable and separatrices  emerge form here and connect  the saddle point at $v=0$ and  $x=0=4\pi$ that straddles the minimum at $(2\pi,0)$, the center of the electron hole.   The eventual boundaries  that delineate the trapped and untrapped particle populations, as the potential saturates into  what appears to be a  final BGK-like state, is what we discuss next.


\noindent\underline{BGK Saturation}

Examination of Figs.~\ref{fig:Two-stream 3D},   \ref{fig:Two-stream 2D}, and \ref{fig:2streamNewModes}  shows the  initial linear growth
 phase, followed by a particle trapping phase,  and  eventually a strong indication of a saturated state with clean contours of $f$,    resulting  at least in part from the small scale averaging inherent in any algorithm.  It is generally believed that this evolution saturates, in some weak sense, to a BGK mode, although no proof exists.  To test this belief we check to see if the contours of $f$ are aligned with contours of the particle energy, $\mathcal{E}=v^2/2-\Phi(x,t)$, which is well-known to be the case for an equilibrium state of the Vlasov equation (e.g.\ \cite{BerGreKru}).

\begin{figure}[ht]
\includegraphics[angle=0,width=0.48\textwidth]{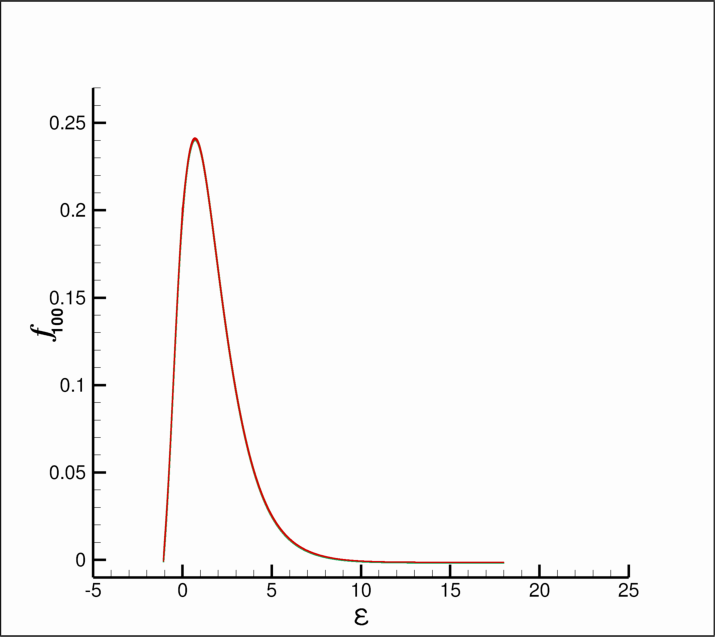}
\includegraphics[angle=0,width=0.48\textwidth]{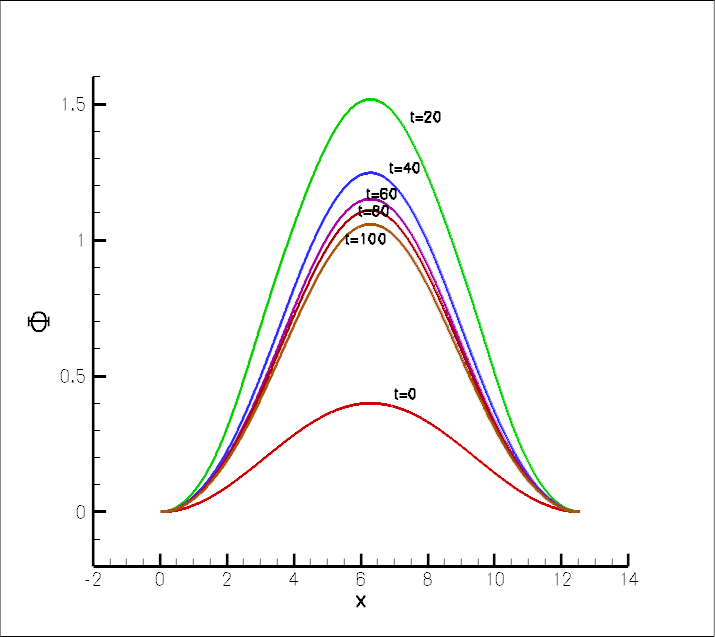}.\\
\caption{(Nonlinear two-stream instability)   Plot  of the  distribution
function at time $t=100$, $f_{100}$,  as a function of the particle energy   $\mathcal{E}=  v^2/2 -\Phi(x,t=100)$.
Green dots correspond to  positive  velocities and  red dots to negative velocities, $v$  {\em (left)}.
Right panel depicts the potential $\Phi(x,t)$ at the times indicated.  $\Phi(x,t=100)$ was used in $\mathcal{E}$ for the plot
of  $f_{100}$  {\em (right)}.
}
\label{fig:tot-energy-vs-pdf}
\end{figure}

To this end we plot $f_{100}(\mathcal{E}(x,v)): = f(x,v, 100)$, where  in  $\mathcal{E}(x,v)=v^2/2-\Phi(x,100)$, the particle energy at  $t=100$.  (Note, we suppress the time variable.) \ \   Figure~\ref{fig:tot-energy-vs-pdf} clearly indicates that a saturated BGK state has been achieved. Here,  in the left panel,  we have plotted $f$ versus $\mathcal{E}(x,v)$,  at $t=100$,  for all 9 million pairs $(x,v)$ of  the  uniformly  distributed mesh over our   phase space.  Observe that to within the thickness of the line, $f_{100}$ appears as  a graph over $\mathcal{E}(x,v)$ and that this is true  even for large values of $\cale$, where $f_{100}>0$ is maintained.    Green and red dots correspond to positive and negative values of the velocity, respectively, and there does not seem to be any systematic directional bias.  Because the  computation was done for piecewise constant element functions,   it is known that DG ensures  $f(x,v,t)$ remains  positive for all times a
 nd this is then the case for  $f_{100}$. The right panel of Fig.~\ref{fig:tot-energy-vs-pdf} is a plot of the electrostatic potential as a function of position for several instances in time as indicated in the figure.  The curve labeled by $t=100$ is taken to be  the saturated electrostatic potential that was used in the calculation of $f_{100}$.

In Figs.~\ref{fig:f_details} and \ref{fig:BGK-near-splitting-birth} high resolution details of $f_{100}$ are depicted.  In Fig.~\ref{fig:cusp} a cusp is seen at the trapping boundary that occurs at $\cale=0$.  A magnified view of this is shown in Fig.~\ref{fig:cusp_zoom}.  In the course of the evolution the presence of the electrostatic force produces a trapped particle population with $\cale<0$.  It is conceivable that trapping, like scooping ice cream, is a non-analytic process and that this cusp represents something real about the mathematical nature of the dynamics.  However, it also may be attributed to the choice of the numerical scheme, yet discontinuities have previously been proposed at trapping boundaries: in  \cite{Sch} a discontinuity is used in making electron hole models and in \cite{Dewar} two kinds of discontinuities are proposed for adiabatic and sudden trapping.  None of these discontinuities match the cusp like feature that we observe, but the traveling wav
 e state treated in  \cite{Dewar} is not the same as that reached by our simulation and it is possible that an analysis similar to that of \cite{Dewar}  could produce a cusp.  Further studies of this effect are  currently being investigated.    Figures~\ref{fig:tip} and \ref{fig:tip_zoom}  examine $f_{100}$ near the minimum value of $\cale$.  Note the clear steepening of $f_{100}$ as it approaches its minimum value. It is possible that there is a universal nature to both this steepening and  to the cusp at the trapping boundary, but we leave an investigation of this to future work.   Figures \ref{fig:max} and \ref{fig:max_zoom} examine $f_{100}$ near its maximum value.   The spread here as well as the other panels of Fig.~\ref{fig:f_details}  give a sense of how close the system has relaxed  to an equilibrium state.

Figure~\ref{fig:BGK-near-splitting-birth} examines the details of $f_{100}$  for higher values of $\cale$.  In Fig.~\ref{fig:split2} a splitting is seen to  occur at  around $\cale=1.3$.  In Figs.~\ref{fig:split_zoom1} and \ref{fig:split_zoom2} we see that this small splitting persists to large values of $\cale$.  It is interesting to note that because the red and green dots are mixed within each band, the splitting is not an effect of velocity direction.  The splitting is within the resolution of the code and is believed to be a real effect, one that seems to indicate that complete saturation has not occurred.

The above tests provide strong indication that the code has relaxed to near a  BGK state.  As further evidence we test to see if the charge density associated with $f_{100}$ is consistent with that for a final BGK state.  To this end we first observe that  $f_{100}$  can be fit reasonably well by a model distribution function of the following form:
\bq
f_{\rm fit}=a(\cale+\Phi_M)(\cale +\cale^*)e^{-\beta \cale}\,.
 \label{roughfit2}
 \eq
Here $\Phi_M$ is the maximum value of $\Phi$, and because $\cale=v^2/2-\Phi$ we see $-\Phi_M$ is the smallest
possible value of $\cale$.  This and positivity of  $f$ imply $\cale^*>\Phi_M$.  Obviously (\ref{roughfit2}) does not account for the fine features discussed above, but it will be sufficient for our purpose.

\begin{figure}[ht]
 \includegraphics[width=.6\textwidth]{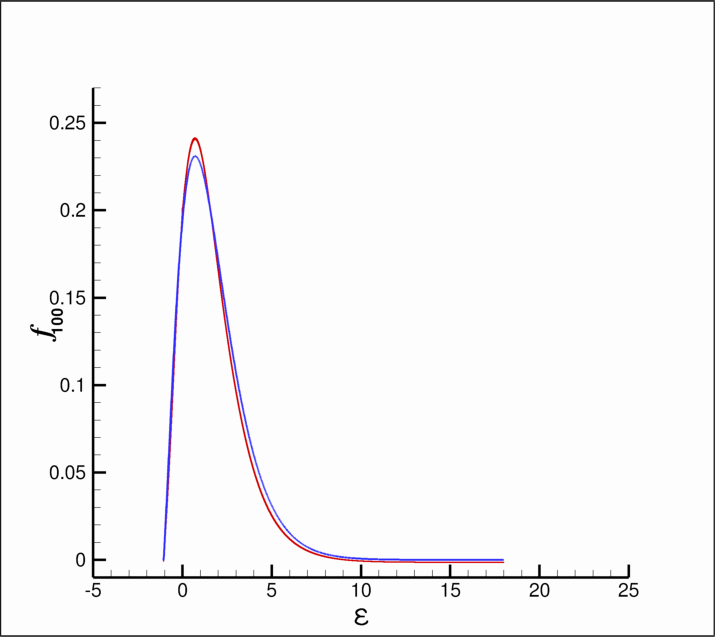}
\caption{(Nonlinear two-stream instability) Model distribution function $f_{\rm fit}$ of Eq.~(\ref{roughfit2})
with $\Phi_M=1.06$,  $\cale^*=1.59$, $\beta=1$, and $a=0.1148$ (solid blue) compared to code results at $t=100$ (red dots).}
\label{12}
 \end{figure}

\begin{figure}[htb]
\centering
\subfigure[]{
\includegraphics[width=0.47\textwidth]{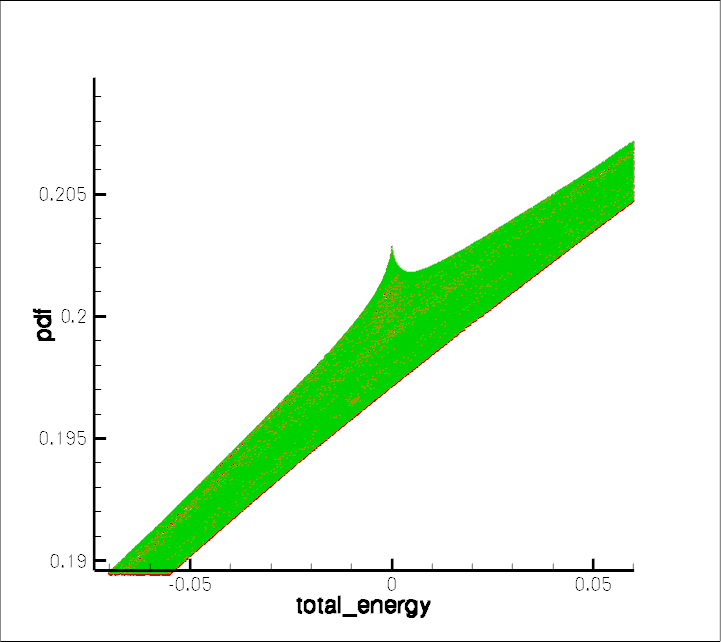}
\label{fig:cusp}
}
\subfigure[]{
\includegraphics[width=0.47\textwidth]{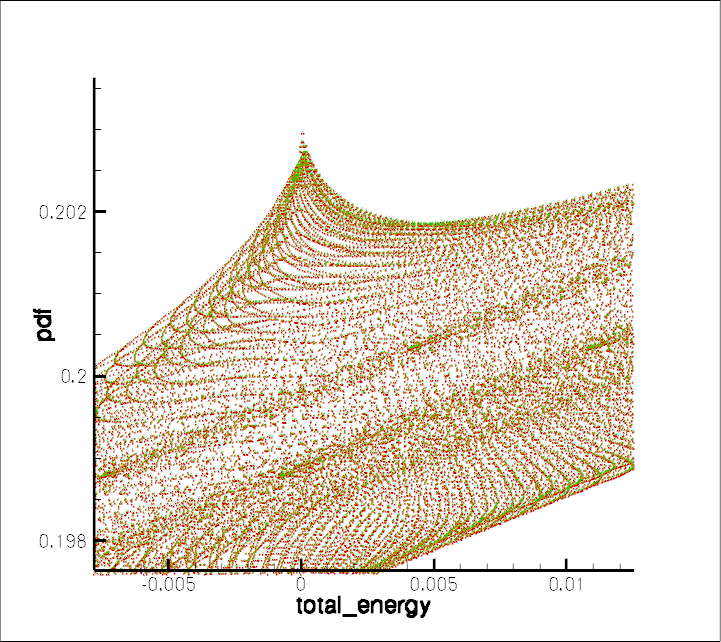}
\label{fig:cusp_zoom}
}
\subfigure[]{
\includegraphics[width=0.47\textwidth]{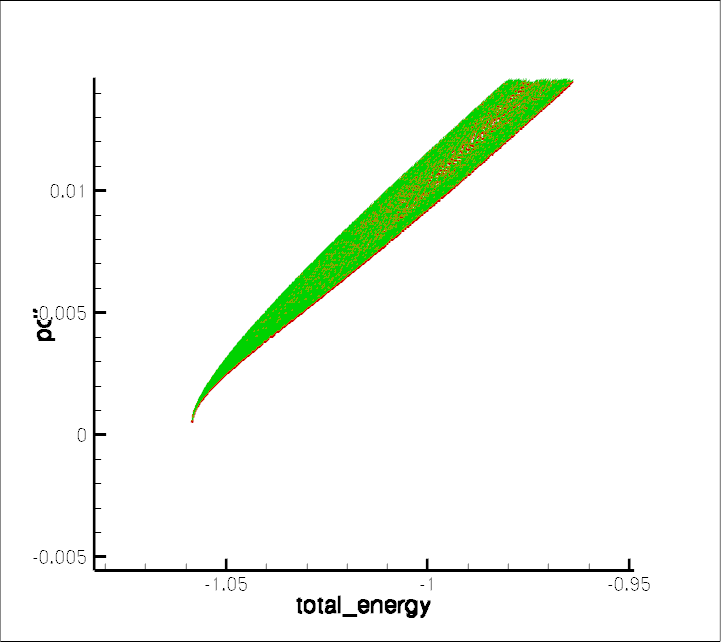}
\label{fig:tip}
}
\subfigure[]{
\includegraphics[width=0.47\textwidth]{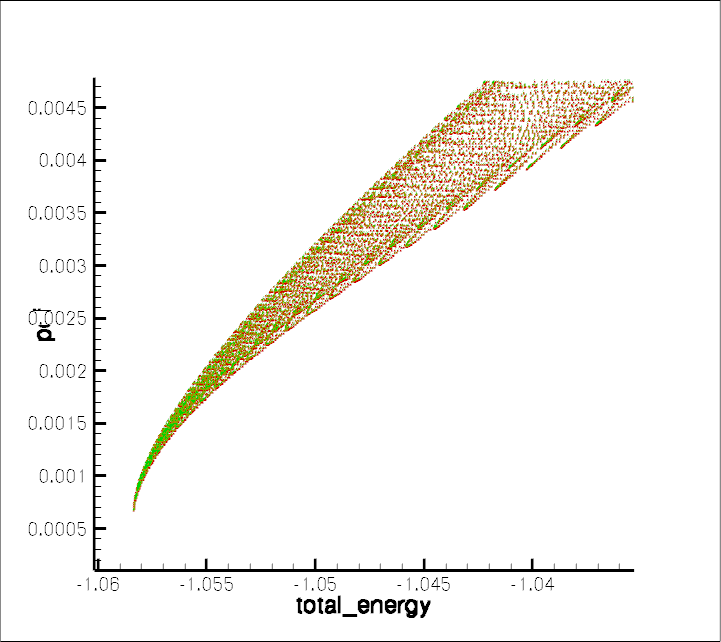}
\label{fig:tip_zoom}
}
\subfigure[]{
\includegraphics[width=0.47\textwidth]{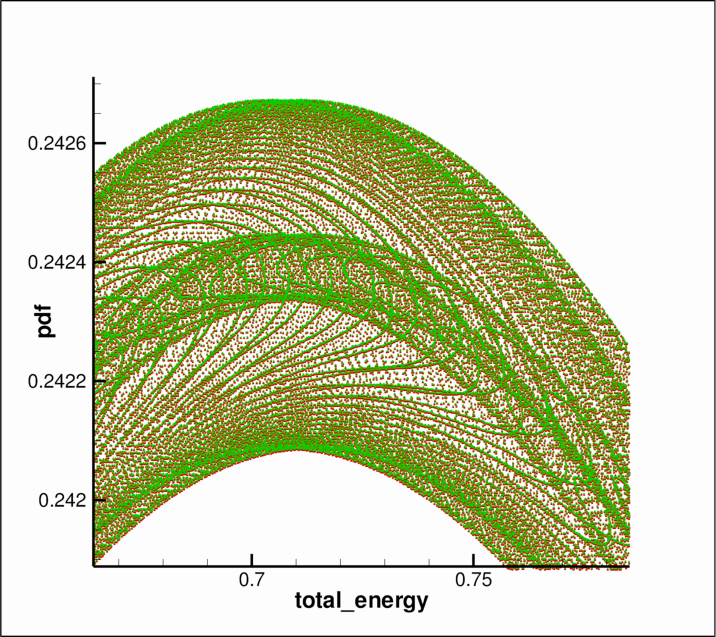}
\label{fig:max}
}
\subfigure[]{
\includegraphics[width=0.47\textwidth]{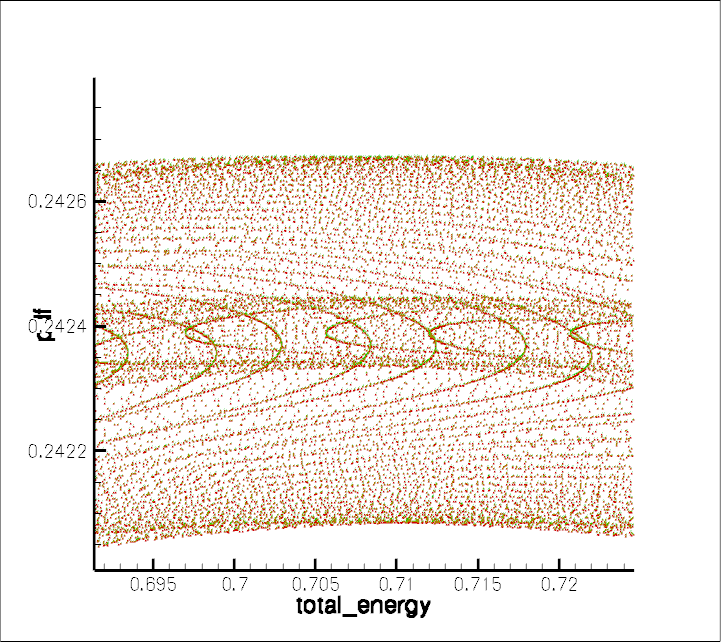}
\label{fig:max_zoom}
}
\caption[details of distribution function]{(Nonlinear two-stream instability)  Fine scale details of the  distribution function $f_{100}(\mathcal{E})$, depicting a cusp formation near the trapping boundary at $\cale=0$ {\em (top left and right)}.    Near the minimum value of $\cale$, $f_{100}$ is observed to greatly steepen {\em (middle left and right)}.  The maximum of $f_{100}$ is achieved at about $\mathcal{E}=0.71$ {\em (bottom left and right)}.   In all plots red dots correspond to  negative velocities and  green dots to positive velocities, $v$.  No asymmetry in the sign of $v$ is evident.
}
\label{fig:f_details}
\end{figure}

\begin{figure}[htb]
\centering
\subfigure[]{
\includegraphics[width=0.47\textwidth]{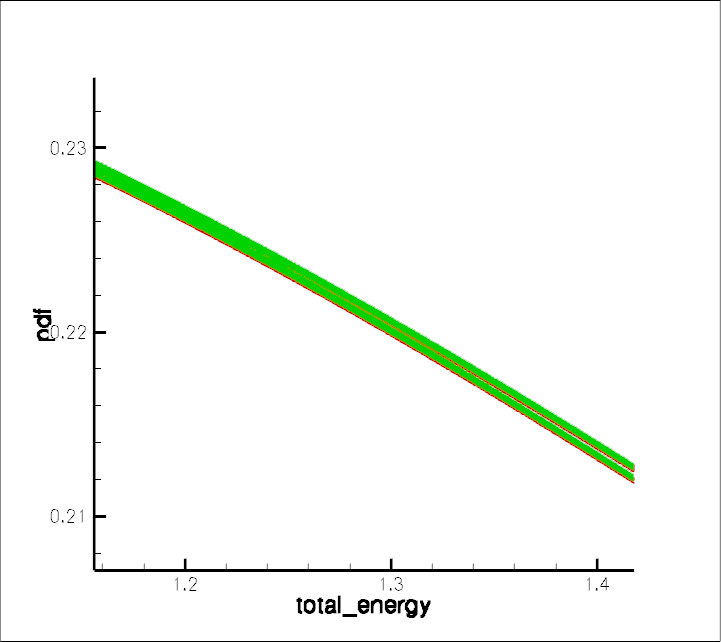}
\label{fig:split1}
}
\subfigure[]{
\includegraphics[width=0.47\textwidth]{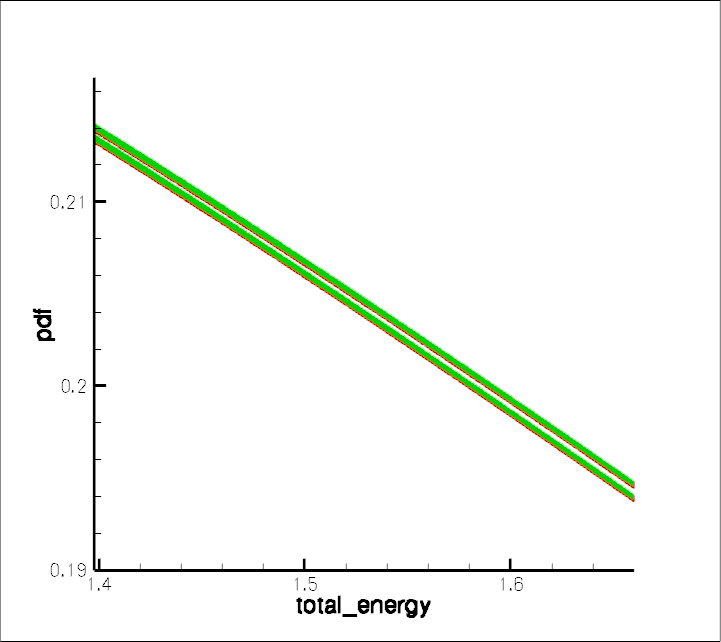}
\label{fig:split2}
}
\subfigure[]{
\includegraphics[width=0.47\textwidth]{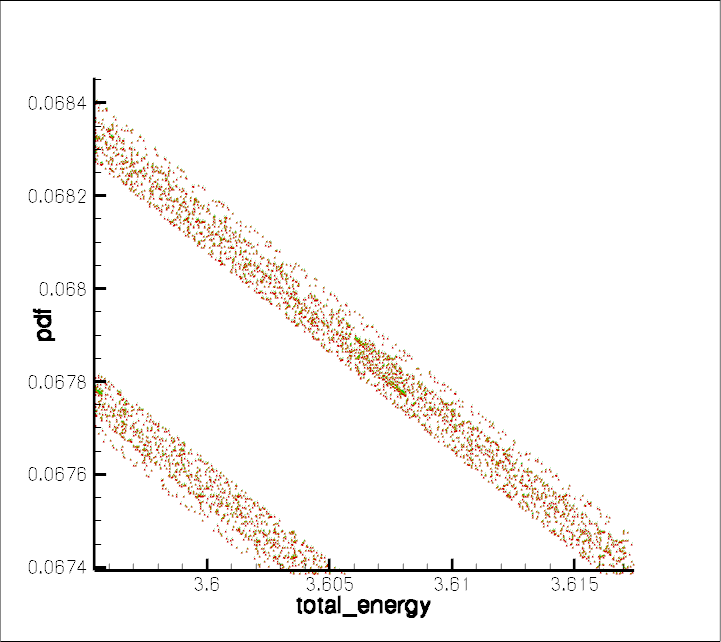}
\label{fig:split_zoom1}
}
\subfigure[]{
\includegraphics[width=0.47\textwidth]{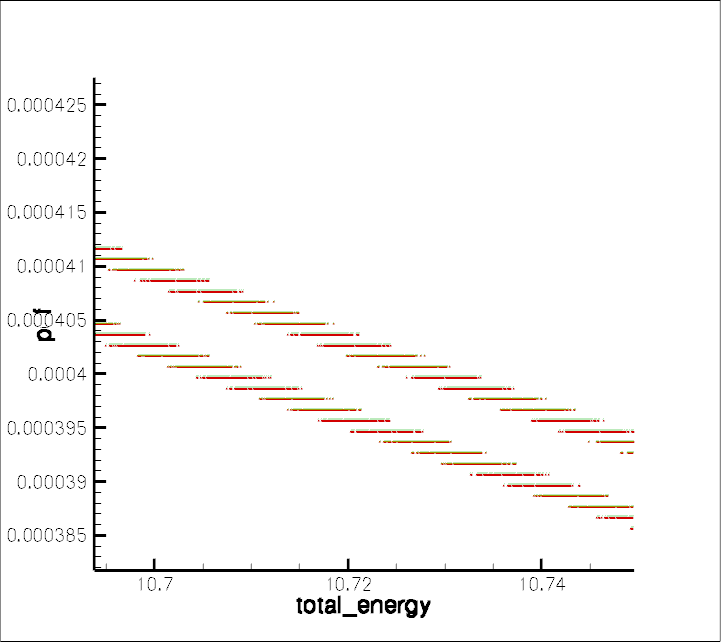}
\label{fig:split_zoom2}
}
\caption[splitting]{(Nonlinear two-stream instability)  Plot of $f_{100}(\mathcal{E})$  near   $\mathcal{E}=1.3$ indicating mild splitting {\em (top left)}, with stronger splitting   seen at  $\mathcal{E}\approx1.5$   {\em (top right)}.  The splitting continues to larger values of $\cale$  {\em (bottom left and right)}.   Red dots correspond to  negative velocities and  green dots to positive velocities $v$, with no evident asymmetry in the sign of $v$.
}
\label{fig:BGK-near-splitting-birth}
\end{figure}

The quantities $a$, $\Phi_M$, $\cale^*$, and $\beta$ are fitting parameters that can be matched to the code output.  We obtain a rough idea of the degree of self-consistency, i.e. the degree to which Eq.~(\ref{roughfit2}) produces the correct electrostatic potential when substituted into Poisson's equation, by using the following values extracted from the code output: $\Phi_M=1.06$ and $\cale^*=1.59$.  For convenience we choose $\beta=1$.  More significant figures in $a$, viz.\  $a=0.1148$, are required to  ensure  that the total net charge is zero. In Fig.~\ref{12} we see these choices   provide a pretty good fit to the data.  However, insufficient charge near the maximum is spread to higher values of $\cale$.   In a forthcoming publication we will examine this sort of modeling in much greater detail.

We note, there is a relationship between $\cale_M$,  
the value   at which $f$ obtains its maximum, and the other parameters.  That is, $f'(\cale_M)=0$  implies
\bq
\cale_M=\frac{2-\beta(\Phi_M+\cale^*)+\sqrt{ 4+\beta^2(\Phi_M-\cale^*)^2}}{2\beta}
\eq
or
\bq
\cale^*=\frac{\Phi_M -\beta \cale_M\Phi_M +  2\cale_M-\beta\cale_M^2}{\beta\cale_M +\beta \Phi_M -1}\,,
\label{estar}
\eq
where the latter is useful when $\cale_M$ is extracted from data.  In fact, the value of $\cale^*=1.59$ used above follows from $\cale_M=0.71$ (cf.\ Fig.~\ref{fig:f_details}).


Given   $f_{\rm fit}$,  the charge density is given by
\bq
\rho_{\rm fit}(\Phi)=1-  \int_{-\infty}^{\infty} \! dv\, f_{\rm fit}=
1-  \int_{-\Phi}^{\infty} \! \frac{d\cale \, f_{\rm fit}(\cale) }{\sqrt{2(\cale +\Phi)}}\,,
\eq
where in the second equality the formula $\cale=v^2/2-\Phi$ is used to change the integration variable from $v$ to $\cale$.
With the choice (\ref{roughfit2}) this quantity can be explicitly integrated to give
\bq
 \rho_{\rm fit}=1-    \frac{ \sqrt{2\pi}\,  a}{\beta^{5/2}}
 \left[
 \frac{3}{4} -\beta^2(\Phi_M-\Phi)(\Phi-\cale^*) + \frac{\beta}{2} (\Phi_M-2\Phi+\cale^*)
 \right]
e^{\beta \Phi}\,.
\eq

Defining a pseudopotential $\mathcal{V}$ by $\p \mathcal{V}/\p \Phi=\rho_{\rm fit}$, and integrating Poisson's equation yields
\bq
\frac{E^2}{2} + \mathcal{V}(\Phi) =  \mathcal{V}_0\,,
\label{E2}
\eq
where $\calv_0$ is a constant and
\begin{eqnarray}\label{V}
\mathcal{V}(\Phi)&=&  \Phi  \\
&-&  \frac{\sqrt{2\pi}  a}{\beta^{7/2}}\left[\frac{15}{4} - \beta^2 (\Phi_M - \Phi) (\Phi - \cale^*) +
   \frac{3}{2} \beta(\Phi_M  - 2\Phi + \cale^*)\right]\, e^{\beta \Phi}
   \nonumber
\end{eqnarray}
is the expression for the pseudopotential.   Here ${E^2}/{2}$ acts as a pseudo kinetic energy.  Thus we can interpret the BGK state by comparison to  a particle dropped in the potential $\calv$ at zero kinetic energy.  Such a fictitious particle returns to a state of zero kinetic energy as $x$ traverses the spatial domain, which must be the case if there is zero net charge.

\begin{figure}[ht]
 \includegraphics[width=1.0\textwidth]{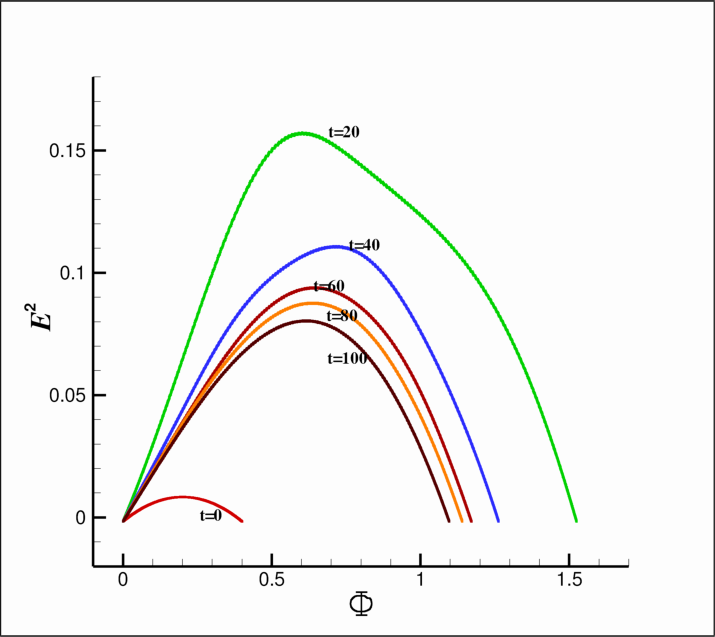}
\caption{(Nonlinear two-stream instability) Plot of $E^2$  versus $\Phi$  for the code output at the times indicated.
 Note, $E^2$ is a graph over $\Phi$ even for the unsaturated states, i.e.\ times $t\le 100$.
}
\label{EEfig}
 \end{figure}

In Fig.~\ref{EEfig},  $E^2$  is  plotted against  $\Phi$ for our simulation data.  In light of (\ref{E2}) and (\ref{V}),  we expect  $E^2$ to be a graph over $\Phi$ at time $t=100$, and from the figure this indeed appears to be the case. Also, since at $t=0$,  $\Phi''= A \cos(x/2)$,     it is easy to see that $E^2(x,t=0)/2 =  A\Phi(x,t=0) - \Phi(x,t=0)^2/8$.  This follows from the  choice of ground for $\Phi$ and the absence of net charge which gives via Gauss' theorem $E(x=4\pi)=E(x=0)$.  With our boundary conditions $E(x=0)=0$.   Thus we expect  the parabolic curve in Fig.~\ref{EEfig} labeled by $t=0$  with maximum $\Phi$ occurring  $\Phi_M = 8A=8\times 0.05=0.40$.  However, it is remarkable that at intermediate times $E^2$  is also a graph over $\Phi$.  This can be shown to be true if $\Phi$ maintains symmetry about $x=2\pi$.  This suggests that $\Phi$ can serve as a good spatial coordinate, an idea we will discuss further in a forthcoming publication.

\begin{figure}[ht]
 \includegraphics[width=.8\textwidth]{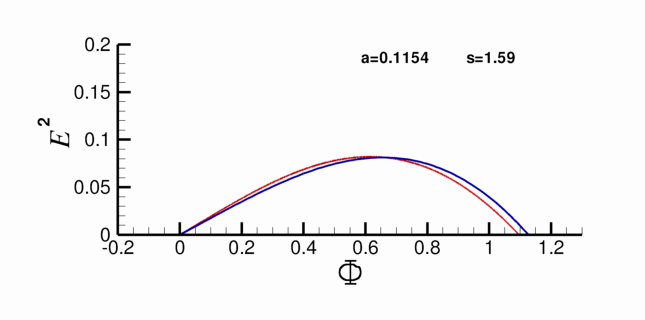}
\caption{(Nonlinear two-stream instability) Plot of $E^2$ vs.\ $\Phi$ as given by
(\ref{E2}) with (\ref{V}) for the model $f_{\rm fit}$ of (\ref{roughfit2}) with the parameters of Fig.~{12},
and code output (red), both at time $t=100$.
}
\label{Vfig}
 \end{figure}

 The data of Fig.~\ref{EEfig} can be compared to the model of Eq.~(\ref{E2}) with (\ref{V}).  Choosing  $\calv_0$ so that $E^2(\Phi=0)=0$ and the parameter values of Fig.~\ref{12},  yields the plot of Fig.~ \ref{Vfig}.  Here we see  a reasonable fit to the data at $t=100$.  Note, $E^2(\Phi_M)\approx 0$ and the maximum value of $E^2$ is a close fit.  We note, a certain level of accuracy  is needed in the parameters because $E^2$ is a sensitive function of $\Phi$.  We have not optimized the fit, but the result of Fig.~\ref{EEfig} is roughly what one might expect for the expansion of $f_{100}$ with only three terms of a Laguerre series, and thus gives a fair indication of self-consistency which was our goal.  As noted above, we will revisit this again in great detail in a future publication.


%

%

%
%


\clearpage

\section{Conclusions}
 \label{sect:conclusions}

 In this paper we have demonstrated the viability of the DG method with upwinding for solving the Vlasov-Poisson system.
 The method was described in detail for general polynomial bases of elements.  Several examples were computed to demonstrate
the utility of the method using a piecewise uniform basis.  A convergence study was performed for the  simple advection problem,
indicating the degree to which  filamentation can be resolved,  and high resolution computations of the linearized Vlasov-Poisson
system were also performed.  For linearization about a Maxwellian equilibrium,   computation results were seen to compare very well
with theoretical calculations of Landau damping.    Computations for Lorentzian equilibria were also presented and the wave number
dependence of the Landau damping rate was verified, apparently for the first time in a simulation.   The problems of nonlinear Landau
damping and the symmetric two-stream instability were then considered, and results were compared to previous calculations.  In both cases, 
constants of motion were monitored and the error was seen to be monotonic, due to nonlinear coupling.   Also,   reasons for the lack of numerical error in the form of recurrence in our results were discussed.   High resolution features  of the distribution function were displayed for the long time BGK state that was reached in the two-stream calculations. Comparisons between theory and code results were made,  particularly for the end BGK state, in an attempt to provide a high standard for truth in numerical code work.

 There are many future directions suggested by this work,  some of which are ongoing.  In a couple of upcoming publications \cite{cheng-gamba-morrison11, cheng-gamba-11} the authors
will report on computations using  higher order polynomial bases with an improved temporal integrator and a limiter that maintains positivity of the distribution function,  as well as a study of numerical recurrence for these schemes. 
Indeed,  \cite{cheng-gamba-morrison11} contains a thorough study of dependence of recurrence 
times and recurrence amplitudes on the mesh size in $x,v$ and the time step $\Delta t$,  for  Vlasov linear advection, as well as  dependence on the order, type, and  choice of basis functions for the DG scheme.  It is quite remarkable, that for the lowest order polynomial space of piecewice constant functions, one can prove that   recurrence  occurs,  but the recurrence amplitude will decay,   thereby  suggesting  value in choosing lower order functions. This fact significantly improves  the performance criteria  originally developed in \cite{CheKno}.

 As noted in the Introduction, the DG method  can be run easily on nonuniform meshes, and in work not reported here we have seen this to be the case.   This is a first step toward producing an adaptive code.  Similarly, parallel implementation awaits.   There are many physical applications within reach, such as the treatment of a plasma diode which requires physical boundary conditions and the inclusion of various collision operators of relevance to plasma dynamics.  Since the Vlasov-Poisson system serves as a test bed  for more sophisticated kinetic theories, it is  our opinion that the DG method proposed here is an attractive alternative for many plasma physics computations.

  \section*{Acknowledgment}
\noindent R.~E.~Heath was supported
by a  Research and Development Grant from  
Applied Research Laboratories, The University of Texas at Austin,
Austin, TX; I.~M.~Gamba   was supported by NSF grants DMS-0807712 and FRG-0757450, while P.~J.~Morrison was supported by U.S.~Dept.\ of Energy Contract \# DE-FG05-80ET-53088. Support from the Institute
from Computational Engineering and Sciences at the University
of Texas at Austin is also gratefully acknowledged.
 We all  thank Clint Dawson for his technical help and  Wenqi Zhao for her technical support 
with data post-processing.  I.~M.~Gamba and P.~J.~Morrison thank  Cedric Villani for interesting discussions related to his recent work on Landau damping.

\appendix

\section{Appendix: Dispersion Relations}
\label{app:Landau}

Here we analyze the dispersion relations for the Lorentzian and two-stream equilibria given
by (\ref{Leq}) and (\ref{TSequil}), respectively.  With the assumption that $f(x,v,t)=f_{eq}(v)+\delta f(x,v,t)$ and
$\delta f(x,v,t)\sim \exp(ikx-i\omega t)$ and the scaling of variables described in Section \ref{sect:VP}, the plasma
`dispersion relation' is given by
\be
\label{eq:def.dispersion}
\epsilon(k,\omega) =  1-\frac1{k^{2}}\,\int\limits_{-\infty}^{\infty} \frac{f_{eq}'(v)}{(v-\omega/k)}\,dv \,,
\ee
where $k$ is assumed to be real and positive and  for bounded systems $k=2\pi n/L$ with $n$ an integer.   Landau damping arises upon  analytically continuing this expression into the lower half $\omega$-plane and thus deforming the contour of integration  \cite{Landau}.  For integration along the real axis, stable and unstable eigenmodes (and embedded modes if they exist) satisfy
\be
 \label{eq:dispersion.relation}
\epsilon(k,\omega)  =0 \,,
\ee
and it is this quantity we wish to investigate for both discrete modes and Landau `modes', the latter  obtained by analytic continuation into the lower half $\omega$-plane.  In the latter case, the solution $\omega(k)$ of   (\ref{eq:dispersion.relation}) characterizes the asymptotic-time
behavior of  mode $E_k(\omega,t)$ of the electric field.  Specifically, if $\omega = \omega_R + i\gamma$,  where $\omega_R$ and $\gamma$ are real-valued, then $\gamma<0$ is the time-asymptotic rate of decay of the mode $E_k(\omega,t)$ and $\omega_R$ is the frequency of oscillation.

For the Lorentz  equilibrium distribution function of  (\ref{Leq}),
\be
\label{eq:derivative.lorentz}
f_{eq}'(v)  = - \frac{2}{\pi}\, \frac{v}{ (v^2 + 1)}\ .
\ee
Upon defining $u=\omega/k$ and substituting (\ref{eq:derivative.lorentz})  into (\ref{eq:def.dispersion}), we obtain
\be
\label{eq:dispersion.lorentzian}
\epsilon(k,ku) =   1  +  \frac{2}{\pi k^2}\,\int_{-\infty}^{\infty}
\frac{v}{(v^2 + 1)^2(v-u)}\,dv \,,
\ee
which expresses $\epsilon$ as a function $u$ and $k$.  Evaluation of this integral is a straightforward application  of Cauchy's theorem.  Assuming  $u$ is in the upper half plane, poles exist at $v=i$ and $v=u$, with corresponding residues of the integrand being $R_i=i/[4(u-i)^2]$ and $R_u=u/(u^2 +1)^2$.  Summing over the residues gives the dispersion function
\be  \label{eq:dispersion.final.form}
\epsilon(k,ku)  = 1 - \frac{1}{k^2(u+i)^2} \,.
\ee
 Upon setting $\epsilon=0$, since  $\omega=ku$, we obtain
 \be
 \omega= \pm1 -ik= \omega_R + i\gamma\,,
 \ee
which demonstrates, contrary to our assumption, that there are no discrete modes in the upper half $u$-plane.  This is consistent with the well-known result that equilibrium distribution functions that are monotonic in $v^2$ possess no discrete growing or damped modes.  However, continuing (\ref{eq:dispersion.final.form}) into the lower half plane gives  Landau damping with $\gamma(k) = -k$ and $|\omega_R(k)| = 1$ for the time asymptotic behavior.

For the two-stream equilibrium of (\ref{TSequil}), (\ref{eq:def.dispersion}) gives rise to instability, i.e., for this case there is in fact a root with $u$ in the upper half plane.   For computational reasons it is convenient to write $\epsilon$ in terms of the plasma $Z$-function which is related to both the Hilbert transform and the error function (see e.g.\ \cite{FriCon}).  Upon inserting  (\ref{TSequil}) and performing some manipulation, (\ref{eq:def.dispersion}) can be written as
\be
\epsilon= 1- \frac{2}{k^2}\left[J_1(z) + J_2(z)\right]\,,
\label{epJs}
\eq
where
\be
J_1(z)=\frac1{\sqrt{\pi}}\int_{-\infty}^{\infty} e^{-w^2} \frac{w \, dw}{w-z}\,,\quad\quad
J_2(z)=\frac1{\sqrt{\pi}}\int_{-\infty}^{\infty} e^{-w^2} \frac{w^3 \, dw}{w-z}\,,
\label{Js}
\ee
and $z=u/\sqrt{2}$.  With the standard definition of the plasma $Z$-function,
\be
Z(z)=\frac1{\sqrt{\pi}}\int_{-\infty}^{\infty} e^{-w^2} \frac{dw}{w-z}=2ie^{-z^2}\int_{-\infty}^{iz} e^{-t^2} dt
\label{Z}
\ee
where in the first expression $\Im(z)>0$ and the value of $Z$ for $\Im(z)<0$ is obtained by analytic continuation, while the second expression is valid for all complex $z$.  The second expression is desirable for computations. Also, sometimes it is convenient to use the derivative formula
\be
Z'=-2(1 + zZ)\,,
\label{Zp}
\ee
which is valid for all $z$.   After some more-or-less standard manipulations and making use of (\ref{Zp}), we obtain
\be
\epsilon= 1-\frac2{k^2}\left[ 1-2z^2 + 2z Z(z)\left(1-z^2\right)\right]\,.
\ee
 We   numerically evaluated this expression and searched for its roots to obtain Fig.~\ref{df}.  Because our system has the size $L=4\pi$, we write $\epsilon(k,\omega)$ with $k$ replaced by $k/2$.



\begin{thebibliography}{00}


\bibitem{Ayuso1} B.~Ayuso, J.~A.~Carrillo, C.-W.~Shu, {\em Discontinuous Galerkin methods for the one-dimensional Vlasov-Poisson system}, Preprint UAB. (2010).

\bibitem{Ayuso2} B.~Ayuso de Dios, J.~A.~Carrillo, C.-W.~Shu, {\em Discontinuous Galerkin methods for the multi-dimensional Vlasov-Poisson problem}, Preprint UAB. (2010). 

\bibitem{armstrong}
{\sc T.~P.~Armstrong, R.~C.~Harding, G.~Knorr, and D.~Montgomery},  {\em Solution of
Vlasov's equation by transform methods}, Methods in Computational Physics,  {\bf 9}, 29-86 (1970).

\bibitem{Babuska1}
{\sc I.~Babu\v{s}ka}, {\em The finite element method with Lagrangian multipliers},
Numerische Mathematik, {\bf 20}, 179-192 (1973).

\bibitem{Babuska2}
{\sc I.~Babu\v{s}ka}, {\em The finite element method with penalty},
Mathematics of Computation, {\bf 27}, 221-228 (1973).

\bibitem{BabZla}
{\sc I.~Babu\v{s}ka and M.~Zl\'{a}mal}, {\em Nonconforming elements in the finite element method with penalty},
SIAM Journal on Numerical Analysis, {\bf 10}, 863-875 (1973).

\bibitem{balescu} {\sc R.~Balescu}  {\em Statistical mechanics of charged particles}  (Wiley-Interscience, New York, 1963). 


\bibitem{BerGreKru}
{\sc I.~B.~Bernstein, J.~M.~Greene, and M.~D.~Kruskal}, {\em Exact nonlinear plasma oscillations}, Phys. Rev. {\bf 108}, 546-550 (1957).

\bibitem{BirLan}
{\sc C.~K.~Birdsall and A.~B.~Langdon}, {\em Plasma physics via computer simulation}
(Institute of Physics Publishing, Bristol/Philadelphia, 1991).

\bibitem{Chen}
{\sc F.~F.~Chen}, {\em Introduction to plasma physics and controlled fusion} (Plenum Press, New York, 1984).

\bibitem{CheKno}
{\sc C.~Z.~Cheng and G.~Knorr}, {\em The integration of the Vlasov equation in configuration space},
Journal of Computational Physics, {\bf 22}, 330-351 (1976).

\bibitem{CGMS-1}
{\sc  Y.~Cheng, I.~M.~Gamba, A.~Majorana and C.-W.~Shu}, 
{\em Discontinuous Galerkin solver for the semiconductor Boltzmann equation}, in 
Simulations of Semiconductor Processes and Devices, Vol.~12,  (Eds.~T.~Grasser and S.~Selberherr)   
(Springer,  Wien,  2007) pp.~257-260. 


\bibitem{CGMS-2} {\sc  Y.~Cheng, I.~M.~Gamba, A.~Majorana, and C.-W.~Shu},  {\em Discontinuous Galerkin solver for Boltzmann-Poisson transients}, Journal of Computational Electronics,  {\bf 7}, 119-123 (2008).

\bibitem{CGMS-cmame2009} {\sc  Y.~Cheng, I.~M.~Gamba, A.~Majorana, and C.-W.~Shu}, 
{\em  A discontinuous Galerkin  solver for Boltzmann Poisson systems in nano devices}, 
Comput. Methods Appl. Mech. Engrg., {\bf 198},  3130-3150 (2009).

\bibitem{CGMS-3} {\sc Y.~Cheng, I.~M.~Gamba, A.~Majorana, and C.-W.~Shu}, 
{\em A discontinuous Galerkin solver for full-band Boltzmann-Poisson models},  in 
13th International Workshop on Computational Electronics Proceedings,  211-214 (2009). 

\bibitem{CGMS-4} {\sc Y.~Cheng, I.~M.~Gamba, A.~Majorana, and C.-W.~Shu}, 
{\em Discontinuous Galerkin methods for the Boltzmann-Poisson systems in semiconductor device simulations
} AIP Conference Proceedings Volume 1333, ISBN: 978-0-7354-0889-0 , pp 890-895 (2010).

\bibitem{cheng-gamba-11} {\sc Y.~Cheng,  I.~M.~Gamba,  } {\em Numerical study of Vlasov-Poisson equations in the
simulation for infinite homogeneous stellar systems. } 
To Appear in Communications in Nonlinear Science and Numerical Simulation, (2011).

\bibitem{cheng-gamba-morrison11}
{\sc Y.~Cheng,  I.~M.~Gamba,  and P.~J.~Morrison}, {\em On Runge-Kutta discontinuous Galerkin schemes for 
Vlasov-Poisson systems},  preprint  (2011).

\bibitem{CGLM11}{ \sc Y.~Cheng,  I.~M.~Gamba,  Fengyan Li and P.~J.~Morrison}, {\em High order positive discontinuous Galerkin schemes for 
Maxwell-Vlasov systems},  in preparation,  (2011).

\bibitem{Cheng-gamba-proft-10} {\sc  Y.~Cheng, I.~M.~Gamba and J.~Proft}, {\em Positive and stable discontinuous Galerkin schemes  
for linear Vlasov-Boltzmann transport equations}, to appear in Mathematics of  Computation, (2011). 

\bibitem{CocDaw}
{\sc B.~Cockburn and C.~N.~Dawson}, {\em Some extensions of the local discontinuous Galerkin method for convection-diffusion equations},
The Proceedings of the Conference on the Mathematics of Finite Elements and Applications, Elsevier, 225-238 (2000).

\bibitem{CocHouShu}
{\sc B.~Cockburn,  S.~Hou, and C.~W.~Shu}, {\em The Runge-Kutta local projection discontinuous Galerkin finite element method for
conservation laws IV: the multidimensional case}, Mathematics of Computation, {\bf 54}, 545-581 (1990).

\bibitem{CocLinShu}
{\sc B.~Cockburn,  S.~Y.~Lin, and C.~W.~Shu}, {\em TVB Runge-Kutta local projection discontinuous Galerkin finite element method for
conservation laws III: one-dimensional systems}, Journal of Computational Physics, {\bf 84}, 90-113 (1989).

\bibitem{CocShu1}
{\sc B.~Cockburn and C.~W.~Shu}, {\em TVB Runge-Kutta local projection discontinuous Galerkin finite element method for
conservation laws II: general framework}, Mathematics of Computation, {\bf 52}, 411-435 (1989).

\bibitem{CocShu2}
{\sc B.~Cockburn and C.~W.~Shu}, {\em The local discontinuous Galerkin method for time-dependent convection-diffusion
systems}, SIAM Journal on  Numerical Analysis, {\bf 35}, 2440-2463 (1998).

\bibitem{CanGazFro}
{\sc J.~Canosa,  J.~Gazdag, and J.~E.~Fromm}, {\em The recurrence of the initial state in the numerical solution of the Vlasov equation}, Journal of Computational Physics, {\bf 15}, 34--45 (1974).

\bibitem{Daw}
{\sc J.~M.~Dawson}, {\em Particle simulation of plasmas},
Reviews of Modern Physics, {\bf 55}, 403-447 (1983).

\bibitem{DawSunWhe}
{\sc C.~Dawson, S.~Sun, and M.~F.~Wheeler}, {\em Compatible algorithms for coupled flow and transport},
Computer Methods in Applied Mechanics and Engineering, {\bf 193}, 2565-2580 (2004).

\bibitem{DenKru} {\sc J.~Denavit and W.~L.~Kruer}, {\em Comparison of numerical solutions of the {V}lasov equation with particle simulations of collisionless plasmas}, Phys.~Fluids {\bf 14}, 1782-1791 (1971).

\bibitem{Dewar} {\sc R.~L.~Dewar}, {\em Frequency shift due to trapped particles}, Phys.~Fluids {\bf 15}, 712-714 (1972).

\bibitem{Eli}
{\sc B.~Eliasson}, {\em Outflow boundary conditions for the {F}ourier transformed one-dimensional {V}lasov-{P}oisson system}, Journal of Scientific Computing, {\bf 16},1--28 (2001).

\bibitem{Evans}
{\sc L.~C.~Evans}, {\em Partial differential equations} (American Mathematical Society, Providence, Rhode Island, 1998).

\bibitem{FilSonBer}
{\sc F.~Filbet, E.~Sonnendr{\"u}cker, and P.~Bertrand}, {\em Conservative numerical schemes for the Vlasov equation},
Journal of Computational Physics, {\bf 172}, 166-187 (2001).

\bibitem{FriCon}
{\sc B.~D.~Fried and S.~D.~Conte}, {\em The plasma dispersion function}
(Academic, London, 1961).


\bibitem{gardner} {\sc  C.~S.~Gardner}, {\em Bound on the energy available from a plasma}, Phys.~Fluids {\bf 6}, 839--840 (1963).

\bibitem{GraFei} {\sc F.~C.~Grant and M.~R.~Feix}, {\em {F}ourier-{H}ermite solutions of the {V}lasov equations in the linearized limit}, Phys.~Fluids {\bf 10}, 696-702 (1967).

\bibitem{Heath}
{\sc R.~E.~Heath}, {\em Numerical analysis of the discontinuous Galerkin method applied to plasma physics}
Ph.D. dissertation, The University of Texas at Austin, Austin, 2007. Available at: 
{\tt  http://repositories.lib.utexas.edu/ bitstream/handle/2152/3065/heathr64379.pdf}.  (Note, this thesis contains a few typographical errors, but the intent is  clear.)

\bibitem{HocEas}
{\sc R.~W.~Hockney and J.~W.~Eastwood}, {\em Computer simulation using particles}
(McGraw-Hill, New York, 1981).

\bibitem{kiv}
{\sc M.~G.~Kivelson and C.~T.~Russel},   {\em Introduction to space physics} (Cambridge University Press, Cambridge, 1996) pp.~37-38.


\bibitem{Kli}
{\sc A.~J.~Klimas}, {\em A numerical method based on the {F}ourier-{F}ourier transform approach for modeling 1-D
electron plasma evolution}, Journal of Computational Physics, {\bf 50}, 270-306 (1983).

\bibitem{KliFar}
{\sc A.~J.~Klimas and W.~M.~Farrell}, {\em A splitting algorithm for Vlasov simulation with filamentation filtration},
Journal of Computational Physics, {\bf 110}, 150-163 (1994).

\bibitem{Landau}
{\sc L.~D.~Landau}, {\em On the vibrations of the electronic plasma},
Journal of Physics U.S.S.R., {\bf 10}, 25, 25-34 (1946).

\bibitem{LesRav}
{\sc P.~LeSaint and P.~A.~Raviart}, {\em On a finite element method for solving the neutron transport equation},
Mathematical Aspects of Finite Elements in Partial Differential Equations, 89-123 (1974).

\bibitem{Lifs-Pita} {\sc  E.~M.~Lifshitz and L.~P.~Pitaevskii}, {\em Physical kinetics: Course of theoretical physics  Vol. 10} ( Pergamon Press, Oxford, 1981). Translated from Russian by J. B. Sykes and R.
N. Franklin.

\bibitem{MaCaTr} {\sc A.~Mangeney, F.~Califano, C.~Cavazzoni, and P.~Tr\'{a}vn\'{i}\u{c}ek}, 
{\em A numerical scheme for the integration of the Vlasov-Maxwell system of equations,}
J.\  Comp.\ Phys., {\bf 179}, 495-538 (2002). 


\bibitem{MasFed}
{\sc V.~P.~Maslov and M.~V.~Fedoryuk}, {\em The linear theory of Landau damping},
Mathematics of the USSR-Sbornik, {\bf 55}, 2, 437-465 (1986).

\bibitem{mere}
{\sc  N.~P.~Meredith,  R.~B.~Horne,  R.~H.~A.~Iles,  R.~M.~Thorne, D.~Heynderickx, and R.~R.~Anderson},
{\em Outer zone relativistic electron acceleration associated with substorm-enhanced whistler mode chorus}, 
J.~Geophys.~Res., {\bf 107},  A7, 1144 (2001).

\bibitem{MonTid}
{\sc D.~C.~Montgomery and D.~A.~Tidman}, {\em Plasma kinetic theory} (McGraw-Hill, New York, 1964).

\bibitem{m80}
{\sc P.~J.~Morrison}, {\em The Maxwell-Vlasov equations as a continuous
Hamiltonian system}, Phys. ~Lett.\ {\bf 80A}, 383-386 (1980).

\bibitem{m82}
{\sc P.~J.~Morrison}, {\em Poisson brackets for fluids and plasmas},  in
Mathematical Methods in Hydrodynamics and Integrability in Dynamical Systems,
eds.\  M.~Tabor and Y.~Treve, American Institute of Physics Conference Proceedings No.~88, New
York (1982) pp.~13--46.

\bibitem{mp}
{\sc P.~J.~Morrison and D.~Pfirsch}, {\em Dielectric energy versus plasma energy,
and Hamiltonian action-angle variables for the Vlasov equation},  Phys.~Fluids
{\bf 4B}, 3038--3057 (1992).

\bibitem{m}
{\sc P.~J.~Morrison}, {\em Hamiltonian description of Vlasov dynamics:
action-angle variables for the continuous spectrum}, Transport  Theory and Statistical
Physics  {\bf 29}, 397--414 (2000).

\bibitem{MouVi} {\sc C.~Mouhot and C.~Villani}, {\em On Landau damping,}  arXiv:0904.2760.
To appear in Acta Mathematica (2011). 

\bibitem{NakYab}
{\sc T.~Nakamura and T.~Yabe}, {\em Cubic interpolated propagation scheme for solving the hyper-dimensional
Vlasov-Poisson equation in phase space},
Computer Physics Communications, {\bf 120}, 122-154 (1999).

\bibitem{Nitsche}
{\sc J.~A.~Nitsche}, {\em \"{U}ber ein {V}ariationsprinzip zur {L}\"{o}sung von {D}irichletproblemen
bei {V}erwendung von {T}eilr\"{a}umen, die keinen {R}andbedingun-gen unterworfen sind,}
Anh. Math. Sem. Universit\"at Hamburg, {\bf 36}, 9-15 (1971).

\bibitem{PohShoKam}
{\sc E.~Pohn,  M.~Shoucri, and G.~Kamelander}, {\em Eulerian Vlasov codes},
Computer Physics Communications, {\bf 166}, 181-93 (2005).

\bibitem{ReeHil}
{\sc T.~R.~Reed and W.~H.~Hill}, {\em Triangular mesh methods for the neutron transport equation},
LA-UR-73-479, Los Alamos Scientific Laboratory, Los Alamos, NM (1973).

\bibitem{Rein}
{\sc G.~Rein}, {\em Collisionless kinetic equations from astrophysics - the Vlasov-Poisson system}, in 
Handbook of Differential Equations, Evolutionary Equations, Vol.~3, (Eds.~C.~M. Dafermos and E.~Feireisl),
(Elsevier, 2007).

\bibitem{RivWheGir}
{\sc B.~Riviere, M.~F.~Wheeler, and V.~Girault}, {\em A priori error estimates for finite elements based on discontinuous approximation spaces for elliptic problems},
SIAM Journal on Numerical Analysis, {\bf 39}, 902-931 (2001).

\bibitem{Sch}
{\sc H.~Schamel}, {\em Electron holes, ion holes and double layers}, Phys. Repts. {\bf 140}, 161-191 (1986).

\bibitem{Sho}
{\sc M.~Shoucri}, {\em The method of characteristics for the solution of hyperbolic differential equations},
Computer Physics Research Trends, (Nova Science, 2007), 1-82.

\bibitem{Stein}
{\sc E.~M.~Stein and G.~Weiss}, {\em Introduction to {F}ourier analysis on {E}uclidean spaces} (Princeton University Press, Princeton, NJ, 1971).

\bibitem{Sun}
{\sc S.~Sun}, {\em Discontinuous Galerkin methods for reactive transport in porous media}
(Ph.D. dissertation, The University of Texas at Austin, Austin, 2003).

\bibitem{SunWhe}
{\sc S.~Sun and M.~F.~Wheeler}, {\em Symmetric and nonsymmetric discontinuous Galerkin methods for reactive transport in porous media},
SIAM Journal on Numerical Analysis, {\bf 43}, 195-219 (2005).



\bibitem{Va} {\sc  F.~Valentini}, 
{\em Nonlinear Landau damping in nonextensive statistics,}
Phys.\ Plasmas, {\bf 12}, 072106 (2005). 



\bibitem{VaDa} {\sc F.~Valentini, and R.~D'Agnosta}, 
{\em Electrostatic Landau pole for $\kappa$-velocity distributions,}
Phys.\ Plasmas, {\bf 14}, 092111 (2007). 


\bibitem{VaTrCaHeMa} {\sc  F.~Valentini, P.~Tr\'{a}vn\'{i}\u{c}ek, F.~Califano, P.~Hellinger, and A.~Mangeney}, 
{\em A hybrid-Vlasov model based on the current advance method for the simulation of collisionless magnetized plasma,}
J.\  Comp.\ Phys., {\bf 225}, 753-770 (2007).  



\bibitem{VaVeMa} {\sc F.~Valentini, P.~Veltri, and A.~Mangeney}, 
{\em A numerical scheme for the integration of the Vlasov-Poisson system of equations, in the magnetized case,}
J.\  Comp.\ Phys., {\bf 210}, 730-751 (2005). 


\bibitem{vk} {\sc N.~G.~Van Kampen}, {\em On the theory of stationary waves in plasmas},  Physica,  {\bf 21}, 949-963 (1955).

\bibitem{Wheeler}
{\sc M.~F.~Wheeler}, {\em An elliptic collocation-finite element method with interior penalties},
SIAM Journal on Numerical Analysis, {\bf 15}, 152-161 (1978).

\bibitem{WhePer}
{\sc M.~F.~Wheeler and P.~Percell}, {\em A local residual finite element procedure for elliptic equations},
SIAM Journal on Numerical Analysis, {\bf 15}, 705-714 (1978).

\bibitem{ZakGarBoy1}
{\sc S.~I.~Zaki, L.~R.~T.~Gardner, and T.~J.~.M.~Boyd}, {\em A finite element code for the simulation of
one-dimensional Vlasov plasmas. I. Theory}, Journal of Computational Physics, {\bf 79}, 184-199 (1988).

\bibitem{ZakGarBoy2}
{\sc S.~I.~Zaki, L.~R.~T.~Gardner, and T.~J.~.M.~Boyd}, {\em A finite element code for the simulation of
one-dimensional Vlasov plasmas. II. Applications}, Journal of Computational Physics, {\bf 79}, 200-208 (1988).

\bibitem{zhang-shu10-1}
{\sc Q.~Zhang and C.~W.~Shu},
{\em Stability analysis and a priori error estimates to the third order
  explicit {Runge-Kutta} discontinuous {Galerkin} Method for scalar conservation laws},
 { SIAM J. Numer. Anal. {\bf 48} (2), 772-795 (2010).}


\bibitem{zhang-shu10-2}
 {\sc Q.~Zhang and C.~W.~Shu},
{\em On maximum-principle-satisfying high order schemes for scalar conservation laws},
{J. Comput. Phys.{\bf 229} 3091-3120, (2010). }

 

\bibitem{ZhoGuoShu}
{\sc T.~Zhou, Y.~Guo, and C.~W.~Shu}, {\em Numerical study on Landau damping}, Physica D, {\bf 157}, 322-333 (2001).


\end{thebibliography}
\end{document}